\documentclass[fleqn,usenatbib]{mnras}

\usepackage[T1]{fontenc}
\usepackage{ae,aecompl}

\usepackage{graphicx}	
\usepackage{amsmath}	
\usepackage{amssymb}	
\usepackage{enumerate}   
\usepackage{color}
\usepackage{ulem}
\usepackage{multirow}   
\usepackage{tabularx}

\usepackage{lipsum} 
\usepackage{newtxtext,newtxmath} 

\defcitealias{keszthelyi2019}{\nobreak{Paper~I\,}} 
\newcommand{\PaperI}{\citetalias{keszthelyi2019}} 
\defcitealias{keszthelyi2020}{\nobreak{Paper~II\,}} 
\newcommand{\PaperII}{\citetalias{keszthelyi2020}} 

\title[The effects of surface fossil magnetic fields on massive star evolution: III]{The effects of surface fossil magnetic fields on massive star evolution: III. The case of $\tau$ Sco}

\author[Z. Keszthelyi et al.]{
Z. Keszthelyi$^{1}$\thanks{E-mail: z.keszthelyi@uva.nl},
G. Meynet$^{2}$,
F. Martins$^{3}$,
A. de Koter$^{1,4}$,
A. David-Uraz$^{5,6,7}$
\\\\  
$^{1}$Anton Pannekoek Institute for Astronomy, University of Amsterdam, Science Park 904, 1098 XH, Amsterdam, The Netherlands \\
$^{2}$Geneva Observatory, University of Geneva, Maillettes 51, 1290 Sauverny, Switzerland \\
$^{3}$LUPM, Universit\'e de Montpellier, CNRS, Place Eug\`ene Bataillon, F-34095 Montpellier, France \\
$^{4}$Institute of Astronomy, KU Leuven, Celestijnenlaan 200D, 3001 Leuven, Belgium \\ 
$^{5}$Department of Physics and Astronomy, University of Delaware, Newark, DE 19716, USA \\ 
$^{6}$Department of Physics and Astronomy, Howard University, Washington, DC 20059, USA \\ 
$^{7}$Center for Research and Exploration in Space Science and Technology, and X-ray Astrophysics Laboratory, NASA/GSFC, Greenbelt, MD 20771, USA \\ 
}

\date{Accepted XXX. Received YYY; in original form ZZZ}

\pubyear{2021}

\begin{document}
\label{firstpage}
\pagerange{\pageref{firstpage}--\pageref{lastpage}}
\maketitle

\begin{abstract}
$\tau$ Sco, a well-studied magnetic B-type star in the Upper Sco association, has a number of surprising characteristics. It rotates very slowly and shows nitrogen excess. Its surface magnetic field is much more complex than a purely dipolar configuration which is unusual for a magnetic massive star. We employ the \textsc{cmfgen} radiative transfer code to determine the fundamental parameters and surface CNO and helium abundances. Then, we employ \textsc{mesa} and \textsc{genec} stellar evolution models accounting for the effects of surface magnetic fields.
To reconcile $\tau$~Sco's properties with single-star models, an increase is necessary in the efficiency of rotational mixing by a factor of 3 to 10 and in the efficiency of magnetic braking by a factor of 10. 
The spin down could be explained by assuming a magnetic field decay scenario. However, the simultaneous chemical enrichment challenges the single-star scenario. Previous works indeed suggested a stellar merger origin for $\tau$~Sco.
However, the merger scenario also faces similar challenges as our magnetic single-star models to explain $\tau$~Sco's simultaneous slow rotation and nitrogen excess. In conclusion, the single-star channel seems less likely and versatile to explain these discrepancies, while the merger scenario and other potential binary-evolution channels still require further assessment as to whether they may self-consistently explain the observables of $\tau$~Sco.

\end{abstract}

 

\section{Introduction} \label{sec:intro}

%
$\tau$ Scorpii (HD~149438) is a magnetic massive B0.2\,V star in the Upper Sco association, which has been well-studied for almost an entire century \citep[e.g.,][]{struve1933,unsold1942,traving1955,aller1966,lamers1978,wolff1985,kilian1992,howk2000}. In addition to a large number of observations in the optical band, there is also a wealth of multiwavelength observations, including X-ray \citep[e.g.,][]{macfarlane1989,cohen1997,cohen2003,dem2001,mewe2003,ignace2010,naze2014,fletcher2018}, ultraviolet \citep[e.g.,][]{walborn1984,peters1985,rogerson1985,cowley1987,snow1994}, and
infrared \citep[e.g.,][]{waters1993,zaal1999,repolust2005}.  

Spectropolarimetric observations of $\tau$ Sco by \cite{donati2006} led to the discovery of a surface magnetic field, which is unusually complex compared to other B-type stars whose field measurements can usually be reconciled with a dipolar configuration \citep[e.g.,][]{petit2013,shultz2018}. Additional observations confirmed these findings, showing that the magnetic energy density indeed resides in higher order spherical harmonic components, which clearly implies that the field is complex \citep{donati2009,shultz22019,shultz32019}.
Based on the circular (Stokes~\textit{V}) spectropolarimetric data set, \cite{kochukhov2016} produced possible magnetic field maps of $\tau$~Sco. However, the specific geometry needs to be known to produce a unique magnetic field map. This relies on further observations using linear polarimetry (Stokes~\textit{Q, U}) to constrain the magnetic field modulus.

Interestingly, despite the non-uniqueness problem, the magnetic field maps produced from the spectropolarimetric observations are quite reminiscent of the magnetohydrodynamic (MHD) simulations of \cite{braithwaite2008} who demonstrated that an initial seed field can relax into a stable non-axisymmetric equilibrium. Indeed, the decade-long spectropolarimetry suggests that the field is stable and of fossil origin. 

Using the magnetic oblique rotator model \citep{stibbs1950}, the rotation period of the star is constrained to 41.033 $\pm$ 0.002 days \citep{donati2006}. Depending on the assumed stellar radius, this yields a surface equatorial rotational velocity of $\approx 5$~km~s$^{-1}$. Considering measurements of the projected rotational velocity ($< 10$~km~s$^{-1}$, e.g.,  \citealt{hardorp1970,mokiem2005,simondiaz2006,nieva2014,cazorla2017a}, and see also Appendix~\ref{sec:app2}), it is clear that the present-day surface rotation of $\tau$~Sco is very slow. 

Several authors, e.g., \cite{kilian1992}, \cite{morel2008}, \cite{przybilla2010}, and \cite{martins2012} measured the surface abundance of CNO elements. Their results  are in excellent agreement, strongly indicating nitrogen excess (log~$[N/H] + 12 = 8.15 \pm 0.15$, $8.15 \pm 0.20$, $8.16 \pm 0.12$, and $8.15 \pm 0.06$ by number fraction, respectively). This means that $\tau$~Sco is highly enriched in core-processed material, by at least a factor of 3 compared to the solar baseline. Helium abundance measurements are more uncertain. Previously obtained values (e.g., He/H = 0.085 by number fraction from \citealt{wolff1985}; He/H = 0.10 $\pm$ 0.025  from \citealt{kilian1992}; $Y ~=~0.28~\pm~0.03$ by mass fraction from \citealt{przybilla2010}) are very close to the solar helium baseline \citep{grevesse1996,asplund2005,asplund2009}, which would be compatible with expectations of a single star in its early main sequence evolution. Although still overlapping within uncertainty, \citealt{mokiem2005} obtained He/H~=~0.12$^{+0.04}_{-0.02}$ by number fraction, which could point towards a slight excess in the surface helium abundance.
 
\cite{nieva2014} measured the stellar parameters of $\tau$~Sco and based on its position on the Hertzsprung-Russell diagram (HRD) concluded that it is a blue straggler star that is much younger than the association it belongs to, suggesting that it could possibly originate from a stellar merger. 
Indeed, \cite{ferrario2009} suggested that the origin of fossil magnetic fields may be stellar merger events. Further work by \cite{schneider2016} explored this scenario and argued that $\tau$ Sco could be rejuvenated via a merger process. 

Recently \cite{schneider2019} presented 3D MHD simulations of stellar mergers and \cite{schneider2020} implemented the results into 1D stellar evolution models to follow the long-term evolution of the post-merger object. These simulations showed that \textit{i)} a seed field can be amplified to a strong magnetic field in the merger process and \textit{ii)} the post-merger object evolves seemingly as a single star, although with an unusual internal chemical composition. Nonetheless, some important observables of $\tau$ Sco (such as the $\approx$~5 km\,s$^{-1}$ rotational velocity and the factor of 3 nitrogen enrichment) are not yet compatible with the current modelling efforts as they predict either modest rotation (50 km\,s$^{-1}$, which is still an order of magnitude larger than the observed value) with no nitrogen excess, or fast rotation (400 km\,s$^{-1}$) with nitrogen excess (see Figures 3 and 5 of \citealt{schneider2020}). The only solution thus far that predicts sufficiently slow rotation (but no nitrogen excess) is a model with a constant surface magnetic dipole moment of $\mu_B = 10^{40}$~G~cm$^3$ corresponding to an approximately 270~kG field at the surface of a star with a radius of 5 R$_\odot$. Such a strong field is reached in the 3D MHD merger simulation \citep{schneider2019}, however, since the field strength remains quasi-constant on the main sequence, it is far too strong compared to any known OBA star and, in particular, to $\tau$~Sco's measured surface field strength of a few hundred G (\citealt{donati2006,shultz2018}, and further details in Appendix~\ref{sec:app2}).

These significant findings pose important questions, for example, \textit{i)} are all fossil magnetic fields generated via a stellar merger event, and \textit{ii)} can we confidently identify signs of a past stellar merger event and its evolutionary consequences when studying a star? This makes $\tau$ Sco a valuable laboratory to further our understanding of magnetic massive stars and for this reason the current discrepancies between models and observations need to be studied in more detail.


The prospect of $\tau$ Sco being a merger product is intriguing. Of course, the validity of this scenario would be strengthened if one can firmly exclude the possibility that the star formed and evolved in isolation, having acquired its fossil field during the assembly process. The purpose of this work therefore is, first, to re-assess stellar and surface properties, including its age and nitrogen abundance, and to comprehensively explore whether these empirical characteristics can be reconciled with single-star evolutionary models that account for surface magnetic field effects; and, second, to discuss implications of these properties for merger models. 

This paper is part of a series in which we aim to explore the effects of surface fossil magnetic fields on massive star evolution. 
In the first paper of the series \citep[][hereafter \PaperI]{keszthelyi2019}, we used the Geneva stellar evolution code \citep{eggenberger2008,ekstroem2012,georgy2013,groh2019,murphy2021} to explore the cumulative impact of magnetic mass-loss quenching, magnetic braking, and field evolution. 
In the second paper \citep[][hereafter \PaperII]{keszthelyi2020}, we implemented and studied massive star magnetic braking in the \textsc{mesa} software instrument \citep{paxton2011,paxton2013,paxton2015,paxton2018,paxton2019}, detailing the magnetic and rotational evolution, and confronting the models with a sample of observed magnetic B-type stars from \cite{shultz2018}. A key finding of \PaperII is that presently slowly-rotating magnetic stars may originate from either slow or fast rotators at the Zero Age Main Sequence (ZAMS).

The paper is organised as follows. In Section \ref{sec:sec4}, we perform atmospheric modelling, particularly focusing on the CNO elements and helium abundance. In Section \ref{sec:sec5}, we carry out numerical experiments with stellar evolution models. In Section \ref{sec:sec6}, we discuss the impact of the initial assumptions and confront empirical evidence with our results. Finally, we summarise our findings in Section \ref{sec:sec7}.

%
%
%
%
%
%
%



%
%
%
\begin{figure*}
\includegraphics[width=0.99\textwidth]{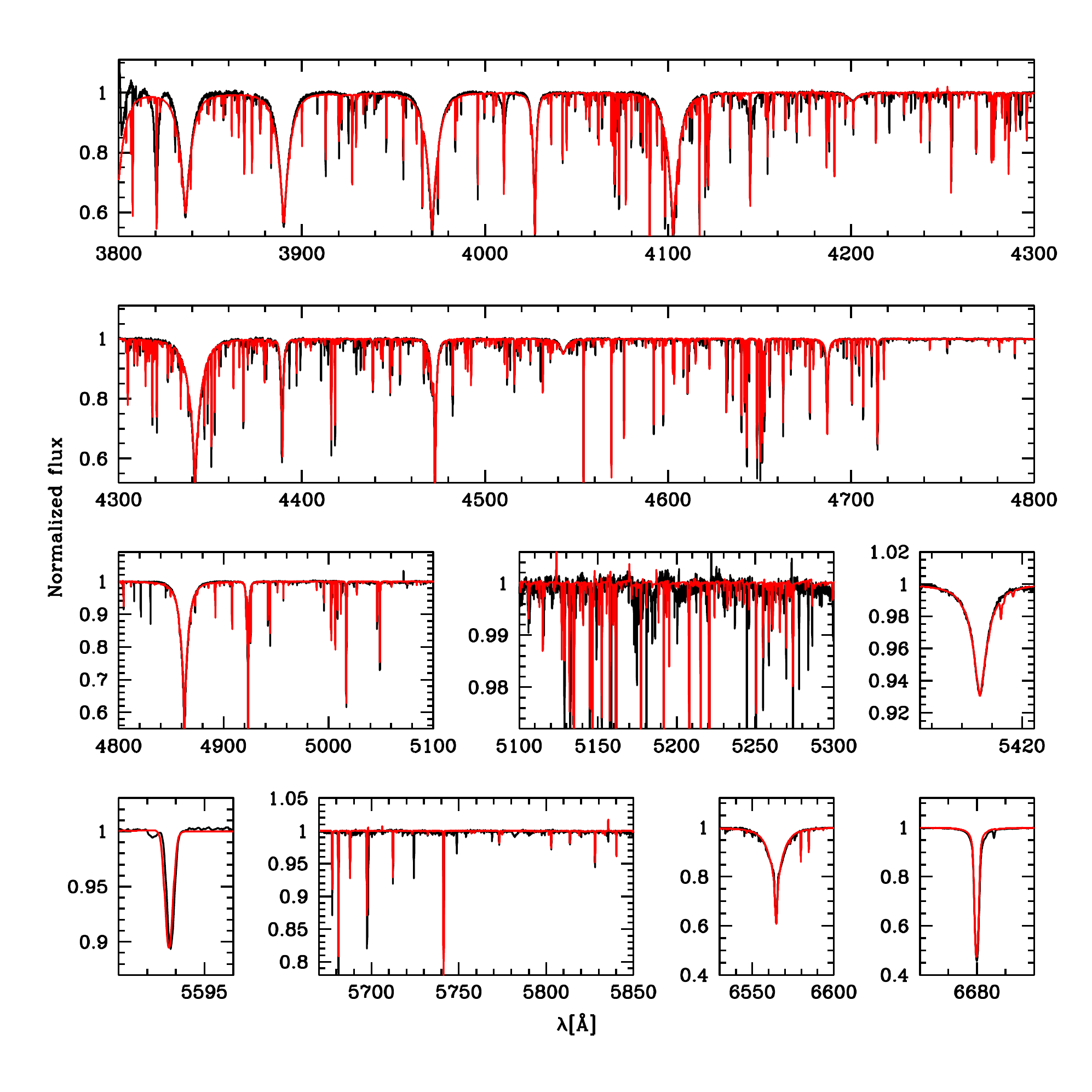}
\caption{Best fit model (red) compared to the observed spectrum (black) of $\tau$~Sco.}\label{fig:spec}
\end{figure*}
%
%
%

%
%
%
%
%
%
%

\section{Atmospheric modelling}\label{sec:sec4}

\subsection{Determination of fundamental parameters} 
\label{s_param}

We have used the code \textsc{cmfgen} \citep{hm98} to determine the fundamental stellar parameters of $\tau$~Sco. \textsc{cmfgen} computes atmosphere models in non-LTE, spherical geometry, and includes outflows and line-blanketing. A complete description of the code is presented in \citet{hm98} and we refer the reader to this publication for further information. 

To the best of our knowledge there is no atmosphere code for hot, massive stars (including non-LTE effects, sphericity, and winds) which includes polarisation in the radiative transfer equation. For lower mass stars, \cite{khan2006} showed that the extra line-blanketing caused by Zeeman splitting had little impact on the emergent spectrum, especially in the field strength regime of $\tau$~Sco. The expected Zeeman splitting of helium and CNO lines is undetectable in individual lines unless the magnetic field is extremely strong (which is why field-detection techniques relying on many lines are used, \citealt{donati2009}), hence the use of non-polarised synthetic spectra is justified. \cite{krticka2018} showed that polarised radiative transfer has little effect on radiative driving so that mass-loss rates should not be affected. However, the presence of a magnetic field confines the expanding wind, leading to non-spherical effects that need to be taken into account when studying the details of stellar winds in magnetic OB stars \citep[e.g.,][]{ud2017}. In the present study, we focus on photospheric parameters which should not be severely affected by the use of "non-magnetic" models.

To determine the main fundamental parameters, we relied on a grid of \textsc{cmfgen} models covering the effective temperatures and surface gravities of hot main sequence stars. We also used the spectrum of $\tau$~Sco presented by \citet{martins2012}. It is an average spectrum built from observations conducted with the spectropolarimeter ESPaDOnS at CFHT. It covers the wavelength range 3800-6800 \AA\ at a spectral resolution of about 65000. In this spectrum, we isolated the classical indicators to determine the effective temperature and surface gravity, namely \ion{He}{i}~4471, \ion{He}{ii}~4542, \ion{He}{ii}~5412, H$\gamma$ and H$\beta$. We subsequently looked for the best fit models using a $\chi^2$ minimisation process. We found $T_{\rm eff}$=31500$\pm$1000~K and $\log g$= 4.2$\pm$0.1 cm\,s$^{-2}$. For consistency, the process was repeated with additional helium and Balmer lines and consistent results were obtained. 

In this process, we used a projected rotational velocity of 9~km\,s$^{-1}$ and a macroturbulence of 10~km\,s$^{-1}$. The former was obtained from the Fourier transform method \citep[see][]{ss07} applied to \ion{O}{iii}~5592 and \ion{He}{i}~4713. To determine the level of macroturbulence we convolved spectra from our grid and with parameters close to the final values with a rotational profile, an instrumental one, and a radial-tangential profile mimicking macroturbulence. We constrained the macroturbulence velocity by fitting \ion{He}{i}~4713. 

To constrain the luminosity of $\tau$~Sco, we computed the bolometric correction from $T_{\rm eff}$ and the relation of \citet{mp06}. The extinction was estimated from the color excess E(B-V) using the intrinsic (B-V)$_0$ color of \citet{mp06}. With the new distance of $d =$~195$\pm$42~pc from Gaia DR2, we finally obtained $\log (L/L_{\odot})=4.56 \pm 0.20$. Using the radius obtained from $T_{\rm eff}$ and $L/L_{\odot}$ as well as the projected rotational velocity, and assuming an inclination of about $90^{\circ}$, we find a rotation period of 36 days. This is in excellent agreement with the 41 days found by \citet{donati2006} from spectropolarimetric measurements. 
\begin{table}
\caption{Stellar parameters and surface abundances in number fractions obtained from atmospheric modelling.}
\centering
\label{tab:t1}
\begin{tabular}{l|c}  
\hline\hline\\[-1.5ex]
$\log (L/L_{\odot})$ & 4.56$\pm$0.20 \\[2pt]
$T_{\rm eff}$ [kK] & 31.5$\pm$ 1  \\[2pt]
log $g$ [cm s$^{-2}$]& 4.2$\pm$0.1 \\[2pt]
\hline\\[-1.5ex]
$\varv \sin i$ [km\,s$^{-1}$] & 9 \\[2pt]
$\varv_{\rm mac}$ [km\,s$^{-1}$] & 10 \\[2pt]
$\varv_{\rm mic}$ [km\,s$^{-1}$] &2 \\
\hline\\[-1.5ex]
C\,/\,H  & $1.8^{+0.9}_{-0.7} \cdot 10^{-4}$  \\[2pt]
N\,/\,H & $1.8^{+0.8}_{-0.6} \cdot 10^{-4}$  \\[2pt] 
O\,/\,H  & $4.1^{+2.0}_{-1.3} \cdot 10^{-4}$  \\[2pt]
He\,/\,H & 0.11$\pm$0.01\\
\hline\hline
\end{tabular} 
\end{table}
\subsection{Surface abundances} 
\label{s_ab}

We determined the surface abundances of carbon, nitrogen, oxygen, and helium. For each element, we computed atmosphere models and synthetic spectra with the set of fundamental parameters derived in Sect.\ \ref{s_param} but different chemical compositions. We then used a set of selected lines of each element to perform a $\chi^2$ minimisation and find the best fitting model. These lines included in the analysis are:

\begin{itemize}

\item Carbon: \ion{C}{iii}~4056, \ion{C}{iii}~4068, \ion{C}{iii}~4070, \ion{C}{ii}~4075 \ion{C}{iii}~4665, \ion{C}{iii}~4667, \ion{C}{ii}~5133,  \ion{C}{ii}~5144, \ion{C}{ii}~5151,  \ion{C}{ii}~5250, \ion{C}{iii}~5254, \ion{C}{iv}~5802, \ion{C}{iv}~5812, \ion{C}{iii}~5827, \ion{C}{iii}~6205, \ion{C}{ii}~6578, \ion{C}{ii}~6583, \ion{C}{iii}~6744.

\item Nitrogen: \ion{N}{ii}~3995, \ion{N}{ii}~4004, \ion{N}{ii}~4035, \ion{N}{iii}~4044, \ion{N}{iii}~4196, \ion{N}{iii}~4216, \ion{N}{ii}~4447, \ion{N}{iii}~4511, \ion{N}{iii}~4515, \ion{N}{ii}~4602, \ion{N}{ii}~4607, \ion{N}{iii}~4615, \ion{N}{ii}~4621, \ion{N}{iii}~4634, \ion{N}{iii}~4640, \ion{N}{ii}~4788, \ion{N}{ii}~4803, \ion{N}{ii}~4995, \ion{N}{ii}~5001, \ion{N}{ii}~5005, \ion{N}{ii}~5011, \ion{N}{ii}~5026, \ion{N}{ii}~5045, \ion{N}{ii}~5678, \ion{N}{ii}~5680.
   
\item Oxygen:  \ion{O}{ii}~3913, \ion{O}{ii}~3954, \ion{O}{ii}~3983, \ion{O}{ii}~4277, \ion{O}{ii}~4278, \ion{O}{ii}~4283, \ion{O}{ii}~4305, \ion{O}{ii}~4316, \ion{O}{ii}~4318, \ion{O}{ii}~4321, \ion{O}{ii}~4354, \ion{O}{ii}~4367, \ion{O}{ii}~4370, \ion{O}{ii}~4397, \ion{O}{ii}~4415, \ion{O}{ii}~4417, \ion{O}{ii}~4453, \ion{O}{ii}~4489, \ion{O}{ii}~4491,  \ion{O}{ii}~4592, \ion{O}{ii}~4597, \ion{O}{ii}~4603, \ion{O}{ii}~4610, \ion{O}{ii}~4662, \ion{O}{ii}~4676, \ion{O}{ii}~4678,
\ion{O}{ii}~4700, \ion{O}{ii}~4705, \ion{O}{iii}~5592.

\end{itemize}

\noindent We obtained the following values of surface abundances in number fractions: C/H=$1.8^{+0.9}_{-0.7} \cdot 10^{-4}$, N/H$=1.8^{+0.8}_{-0.6} \cdot 10^{-4}$ and O/H$=4.1^{+2.0}_{-1.3} \cdot 10^{-4}$. For these determinations we used a microturbulent velocity of 2 km\,s$^{-1}$. A larger value would lead to excessive broadening, as, for instance, evidenced in the \ion{N}{ii} doublet at 5001 \AA. The reduced microturbulence compared to the study of \citet{martins2012} explains the slightly larger N/H ($1.8 \cdot 10^{-4}$ in the present study versus $1.4 \cdot 10^{-4}$ by \citealt{martins2012}). The best fit synthetic spectrum is shown in Figure\ \ref{fig:spec}.

We also investigated the possibility that $\tau$~Sco is enriched in helium. To this extent, we ran models with He/H between 0.1 (the reference value) and 0.2. We used the following lines to estimate the goodness of fit: 

\begin{itemize}
\item Helium: \ion{He}{i}~4026, \ion{He}{ii}~4200, \ion{He}{i}~4388, \ion{He}{i}~4471, \ion{He}{ii}~4542, \ion{He}{i}~4713, \ion{He}{i}~4921, \ion{He}{ii}~5412, \ion{He}{i}~6680.
\end{itemize}

The best fits were obtained for models with He/H\,$=$\,0.11$\pm$0.01. This corresponds to Y\,$=$\,0.30$\pm$0.02 in mass fraction, implying a modest enrichment compared to the baseline value of Y$
_{\rm ini}$\,$=$\,0.266. The results are summarised in Table~\ref{tab:t1}.

%
%
%
%
%
%

\section{Evolutionary modelling}\label{sec:sec5}

Evolutionary models were computed with two different codes, Modules for Experiments in Stellar Astrophysics (\textsc{mesa}, \citealt{paxton2011,paxton2013,paxton2015,paxton2018,paxton2019}) and the Geneva stellar evolution code (\textsc{genec}, \citealt{eggenberger2008,ekstroem2012,georgy2013}; \PaperI). The two codes are sufficiently different that they provide independent tests. In particular, the implementations of angular momentum transport and loss are different (see \PaperII{} Appendix A).

%
%
\subsection{\textsc{mesa} modelling: Setup}

The general model parameters we adopt in \textsc{mesa} are as follows. A solar metallicity of $Z = 0.014$ is adopted with the \cite{asplund2009} mixture of metals; isotopic ratios are from \nobreak\cite{lodders2003}. The mixing efficiency in the convective core is treated by adopting $\alpha_{\rm MLT} = 2.0$. Exponential overshooting is used above the convective core with $f_{\rm ov} = 0.034$ and $f_0 = 0.006$ \citep{herwig2000,paxton2013}. This would approximately correspond to extending the convective core by 25\% of the local pressure scale height (thus $\alpha_{\rm ov} = 0.25$) in a non-rotating model\footnote{A range of $\alpha_{\rm ov} = 0.1 - 0.5$ is presently considered plausible based on asteroseismic measurements as well as multidimensional hydrodynamic simulations \citep[e.g.,][]{moravveji2015,papics2017,herwig2000,meakin2007}. See recently \cite{kaiser2020} and references therein. We test other values in Appendix~\ref{sec:app1}, nonetheless, the precise choice of the overshooting parameter does not play an important role here as without efficient envelope mixing, the surface composition will remain unaltered.}. The OPAL opacity tables are used \citep{rogers1992}. The mass-loss scheme is adopted from \cite{dej1988} \footnote{This specific choice is made because the \citet{dej1988} mass-loss rates in this mass and temperature range are slightly higher than the \citet{vink2001} rates, consequently they (very minimally) also aid the spin-down of the star \citep[e.g.,][]{keszthelyi2017b}. However, we tested models with both mass-loss prescriptions and found that they lead to completely negligible differences in this study. This is expected since the mass-loss rates during the early evolution of the models are on the order of $\dot{M} \approx 10^{-9} - 10^{-8}$~M$_\odot$\,yr$^{-1}$.}. This mass-loss rate is then systematically scaled by the magnetic mass-loss quenching parameter $f_B$ and rotational enhancement $f_{\rm rot}$ (see \PaperII). 

To account for rotationally-induced instabilities which lead to chemical mixing, we use \textsc{mesa}'s default parametrisation in a fully diffusive approach. The diffusion coefficients arising from dynamical and secular shear instabilities \citep{endal1978,pin1989}, Eddington-Sweet circulation \citep{eddington1925,sweet1950}, Solberg-H\o iland \citep{solberg1936,hoiland1941}, and Goldreich-Schubert-Fricke instabilities \citep{goldreich1967,fricke1968} are included. The Spruit-Tayler dynamo \citep{spruit2002,tayler73} is not adopted. To consider the inhibiting impact of composition gradients, $\nabla \mu$ is scaled by $f_\mu = 0.05$ when calculating stability criteria for chemical mixing \citep{yoon2006,brott2011,paxton2013}. 
%

%
%
In \textsc{mesa}, angular momentum transport is modelled as a diffusive process. This means that angular momentum is only transported in one (outward) direction. The stellar core has a constant angular velocity profile. Close to the core-envelope boundary (at $q = 0.4$, i.e. at the layer encompassing 40\% of the total mass), we set the lower boundary to apply magnetic braking (see below). Therefore we assume that the magnetic field spreads through the stellar envelope, leading to a very efficient angular momentum transport. An appropriate transport equation, however, relies on the properties of the internal magnetic field (e.g., strength, geometry, obliquity), which are unknown. Instead, we adopt a uniformly high diffusion coefficient ($D = 10^{16}$~cm$^2$~\,s$^{-1}$) throughout the stellar envelope (from the photosphere down to $q = 0.4$) to account for the putative effect of the magnetic field\footnote{This choice has a negligible impact on the computations as the nominal diffusion coefficients in \textsc{mesa} lead to near solid-body rotation on the main sequence anyway. As the core shrinks and becomes less heavy during the evolution, the fixed boundary $q = 0.4$ shifts a little farther from the core.}. In the intermittent region between the core-envelope boundary and $q = 0.4$, the diffusion coefficient for angular momentum transport is given by rotationally-induced instabilities, dominated by the Eddington-Sweet circulation and shear instabilities. (In Appendix \ref{sec:app1}, we test whether different overshooting parameters result in any significant changes due to more efficient mixing in this region.) In practice, the initial configuration is very close to solid-body rotation, however, some differential rotation develops between the stellar core and the surface over time.

%
%
To parametrise the impact of the surface magnetic field, we use the \texttt{run\_star\_extras} file developed in \PaperII{} and shared through zenodo at \url{https://doi.org/10.5281/zenodo.3250412} and \url{https://doi.org/10.5281/zenodo.3734209}. In brief, this extension accounts for mass-loss quenching, magnetic field evolution, and magnetic braking (see \citealt{petit2017}, \PaperI{}and \PaperII{}, and references therein).

For our model computations in Sections~\ref{sec:sec41} to \ref{sec:sec44}, we set the equatorial magnetic field strength to $B_{\rm eq} = 300 $~G (corresponding to a 600 G polar field strength) and consider it constant in time. This field strength is chosen to match the currently measured average field of $\tau$~Sco (see more discussion in Appendix~\ref{sec:app2}). In contrast, in Section~\ref{sec:45}, we change the initial field strength and investigate magnetic field evolution.

In the present work, we focus on the efficiency of magnetic braking, therefore we introduce the arbitrary scaling factor $f_{\rm MB}$ to the formalism used in \PaperII{}, such that magnetic braking is considered via changing the specific angular momentum as:
\begin{equation}\label{eq:eq1}
    \sum_{k=1}^{k=x} \frac{\mathrm{d j_{\rm B}}}{\mathrm{d} t} = - f_{\rm MB} \frac{J_{\rm brake}}{J_{\rm envelope}} \sum_{k=1}^{k=x}  \frac{\mathrm{d}j}{\mathrm{d}t} \, , 
\end{equation}
\noindent where $\mathrm{d}j_{\rm B}/\mathrm{d} t$ is the rate of specific angular momentum change (dubbed as "\texttt{extra\_jdot}" in \textsc{mesa}), the negative sign is added to reduce the reservoir (i.e., to account for loss), $J_{\rm brake}$ is the total angular momentum lost per unit time, $J_{\rm envelope}$ is the angular momentum reservoir of the stellar envelope, $j$ is the specific angular momentum of a layer (called "\texttt{j\_rot}" in \textsc{mesa}), and d$t$ is one time-step in the computation\footnote{We use a time-step control, specified in \PaperII{}, which prevents the star model from fully exhausting specific angular momentum in any layer. Furthermore, we set the \textsc{mesa} control \texttt{max\_years\_for\_timestep~=~8.d3} to avoid large time steps.}. The summation goes over all layers of the stellar envelope from the surface ($k=1$) to the lower boundary close to the stellar core ($x \approx 1500$ zones out of typically 2000 zones of the stellar structure model, corresponding to the location where $q = 0.4$). We experimented with changing the value of $q$, and found that it plays an insignificant role on our \textsc{mesa} results within the present setup. Nonetheless, we note that this assumption is different than the ones taken in \PaperII{}, where the torque was either applied for the entire star or only to a very small near-surface reservoir. In the \textsc{genec} models, only the surface is ascribed to brake its rotation by the magnetic field and we will contrast the \textsc{mesa} models with those.
The quantity $J_{\rm brake}$ is obtained using the prescription of \citealt{ud2009}, such that:
\begin{equation}
J_{\rm brake} = \frac{\mathrm{d}J_{\rm B}}{\mathrm{d}t}  \mathrm{d}t = \frac{2}{3} \dot{M}_{B=0} \, \Omega_\star R_{A}^2 \mathrm{d}t \, ,
\end{equation}
where $\dot{M}_{B=0}$ is the mass-loss rate the star would have in absence of a magnetic field and, $\Omega_\star$ is the surface angular velocity, $R_{A}$ is the Alfv\'en radius, and the rate of angular momentum loss $\mathrm{d}J_{\rm B}/\mathrm{d}t$ has been found to be in good agreement with the formalism developed by \cite{weber1967}.

%
%
%
%
%
\begin{figure*}
\includegraphics[width=7cm]{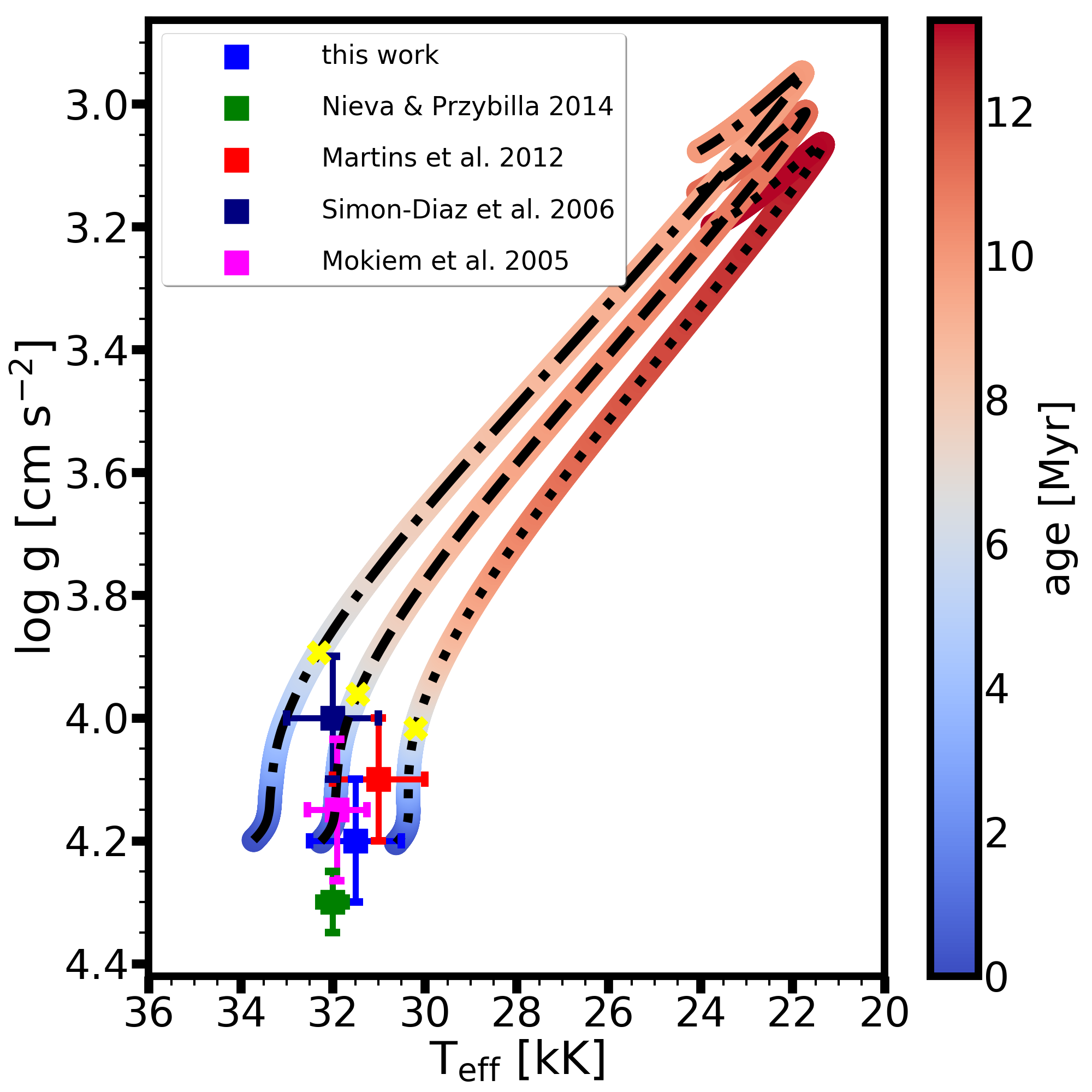}\hspace{1em}\includegraphics[width=7cm]{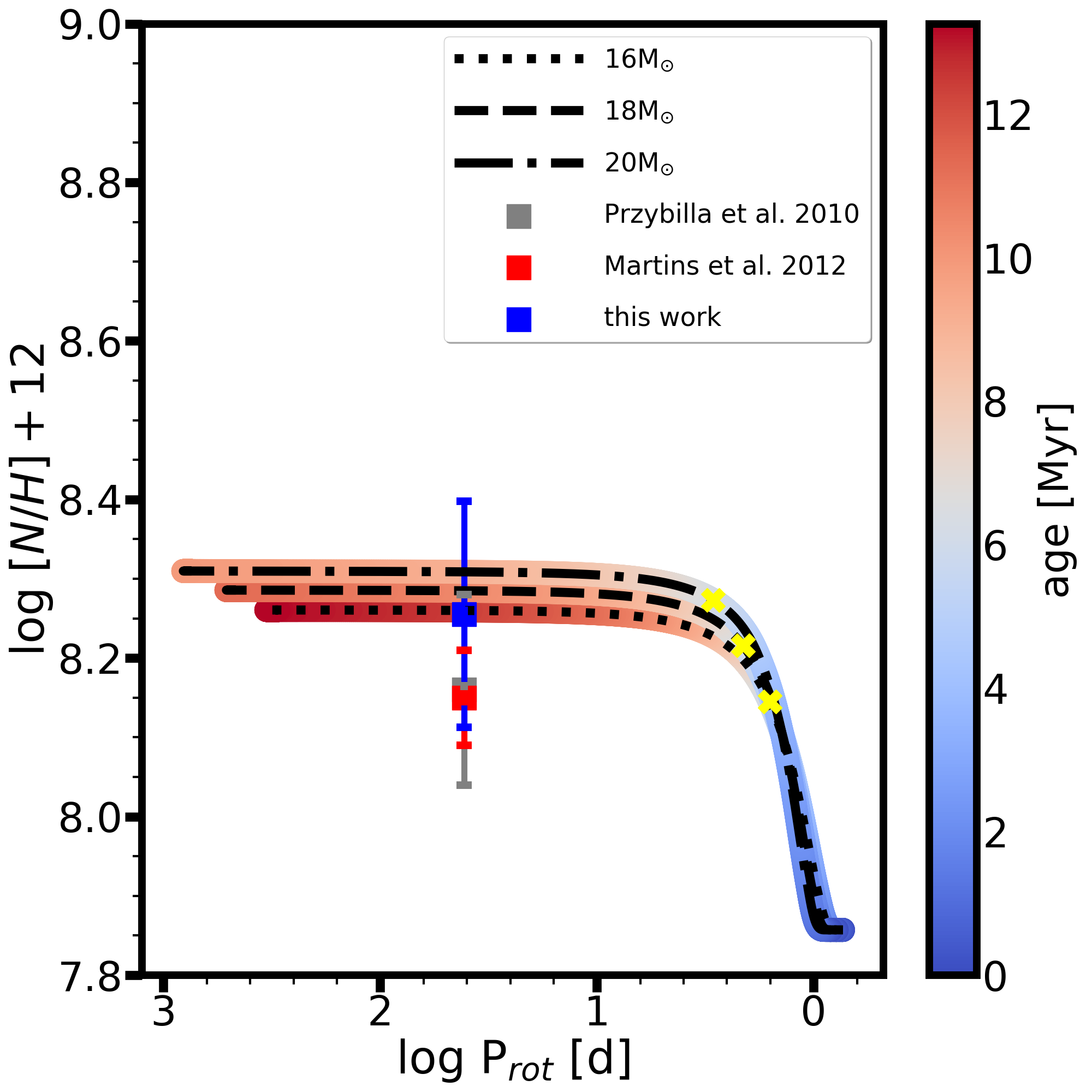}
\includegraphics[width=7cm]{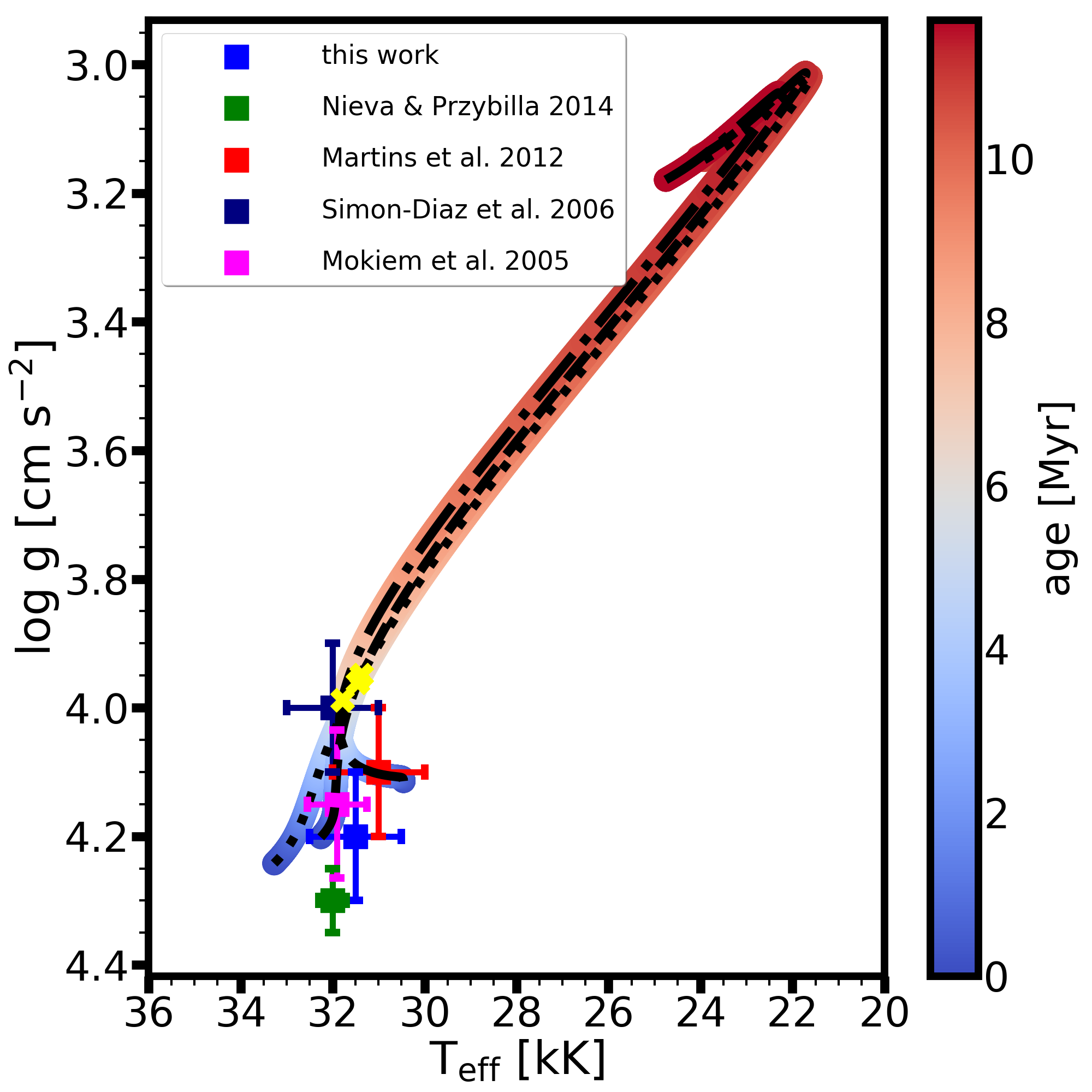}\hspace{1em}\includegraphics[width=7cm]{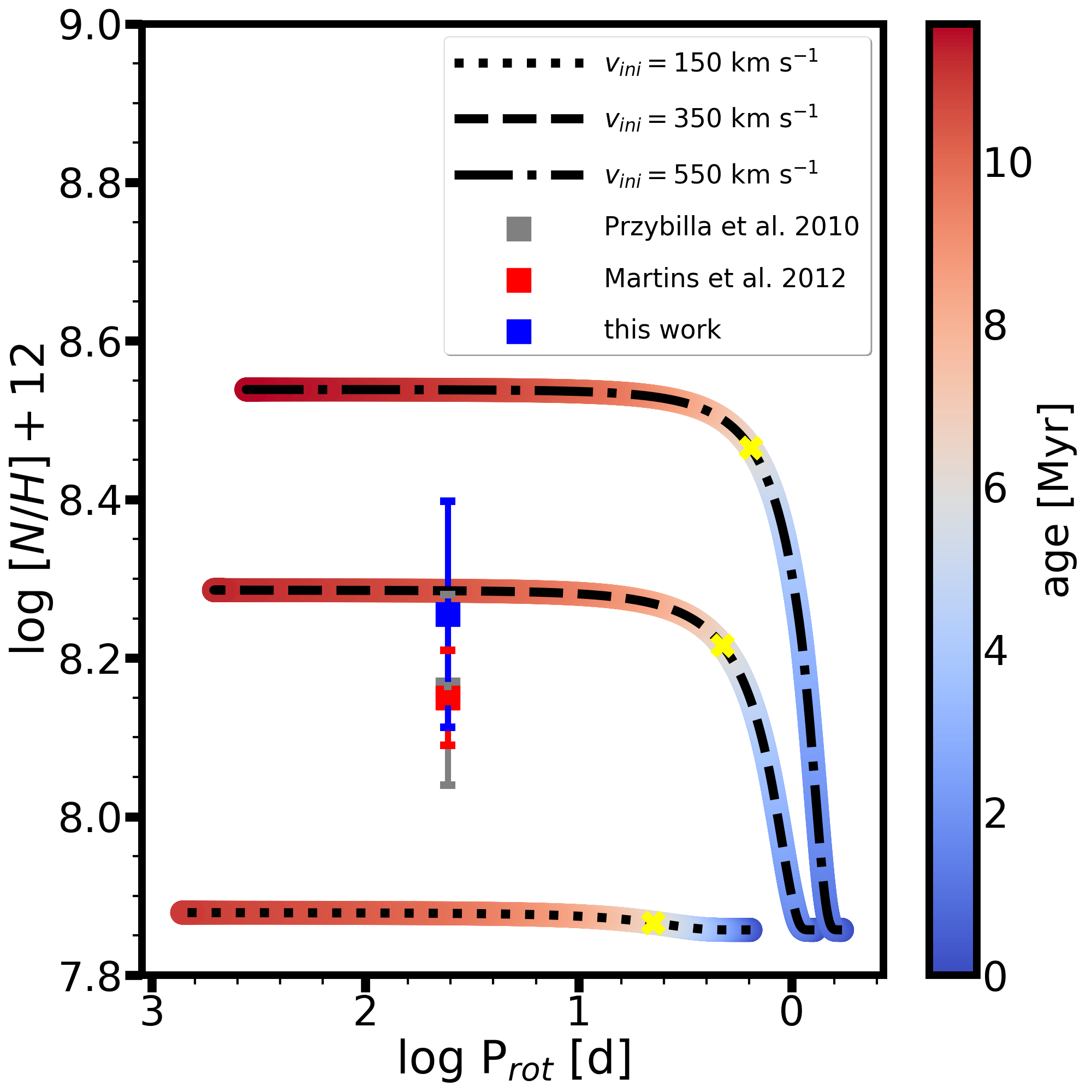}
\includegraphics[width=7cm]{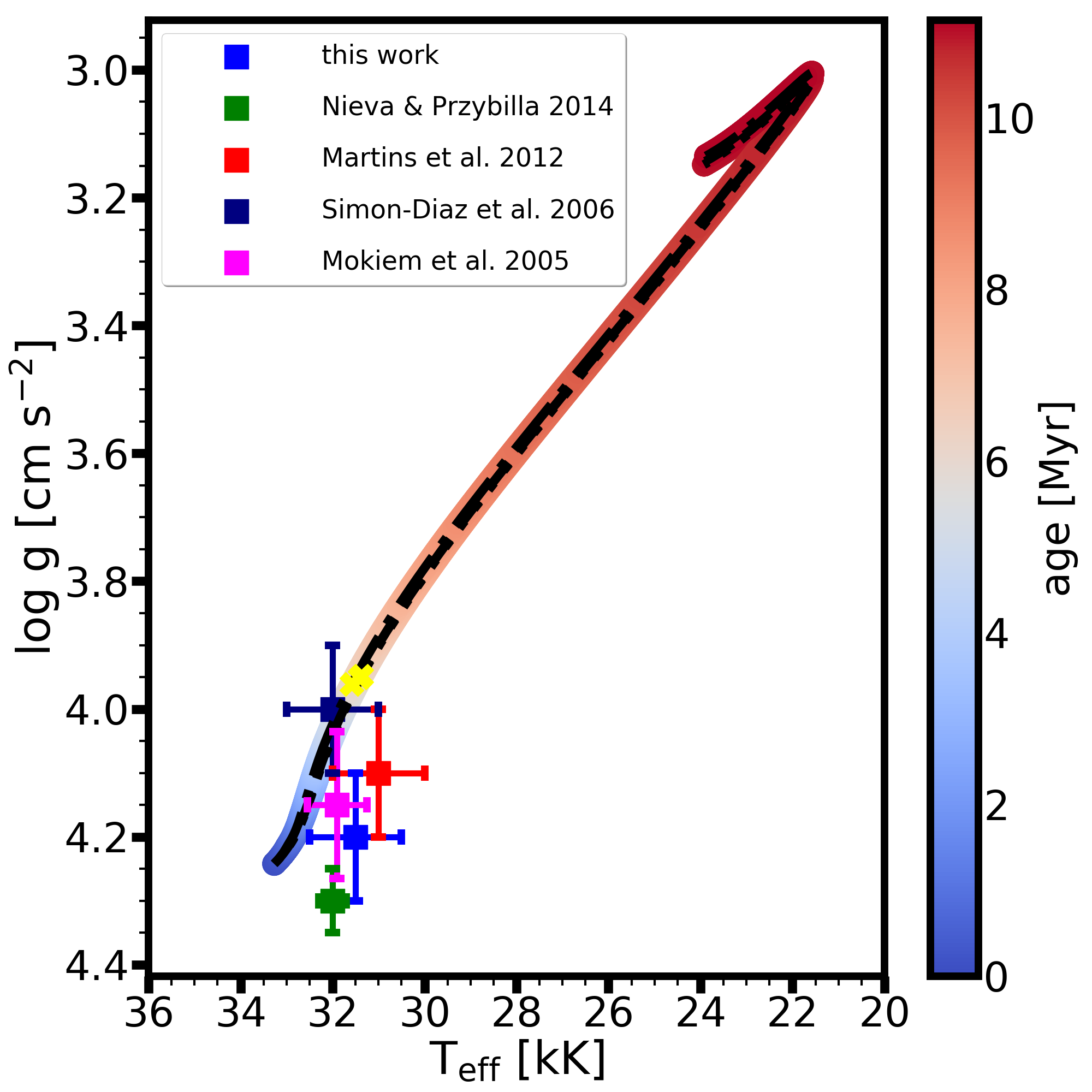}\hspace{1em}\includegraphics[width=7cm]{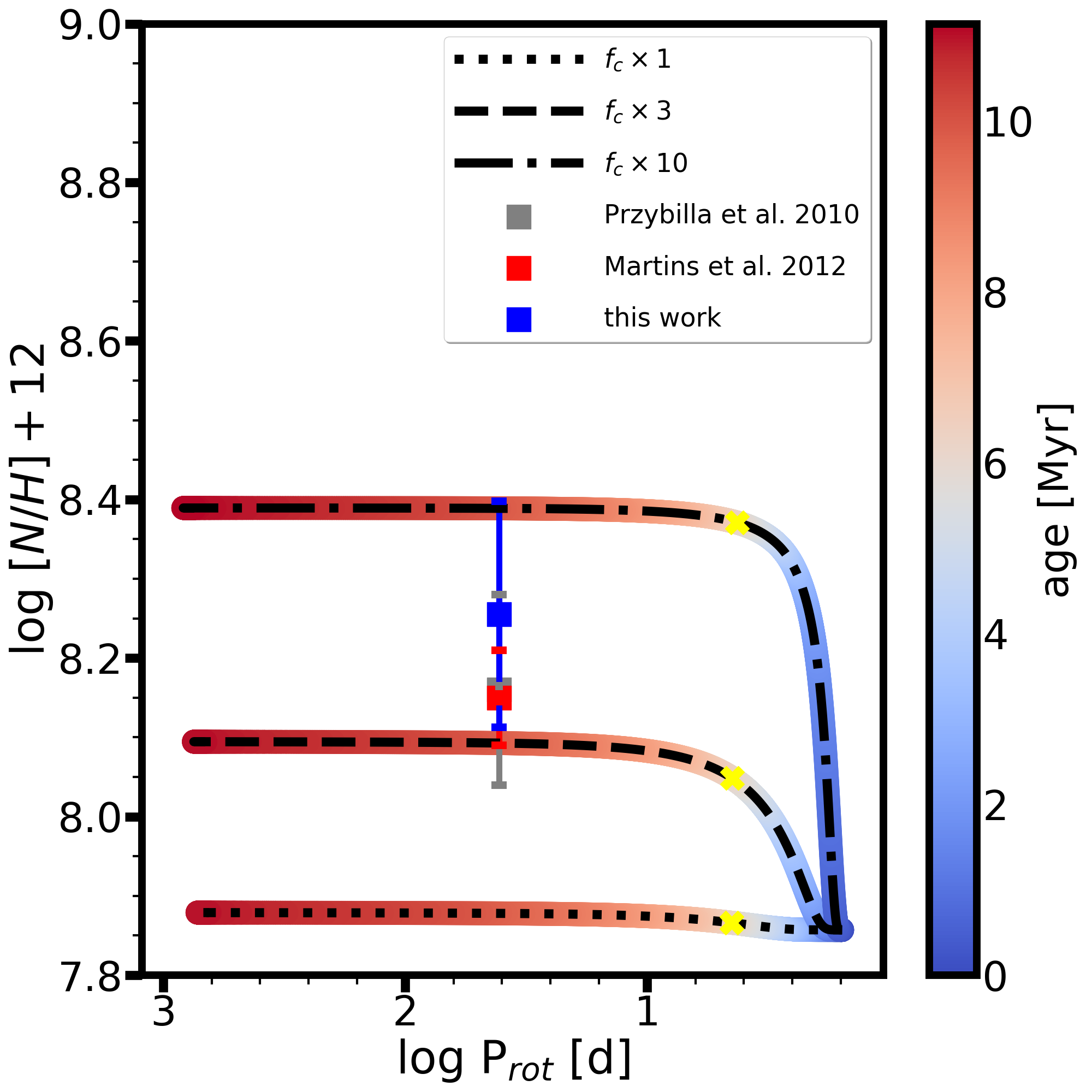}
\caption{\textsc{mesa} models are shown on the Kiel (left) and Hunter-P (right) diagrams. \textit{Upper panels}: varying the initial mass for $\varv_{\rm ini} = 350$ km\,s$^{-1}$. \textit{Middle panels}: varying the initial rotation rates for $M_{\rm ini} = 18$~M$_\odot$. \textit{Lower panels}: adopting slow rotation with $\varv_{\rm ini} = 150$km\,s$^{-1}$ while varying mixing efficiency for $M_{\rm ini} = 18\,$M$_\odot$. The colour-coding scales with stellar age. Yellow crosses indicate an age of 6 Myr. Observations from various authors are shown. The rotation period (here and hereafter, $P_{\rm rot} = 41.033 \pm 0.002$ days) is adopted from \protect{\citet{donati2006}}.}\label{fig:f2}
\end{figure*}
%
%
%
%
\begin{figure*}
\includegraphics[width=7cm]{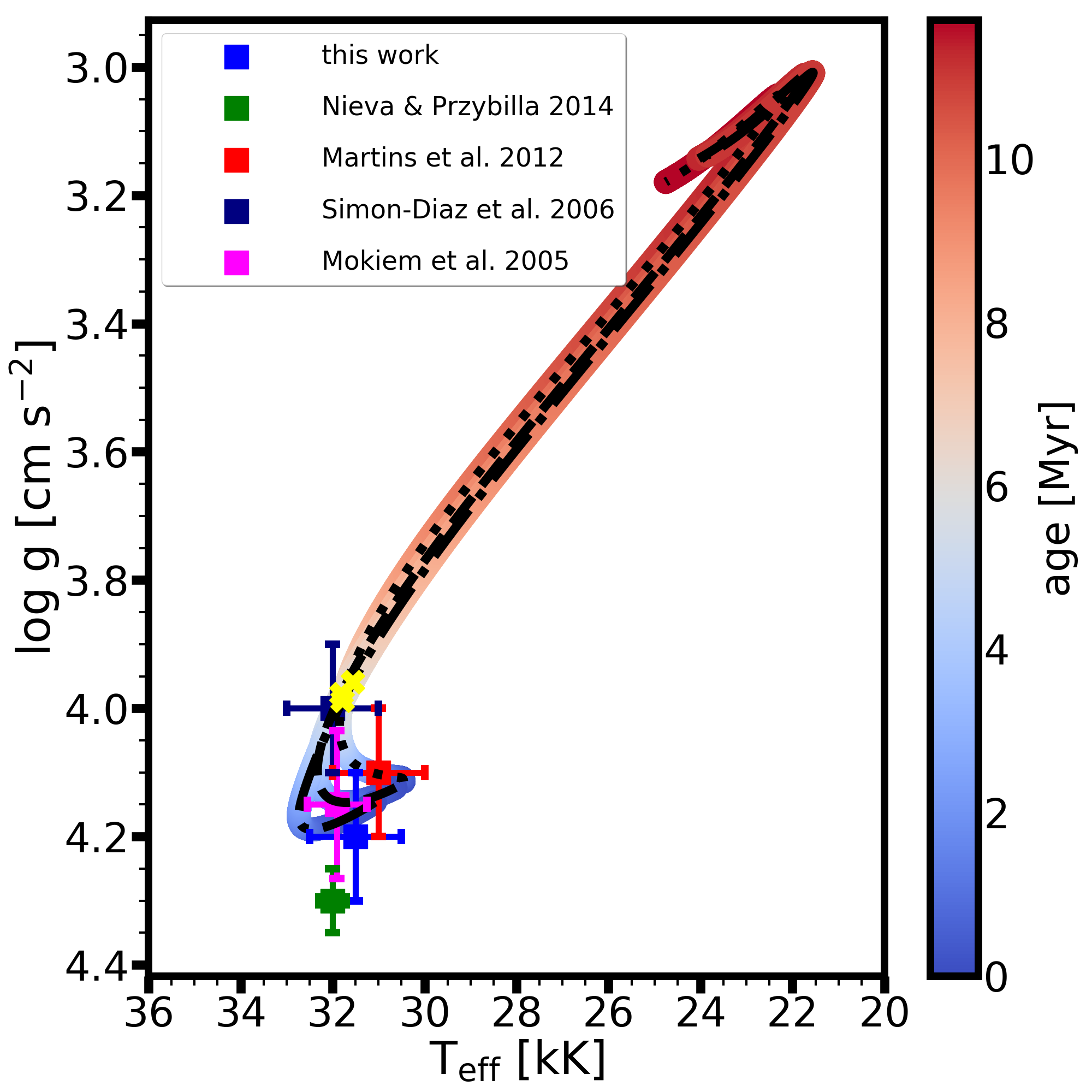}\hspace{1em}\includegraphics[width=7cm]{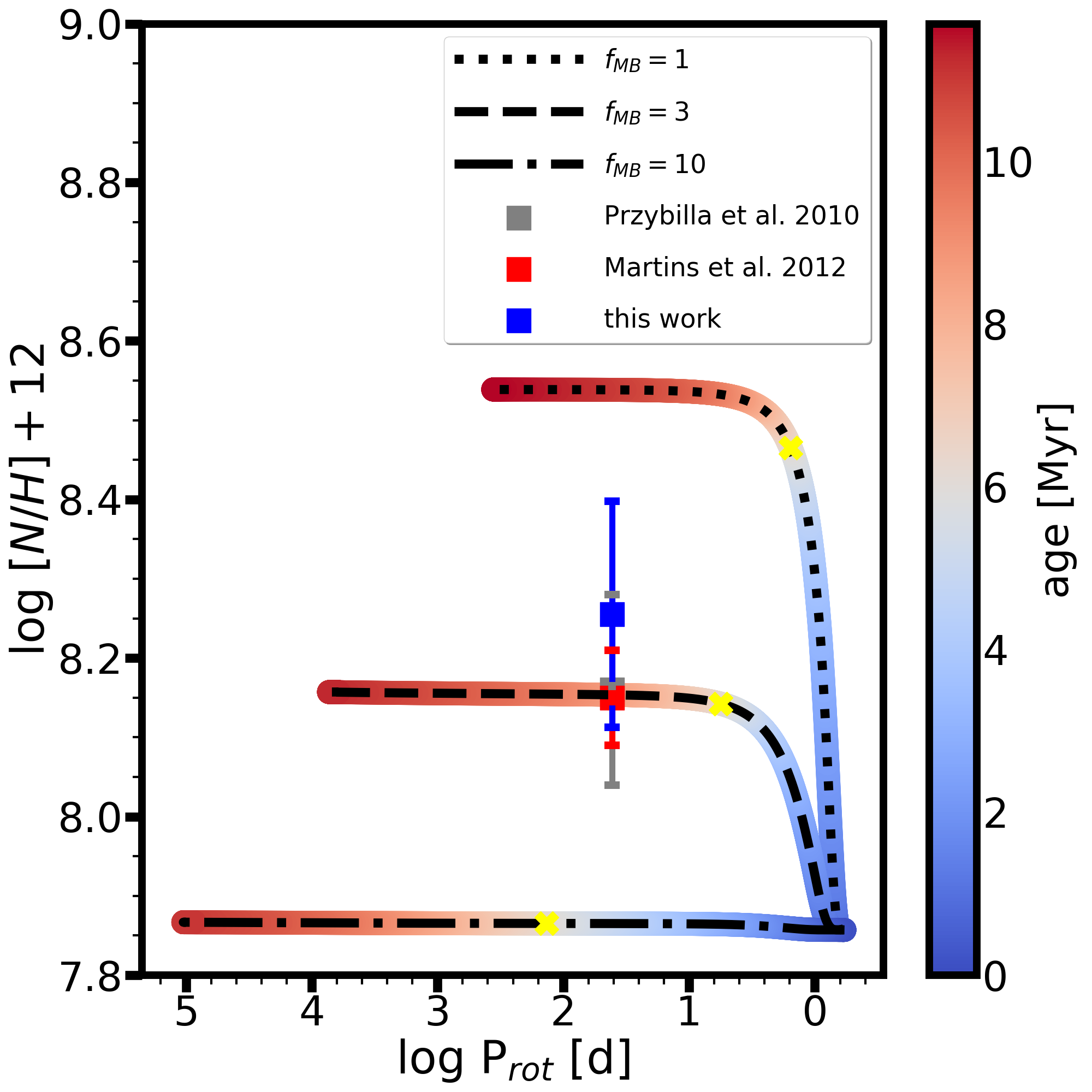}
\includegraphics[width=7cm]{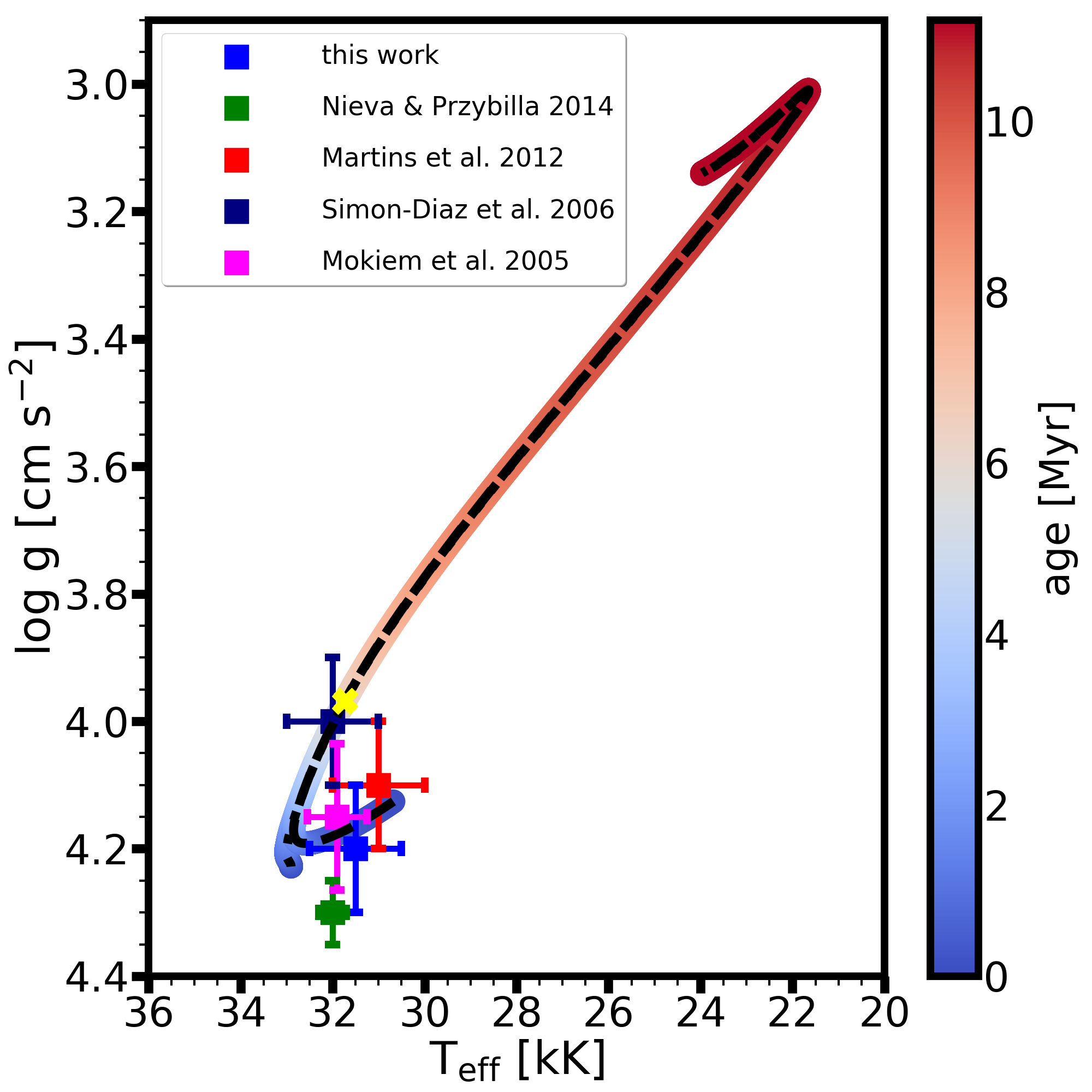}\hspace{1em}\includegraphics[width=7cm]{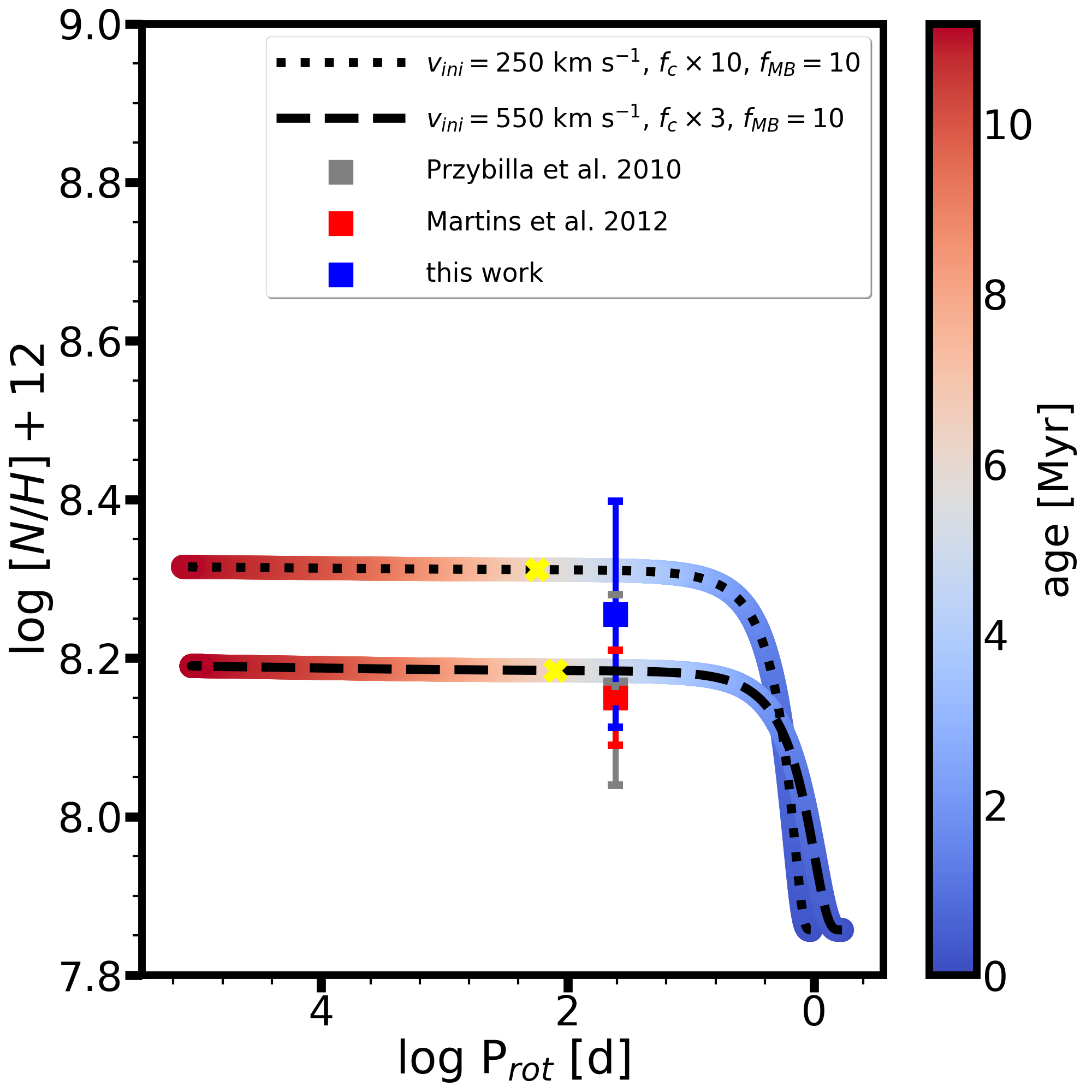}
\includegraphics[width=8cm]{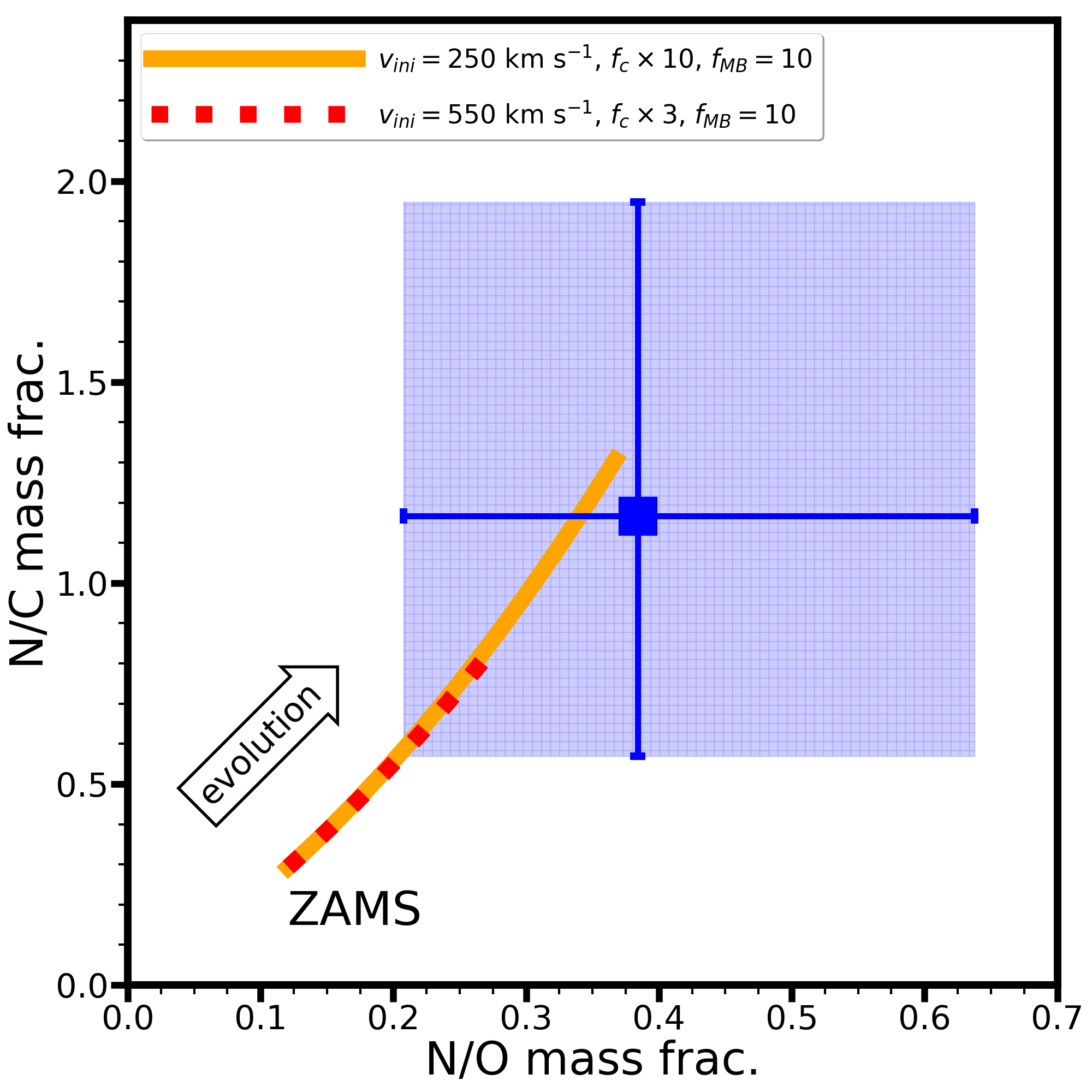}
\caption{\textit{Upper panels}: same as Figure~\ref{fig:f2} but with $M_{\rm ini} = 18$M$_\odot$ and adopting fast rotation with $\varv_{\rm ini} = 550$ km\,s$^{-1}$ while varying the braking efficiency. \textit{Middle panels}: with $M_{\rm ini} = 18$M$_\odot$ and $\varv_{\rm ini} = 250$ km\,s$^{-1}$, $f_{\rm c} = 0.33$ and $f_{\rm MB} = 10$ (dotted), and $\varv_{\rm ini} \approx 550$ km\,s$^{-1}$, $f_{\rm c} = 0.1$ and $f_{\rm MB} = 10$ (dashed). \textit{Lower panel}: $^{14}$N/$^{12}$C mass fraction as a function of $^{14}$N/$^{16}$O mass fraction. The derived CNO abundance ratios are shown with blue marker. The evolutionary models (same as in the middle panels) which can mix core material to the surface evolve towards the top right of the diagram. The end point of the model is where surface rotation is slow enough ($P_{\rm rot} \approx$10 d, see middle panel) that the surface abundance does not change anymore. Since this is reached in about the first 2-3 Myr, the further time evolution leads to no changes on this diagram. }\label{fig:f3}
\end{figure*}

\subsection{\textsc{mesa} modelling: Results and Analysis}

With the model computations we aim to consistently (with the same age) match four strict and well-determined observables of $\tau$~Sco, namely, the surface gravity, effective temperature, rotation period, and nitrogen abundance. Therefore we seek to reconcile models with observations on the Kiel diagram ($\log g$ vs. $T_{\rm eff}$) and the modified Hunter diagram (nitrogen abundance plotted against rotation period instead of projected rotational velocity; referred to as Hunter-P diagram hereafter\footnote{The rotation period is very accurately known from observations and consequently the Hunter-P diagram provides a much more strict constraint than the classical Hunter diagram.}).

%
\subsubsection{Impact of initial mass}\label{sec:sec41}

The upper panels of Figure \ref{fig:f2} show the impact of varying the initial mass for models with a 600 G polar field strength constant in time. 
The primary consequence of increasing the initial mass is an increase in effective temperature. In this mass range, neither the nitrogen enrichment nor the spin down are significantly affected.

Observations on the Kiel diagram indicate a young age, depending on the model assumptions and observational uncertainties, up to a maximum of 6 Myr. Throughout the paper, we will use this maximum age of 6 Myr as our criterion to obtain a self-consistent solution with the observables of $\tau$~Sco. It is important to note that this 6 Myr, which is roughly half of the main sequence lifetime of the models (more precisely depending on the assumed initial parameters), is spent in a very narrow $\log g$ and $T_{\rm eff}$ range (approx. 0.2 dex and 2 kK, shown with the blue part of the colourbar on these figures), practically encompassed by the various observational results. As the model evolves towards lower $\log g$ and $T_{\rm eff}$, a more precise age estimate could be possible since the second 6 Myr of the evolution is spent over a range of approx. 1 dex in log~$g$ and 8~kK in $T_{\rm eff}$ in these models.

Observations on the Hunter-P diagram indicate proximity to the TAMS, i.e., an age well above 6 Myr. In principle, the observed surface nitrogen enrichment is achieved in about 6 Myr, however, the spin down of the model is not efficient enough. 

%
\subsubsection{Impact of initial rotation}

In the middle panels of Figure \ref{fig:f2}, the initial rotational velocities are varied for a \textsc{mesa} model with an initial mass of 18 M$_\odot$.
The main difference for higher rotation is a shift in the ZAMS position on the Kiel diagram and a more efficient chemical mixing on the Hunter-P diagram. Even though this mixing leads to a more rapid nitrogen enrichment, the spin down of the model is not sufficient to reproduce the long rotation period in less than 6 Myr.

Models with much lower initial rotational velocity may, in principle, better approximate the currently observed long rotation period within the timescale inferred from the position of $\tau$~ Sco on the Kiel diagram. However, in that case the efficiency of rotational mixing is not sufficient to reproduce the observed nitrogen abundance.

\subsubsection{Increasing the mixing efficiency in a slow rotator model}

These experiments show that rotational mixing and magnetic braking tend to work against each other. In \textsc{mesa}'s diffusive scheme, when magnetic braking is efficient, the (surface and thus internal) rotational velocity is lower, and thus chemical mixing is reduced.

To simulate a more efficient mixing for fixed stellar parameters, we increase the scaling factor $f_{\rm c}$ which multiplies the sum of all diffusion coefficients considered for chemical mixing\footnote{The Eddington-Sweet circulation is the dominant term throughout the stellar envelope.} (see \citealt{heger2000,brott2011,paxton2013}). 

Figure \ref{fig:f2} (lower panels) shows this parameter test for an 18\,M$_\odot$ model with slow rotation, adopting $\varv_{\rm ini} =~150$~km\,s$^{-1}$. The model with "standard" mixing efficiency ($f_{\rm c} = 0.033$, calibrated by \citealt{heger2000}) does not lead to any nitrogen enrichment. When $f_{\rm c}$ is increased by a factor of 3, there is some enrichment which is compatible with the lower limit of the observations. Increasing $f_{\rm c}$ by an order of magnitude allows to reach the upper limit of the measured nitrogen abundance of $\tau$ Sco - in this particular model setup. This means that for even lower initial rotation rates than we considered here, an extremely (likely unphysically) high mixing efficiency would need to be invoked. Thus we see no feasibility to decrease the initial rotation rate below $\varv_{\rm ini} \approx 150$~km\,~s$^{-1}$.

%
\subsubsection{Increasing the braking efficiency in a fast rotator model}

A major shortcoming of all the previous parameter tests has been that the current long rotation period of $\tau$ Sco -- together with the position in the Kiel diagram and the observed surface nitrogen abundance -- could not be self-consistently (i.e., with the same age) reproduced by the models. Therefore, we continue our thought experiment with testing whether a more efficient braking mechanism may help to overcome this problem. 
In the upper panels of Figure \ref{fig:f3}, we show models with initially $M_{\rm ini} = 18$~M$_\odot$ and fast rotation\footnote{The exact values of $\varv_{\rm ini}$ are somewhat different at the ZAMS as a consequence of the model relaxation, applied magnetic braking, and the ZAMS definition itself. We use the criteria that 0.3\% of core hydrogen has already burnt to define the ZAMS. Nonetheless, the exact initial rotational velocities do not significantly affect the trend that we show with this parameter test.} with $\varv_{\rm ini}~>~500$~km\,~s$^{-1}$, using a "usual" ($f_{\rm MB} = 1$, c.f. Equation~\ref{eq:eq1}), a factor of 3, and a factor of 10 more efficient magnetic braking (lower panels of Figure~\ref{fig:f3}). 
When magnetic braking is more efficient by a factor of 10, the observed spin rate of $\tau$~Sco can be recovered within 6 Myr. However, the efficient spin down leads to very inefficient rotational mixing and thus no nitrogen is seen on the stellar surface. When $f_{\rm MB} = 3$, chemical mixing can still remain efficient during the early evolution and thus the observed values are well-approximated by the model. However, the spin rate of this model after 6 Myr is still an order of magnitude larger than the observed one.

%
%
%
%
\begin{figure*}
\includegraphics[width=7cm]{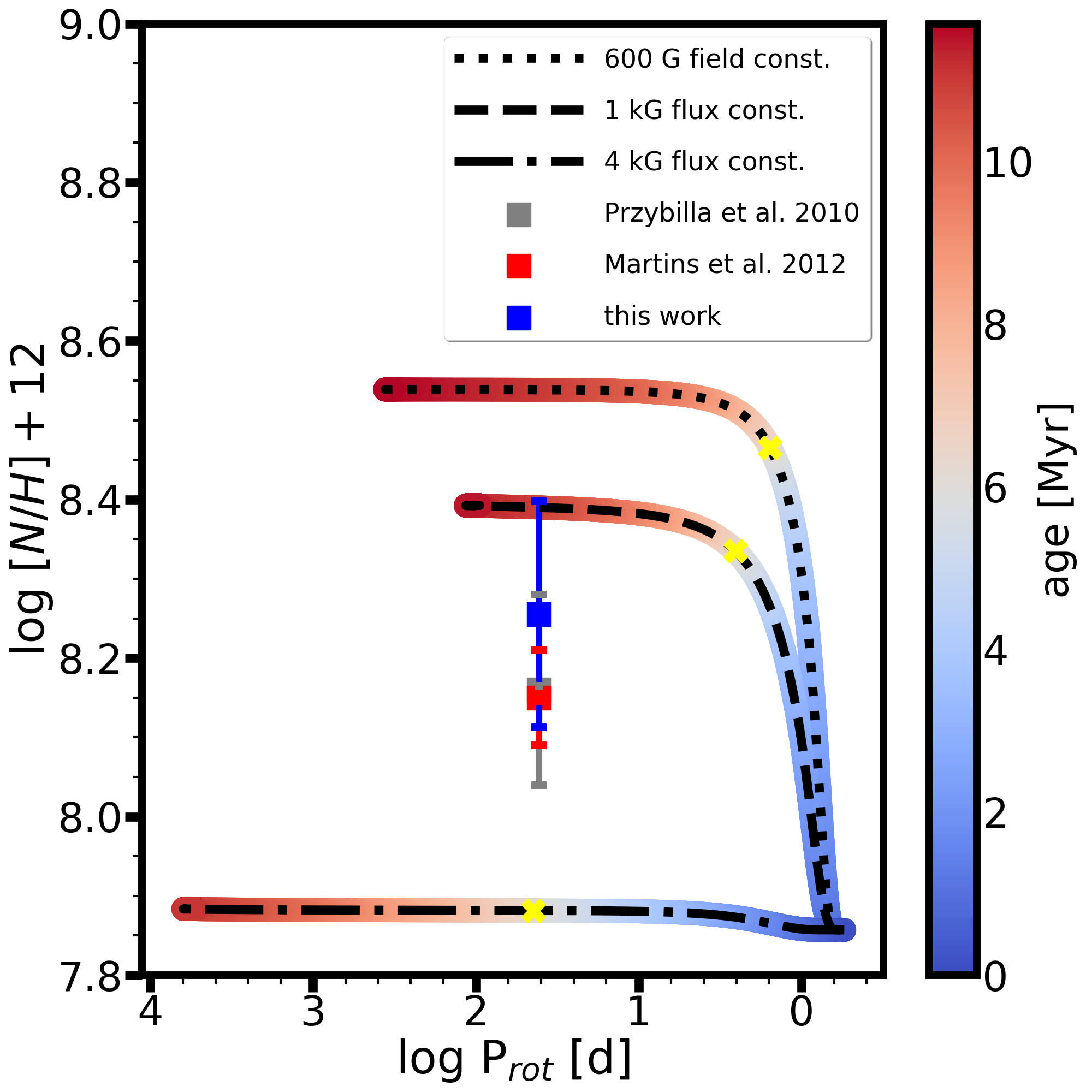}\hspace{1em}\includegraphics[width=7cm]{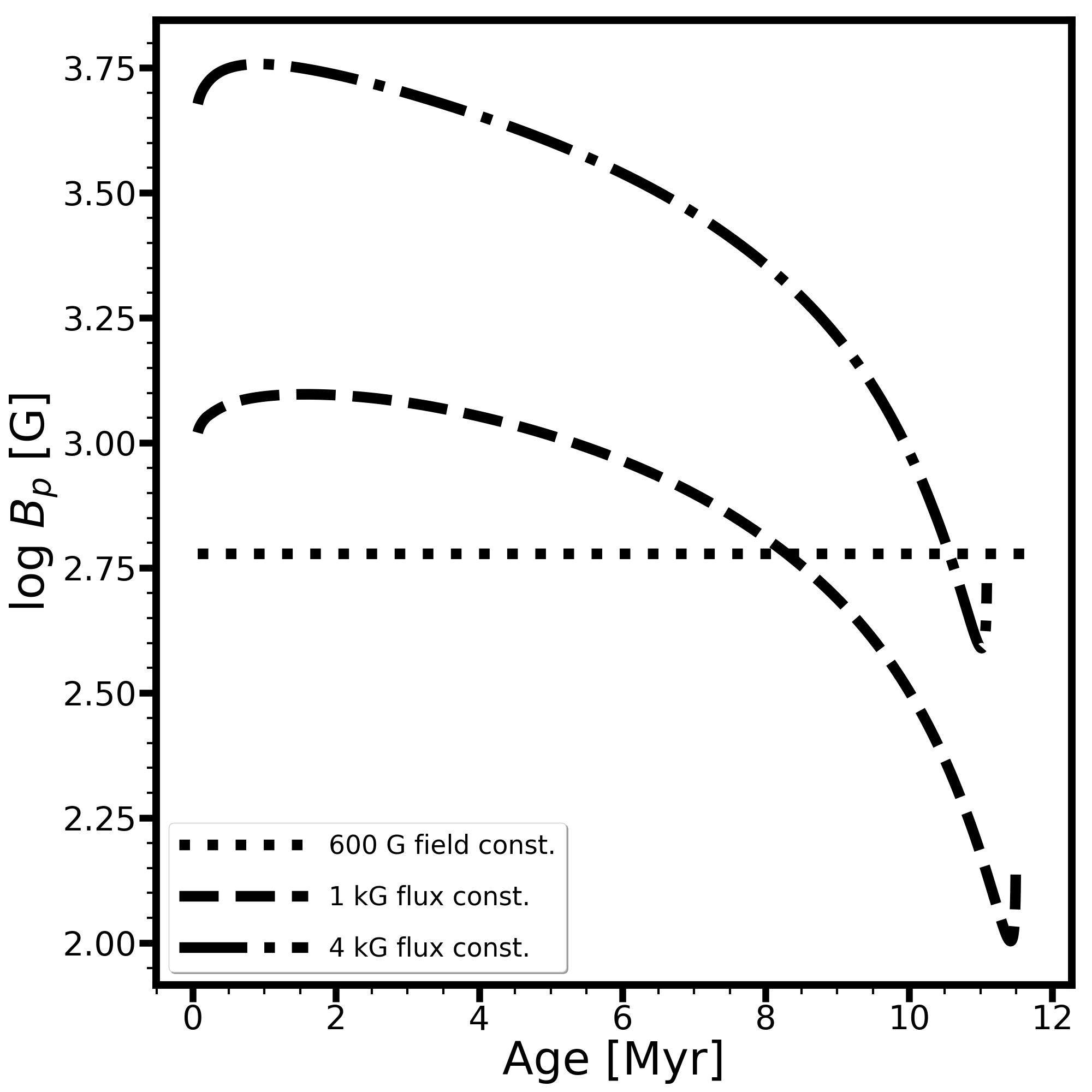}
\includegraphics[width=7cm]{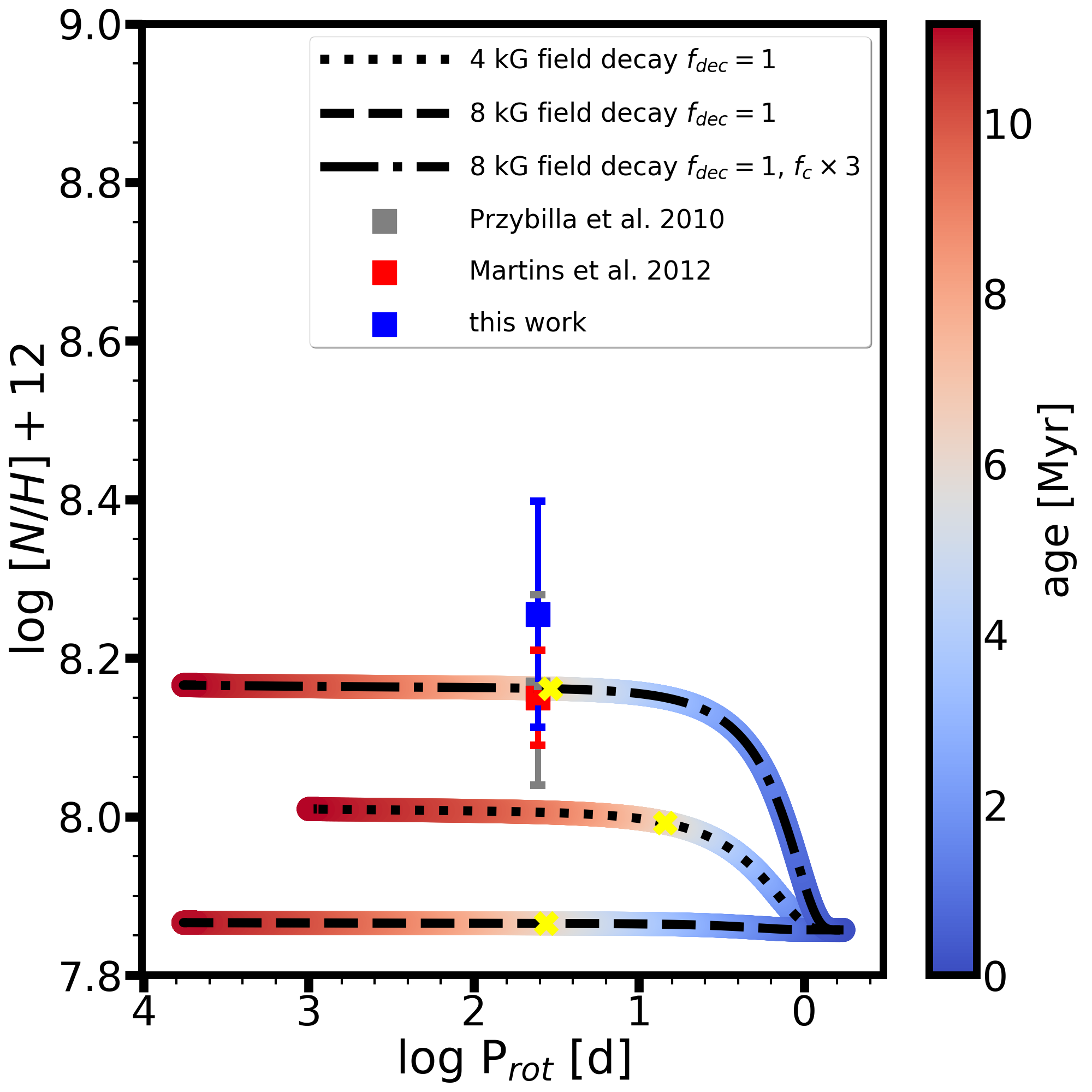}\hspace{1em}\includegraphics[width=7cm]{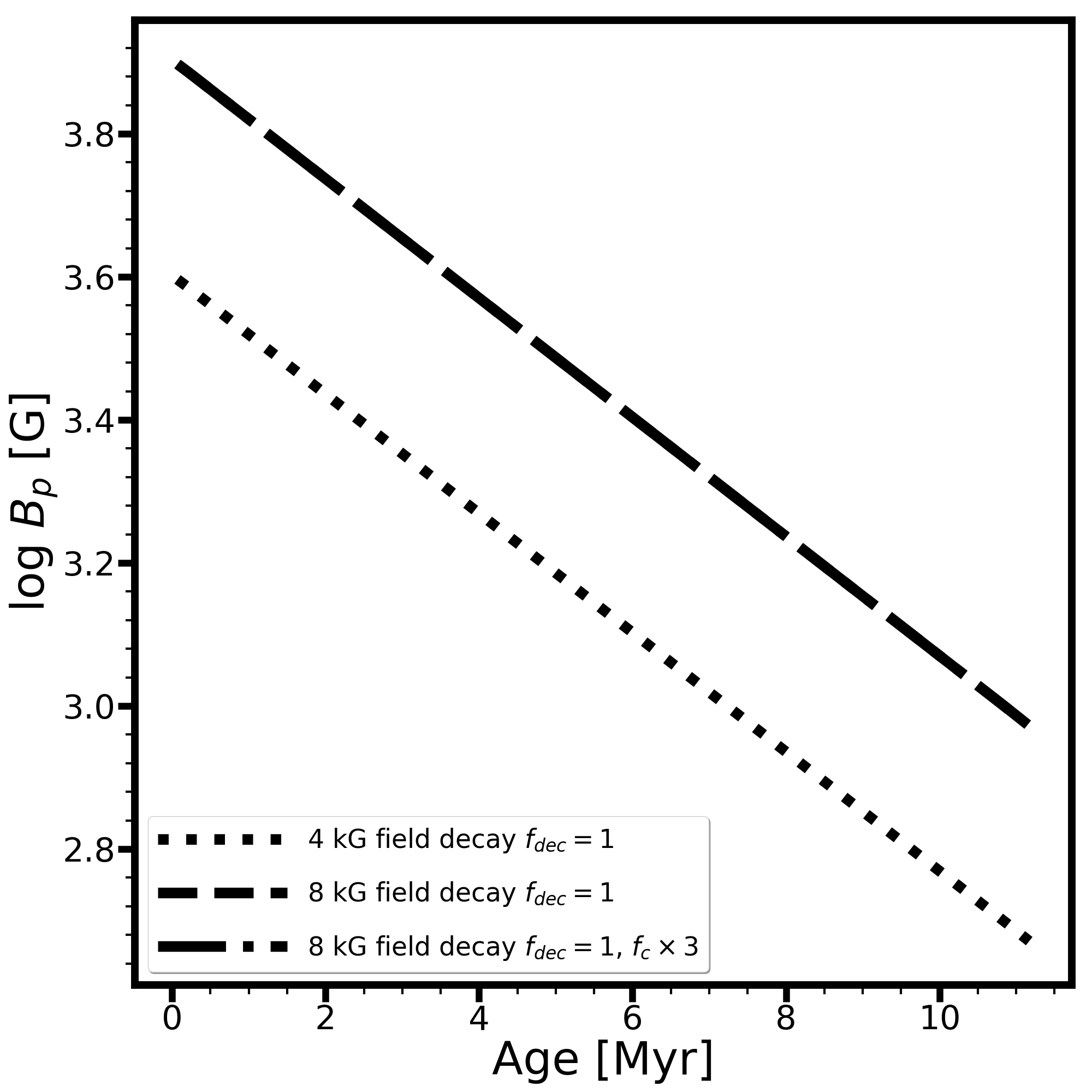}
\includegraphics[width=7cm]{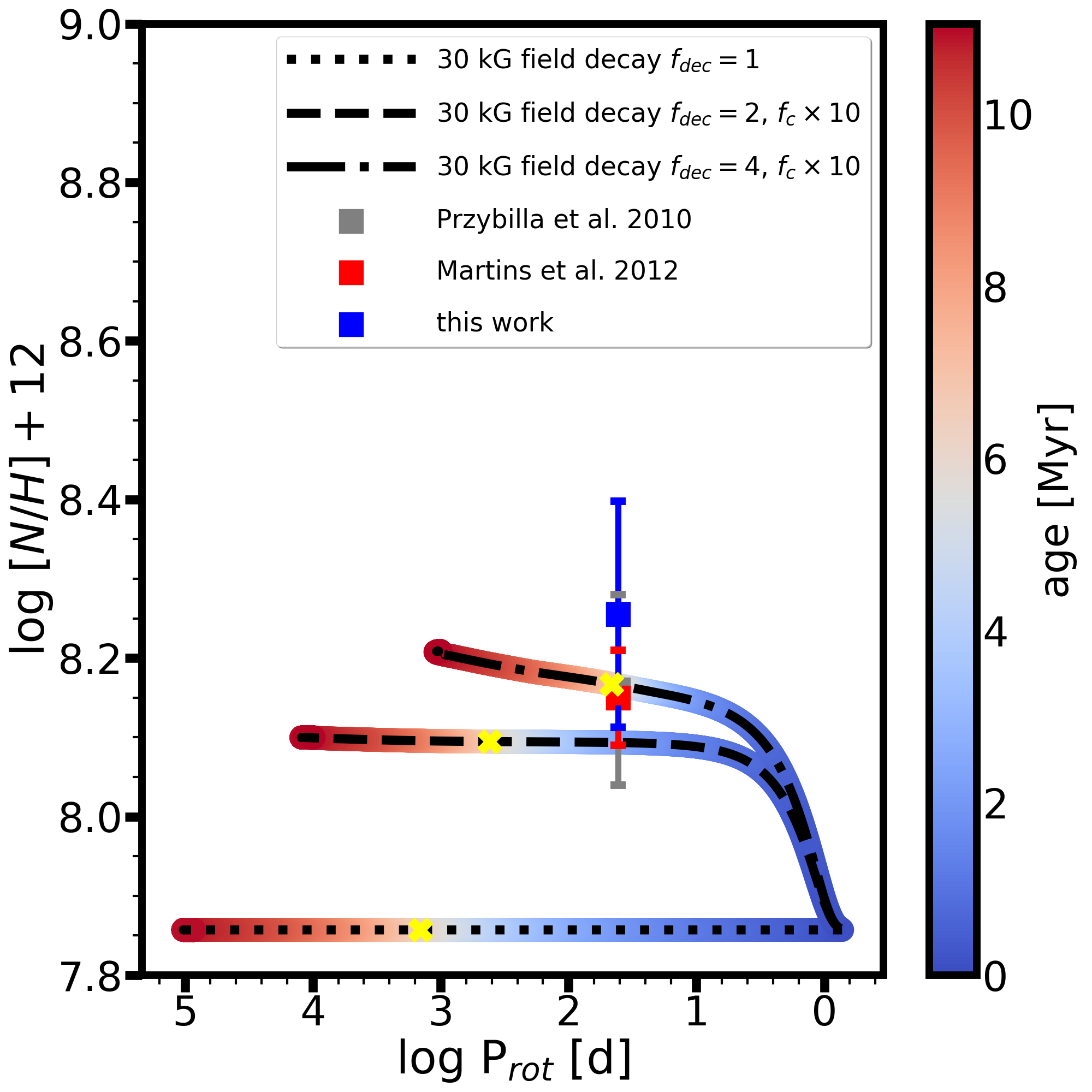}\hspace{1em}\includegraphics[width=7cm]{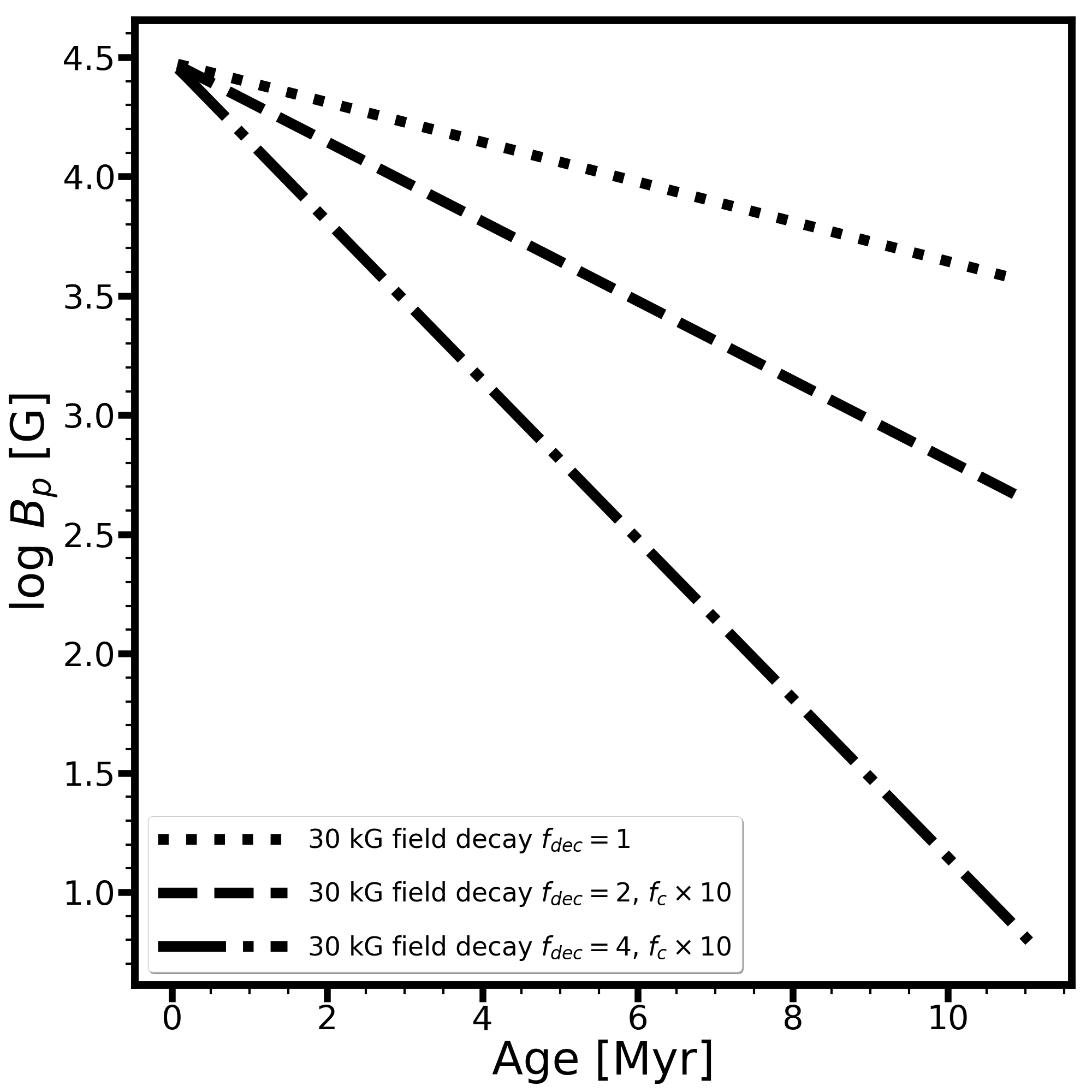}
\caption{\textsc{mesa} models with $M_{\rm ini}=$18\,M$_\odot$ and $\varv_{\rm ini} >500$ km\,s$^{-1}$ are shown. \textit{Upper panels}: Models with magnetic flux conservation compared to the reference model ($B_{\rm p}=$\,600 G). Standard efficiencies are used for $f_{\rm c}$ and $f_{\rm MB}$. \textit{Middle and lower panels}: Models with magnetic field decay. The magnetic braking efficiency is set with $f_{\rm MB}= 1$ but the decay efficiency $f_{\rm dec}$ and rotational mixing efficiency $f_{\rm c}$ are varied (c.f. Equations \ref{eq:eq1} and \ref{eq:decay}). The two models with initially 8 kG fields in the middle right panel overlap.}\label{fig:f5}
\end{figure*}

%
\subsubsection{Varying both the mixing and braking efficiencies}\label{sec:sec44}

The only experiment which succeeded to reproduce the long rotation period of $\tau$~Sco within 6 Myr (which is our criterion to obtain a self-consistent solution to match all key observables) is the one where the magnetic braking efficiency is increased by a factor of 10. To see if it is possible at all -- within this thought experiment -- to obtain nitrogen excess in this case, we compute models where additionally the mixing efficiency $f_{\rm c}$ is increased by a factor of 10 for a model with $\varv_{\rm ini} = 250$ km\,s$^{-1}$, while it is increased by a factor of 3 for a model\footnote{In this fast-rotating model too, the rotational velocity undergoes some adjustment, therefore the initial value stated (at our ZAMS definition) should not be considered as an exact limit.} with $\varv_{\rm ini} \approx 550$ km\,s$^{-1}$ (Figure \ref{fig:f3}, middle panels). 

In the lower panel of Figure \ref{fig:f3} we also show the measured $^{14}$N/$^{12}$C mass fraction as a function of $^{14}$N/$^{16}$O mass fraction. This confirms that the nitrogen abundance alone is not anomalous, instead the overall CNO abundances are consistent with the surfacing of stellar core material. That is, initially carbon becomes depleted at the expense of nitrogen, whereas oxygen remains approximately constant \citep{przybilla2010,maeder2014a}. After the CN cycle is in equilibrium, a slight oxygen depletion takes place leading to a further increase of the nitrogen abundance. It is thus clear that these observations cannot be reconciled with an initially slowly-rotating single star model where the surface does not reflect the core abundance (see the Hunter-P diagram in Figure~\ref{fig:f2}). The evolutionary models begin with approximately N/C~$=$~0.3 and N/O~$=$~0.1, however, the tracks only represent the time evolution until mixing is efficient. When rotation becomes slow ($P_{\rm rot} \gtrsim$~10~days in our experiment), the surface CNO abundance ratios remain unaltered. Thus main sequence models with inefficient mixing would appear as single points on this diagram, close to the ZAMS values. 

These artificially-engineered models, for the first time, can approximately reproduce not only the four strict observables but also the overall CNO abundance ratios with a consistent stellar age of less than 6 Myr. The question therefore is whether there might be a physically meaningful reason to justify large departures in the efficiency of commonly-used prescriptions for rotational mixing and magnetic braking.

%
\subsubsection{Tests with different magnetic field evolution}\label{sec:45}

Thus far, we made the simplifying assumption that the surface magnetic field strength did not vary in time. However, magnetic field evolution is presently not well-constrained and therefore we now experiment with cases where the surface magnetic field strength can change in time. 

The usual, first order estimate is based on magnetic flux conservation (following Alfv\'en's theorem, \citealt{alfven1942}), 
\begin{equation}\label{eq:flux}
    F \propto B_{\rm surf} (t) \, R_{\star}^{2} (t) = const. =  B_{\rm surf} (t = 0) \, R_{\star}^{2} (t = 0)  \, ,
\end{equation}
\noindent where $B_{\rm surf} (t = 0)$ and $R_{\star} (t=0)$ are the ZAMS magnetic field strength (which is assumed) and stellar radius (which is calculated). In this case, the magnetic field strength only varies as a function of the stellar radius.

Since an initially stronger magnetic field aids the spin down of the star, considerations have been given to a magnetic field decay scenario, which we assume to have the form of 
\begin{equation}\label{eq:decay}
    B_{\rm surf} (t) = B_{\rm surf} (t = 0) \, \mathrm{exp} \left(- f_{\rm dec} \, t / \tau \right) \, , 
\end{equation}
\noindent where $f_{\rm dec}$ is an arbitrary scaling factor to which we refer to as the decay efficiency, $t$ is the time, and $\tau$ is a characteristic timescale. We set $\tau =$~12~Myr, which is the approximate main sequence lifetime of our models, so that the magnetic field strength would weaken to roughly 60\% of its initial value in 6 Myr with a decay efficiency of unity. With $f_{\rm dec} = 2$ and 4, the field strength becomes about 30\% and 15\%, respectively, of its initial value in 6 Myr. We note that since the change in stellar radius is very modest in the first half of the main sequence, these arbitrary \textit{field decay} scenarios are practically identical to a (perhaps more commonly known) \textit{flux decay} scenario. Contrasted with the flux conservation scenario (Equation~\ref{eq:flux}), we now have:
\begin{equation}\label{eq:decay2}
    F \propto B_{\rm surf} (t) \, R_{\star}^{2} (t) =  B_{\rm surf} (t = 0) \, \mathrm{exp} \left(- f_{\rm dec} \, t / \tau \right) \, R_{\star}^{2} (t) \, .
\end{equation}

In Figure \ref{fig:f5}, models with different magnetic field evolution scenarios are shown. The top panels show models with magnetic flux conservation (Equation \ref{eq:flux}), whereas the middle and lower panels show models with magnetic field decay (Equations \ref{eq:decay}-\ref{eq:decay2}). 

With magnetic field evolution, the models can have an initially  stronger magnetic field and consequently initially more efficient magnetic braking. Although time dependent, this is somewhat similar to the previous tests where $f_{\rm MB}$ was used to obtain a more efficient magnetic braking. This means that magnetic field evolution, in principle, allows for adopting a higher initial rotational velocity and thus a higher initial ("natural") efficiency of rotational mixing. 

Assuming magnetic flux conservation, a field with an initial strength of 1 kG (assumed dipolar) well approximates the present-day field strength of $\tau$~Sco, however, leads to a small impact in its overall evolution compared to the reference model (600 G polar field strength constant in time). An initial 4 kG field produces efficient braking albeit no nitrogen enrichment. This model also leads to a magnetic field strength over the entire evolution (except close to the TAMS) which is far too strong ($>$ ~1~kG) to be compatible with spectropolarimetric measurements (see Appendix~\ref{sec:app2}).


When magnetic field decay is considered, the model with an initial 3 kG field strength (assumed dipolar) and decay efficiency $f_{\rm dec}$ of unity produces sufficient nitrogen excess, however, does not spin down fast enough (middle panels). The model with an initial 9 kG field strength does produce efficient braking, and while the long rotation period is recovered, no mixing is achieved. Increasing the mixing efficiency in this model by a factor of 3 leads to an acceptable, self-consistent match with the observables of $\tau$~Sco. However, the observable field strength would remain above 3 kG in the first 6 Myr, which is at odds with the observations. 

Experiments with 30 kG initial field strengths allow to reproduce the current long rotation period (lower panels). Increasing $f_{\rm c}$ by a factor of 10 is required to reach nitrogen excess. However, a decay efficiency of 2 still yields a magnetic field strength well above 1 kG in 6 Myr. $f_{\rm dec} = 4$ results in an acceptable self-consistent solution.

From these tests we conclude that while magnetic field evolution can, to some extent, alleviate the large magnetic braking efficiency ($f_{\rm MB }= 10$) that was needed in the model with a constant 600 G field strength, it faces the challenge to reach a sufficiently low, sub-kG field strength in 6 Myr. This requires a high magnetic field decay efficiency. Importantly, in all tests, the need for a more efficient chemical mixing is still present.

%
%

%
%
%
%
%
%
\begin{figure*}
\includegraphics[width=8cm]{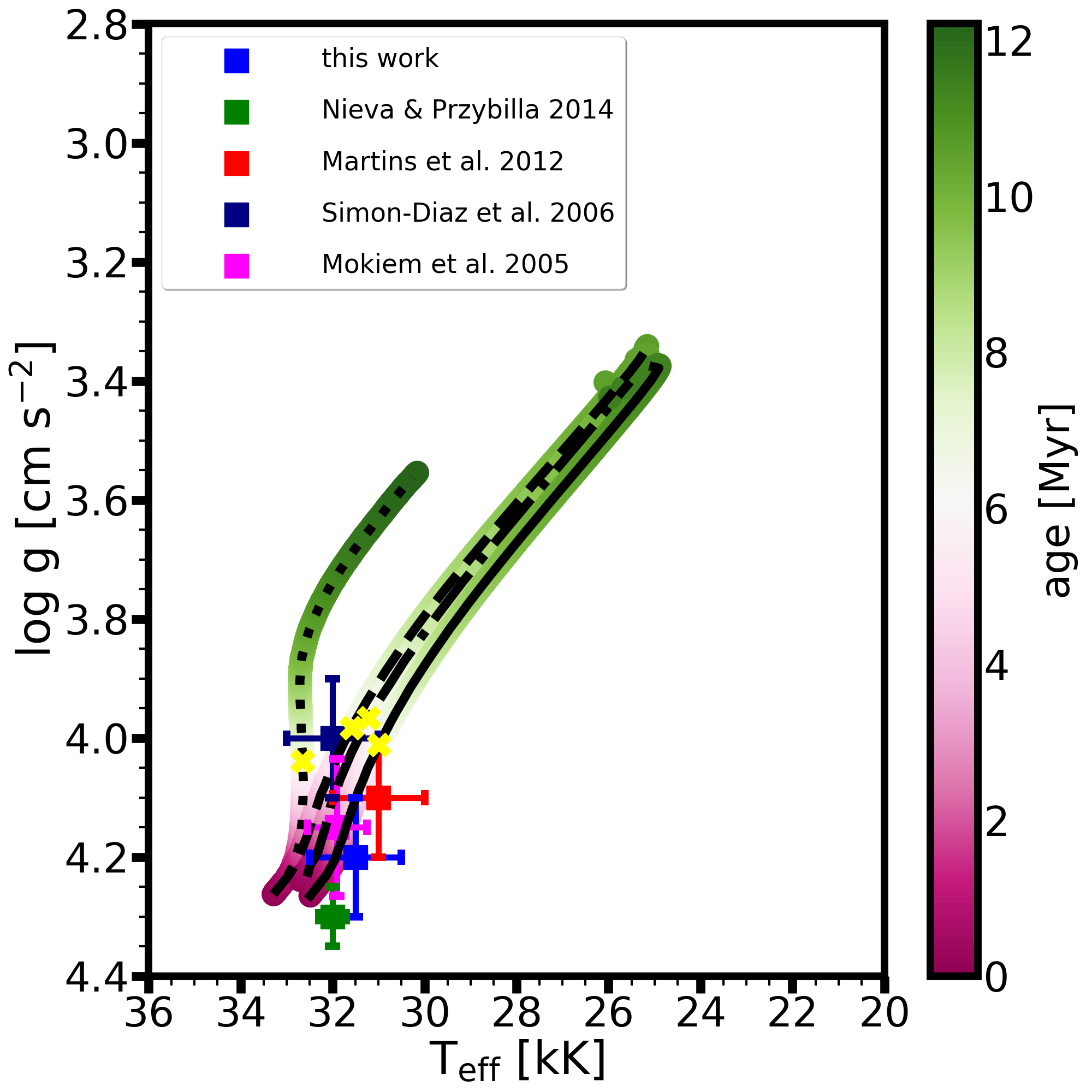}\hspace{1em}\includegraphics[width=8cm]{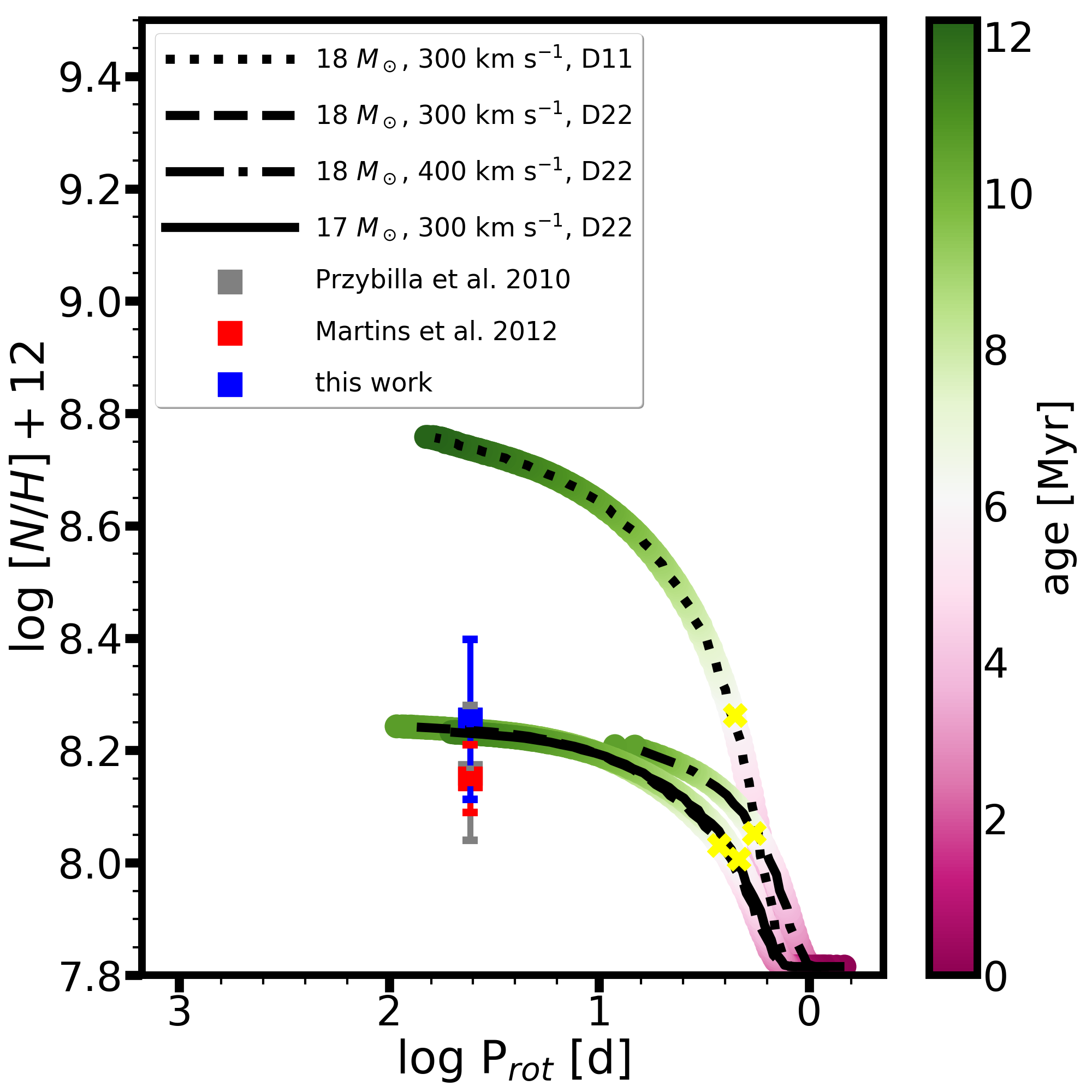}
\caption{Shown are models computed with the Geneva code on the Kiel (left) and Hunter-P diagrams (right). The colour-coding scales with stellar age. Yellow crosses indicate an age of 6 Myr.}\label{fig:f6}
\end{figure*}

\subsection{\textsc{genec} modelling: Setup}

In \textsc{genec}, we adopt similar modelling assumptions as \cite{ekstroem2012}. A solar metallicity of $Z = 0.014$ is used with the \cite{asplund2009} mixture of metals except for neon \citep{ekstroem2012}, and isotopic ratios are from \nobreak\cite{lodders2003}. The adopted mixing efficiency in the convective core is $\alpha_{\rm MLT} = 1.6$. A step overshooting method is applied with $\alpha_{\rm ov} = 0.1$. The opacity tables are adopted from OPAL \citep{rogers1992}. Mass-loss rates are calculated following the prescription of \citealt{vink2000} and \citealt{vink2001}, multiplied by a factor of 0.85. \\

Chemical element transport is modelled as a diffusive process \citep{pin1989}, adding up from three terms: vertical shear, horizontal turbulence, and meridional currents. Following \cite{zahn92}, the latter two are combined into one effective diffusion coefficient, $D_{\rm eff}$ (see \citealt{meynet2013} and references therein). In this work we test two prescriptions. In one case (which will be referred to as D11 hereafter), the shear is adopted from \cite{maeder1997} as 
\begin{equation}
    D_{\rm shear} = \frac{H_{\rm p}}{g \delta} \frac{K}{\left[ \frac{\varphi}{\delta} \nabla_{\mu} + (\nabla_{\rm ad} - \nabla_{\rm rad}) \right]} \left( \frac{9 \pi}{32} \Omega \frac{\mathrm{d} \ln \Omega}{\mathrm{d} \ln r} \right)^2 \, ,
\end{equation}
\noindent where $H_{\rm p}$ is the local pressure scale height, $g$ is the local gravitational acceleration, $\delta$ and $\varphi$ are derivatives from the equation of state, $K$ is the thermal diffusivity, $\Omega$ is the angular velocity, and $r$ is the distance from the centre. The horizontal turbulence is adopted from \cite{zahn92} as %
\begin{equation}
    D_{\rm h} = r \vert 2 V(r) - \alpha U(r) \vert \, , 
\end{equation}
\noindent where $\alpha = \frac{1}{2} \frac{\mathrm{d} \ln (r^2 \Omega)}{\mathrm{d} \ln r}$ and $U(r)$ and $V(r)$ are the horizontal and vertical components of the meridional circulation.  

In the second case (which we will refer to as D22 hereafter), the shear is adopted from \cite{talon1997} as 
\begin{equation}
    D_{\rm shear} = \frac{H_{\rm p}}{g \delta} \frac{K + D_{\rm h}}{\left[ \frac{\varphi}{\delta} \nabla_{\mu} \left( 1 + \frac{K}{D_{\rm h}} \right)  + (\nabla_{\rm ad} - \nabla_{\rm rad}) \right]} \left( \frac{9 \pi}{32} \Omega \frac{\mathrm{d} \ln \Omega}{\mathrm{d} \ln r} \right)^2 \, ,
\end{equation}
\noindent and the horizontal turbulence is adopted from \cite{maeder2003b} as 
\begin{equation}
    D_{\rm h} = A r \left( r \Omega (r) V(r) \vert 2 V(r) - \alpha U(r) \vert \right)^{1/3} \, ,
\end{equation}
\noindent where $A = (3 / 400 n \pi)^{1/3}$ with $n$ being the number of axial rotations \citep{maeder2003b}.

Angular momentum transport is modelled by using an advecto-diffusive equation which accounts for the radial component of meridional currents (the advective term) and shears (the diffusive term). The meridional currents are an advective process by nature. The shear term is modified when using the D11 ($D_{\rm shear}$ \, adopted from \citealt{maeder1997}) or D22 ($D_{\rm shear}$ adopted from \citealt{talon1997}) schemes in the models. Both of these cases, without internal magnetic fields, allow for radial differential rotation and lead to a weaker core-envelope coupling than in solid-body rotating models. We do not test solid-body rotating models here because it was shown in previous works that it leads to less surface enrichment than differentially-rotating models (\citealt{meynet2011}, \PaperI).

The effects of the surface magnetic field are modelled via magnetic braking, which is implemented as a boundary condition for internal angular momentum transport (\citealt{meynet2011,georgy2017}; \PaperI, \PaperII{}). We refer the reader to \PaperII{} Appendix B, where the \textsc{mesa} and \textsc{genec} implementations of magnetic braking are detailed and contrasted. Since in \textsc{genec} only the outermost layers are ascribed to lose specific angular momentum, the use of angular momentum transport without a strong coupling means that significant shears can develop in the outer part of the stellar envelope, while meridional currents remain efficient to transport chemical elements close to the stellar core. (Note that in \textsc{mesa} we model the opposite scenario: shears remain efficient close to the core but (weak) meridional circulation dominates the transport in the outer envelope.) The equatorial magnetic field strength is set to $B_{\rm eq} = 300 $~G (corresponding to 600 G polar) and is kept constant over time. 

\subsection{\textsc{genec} modelling: Results and Analysis }

Some of the major differences between \textsc{genec} and \textsc{mesa} are the treatment of angular momentum transport and the way magnetic braking is applied. We seek to probe here whether the same four strict observables of $\tau$ Sco could be reconciled when using the modelling assumptions as described above. Here we do not introduce efficiency parameters, instead we test whether a change in mixing prescription alone could remedy the discrepancies.

%
%
\subsubsection{Impact of initial mass, rotation, and mixing prescription}

Figure \ref{fig:f6} shows a reference model (dashed line) with $M_{\rm ini}=$18 M$_\odot$ and $\varv_{\rm ini} = 300$ km\,s$^{-1}$, using the D22 scheme (see above) and assuming a 600 G polar field strength constant in time. The colour-coding scales with the stellar age. %
When decreasing the initial mass, only modest differences are seen (primarily, a shift in the effective temperature to lower values).
Likewise, when the initial rotational velocity is increased (by 100 km\,s$^{-1}$ here) compared to the reference model, the impact remains small.
Changing the mixing prescription, however, leads to a large difference. The model with the D11 scheme allows for a far more efficient chemical mixing and due to the increased mean molecular weight on average inside the star a more vertical evolution on the Kiel diagram. 

From these model computations, we can conclude that within observational and modelling uncertainty the Kiel diagram predicts an age of at most 6 Myr\footnote{The values obtained from \citet{simondiaz2006} may allow for a somewhat higher age, although the values we derive in this study would point to a lower age and perhaps very slightly lower initial mass than 17 M$_\odot$, based on the \textsc{genec} models.}. This is at odds with the model ages predicted from the Hunter-P diagram where we encounter the same problem as before. Namely, sufficient nitrogen enrichment and low spin velocity are not obtained with a self-consistent solution when using standard formulas for chemical mixing and magnetic braking. This is consistent with findings of \citealt{meynet2011} and \PaperI, where the model predictions typically yield notable surface enrichment in the second half of the main sequence only.

%
%
%
%
%
%
%

\section{Discussion}\label{sec:sec6}

We have shown that \textit{standard} single-star evolutionary models \nobreak{cannot} simultaneously reproduce the observed log $g$, $T_{\rm eff}$, N/H, and $P_{\rm rot}$ of $\tau$~Sco. 

Single-star models matching the observations on the Kiel diagram (fitting log $g$ and  $T_{\rm eff}$) can either reproduce the N/H ratio but not the observed rotation, or, inversely can reproduce the observed rotation but not the surface enrichment. This is precisely the same problem encountered by the post-merger models of \cite{schneider2020}.

Possible resolutions, in the frame of single-star models, is to consider a more efficient chemical mixing and/or a more efficient magnetic braking. The latter one hypothetically could be replaced by a magnetic field decay scenario. Overall, we find that in order to reconcile all observables of $\tau$~Sco, rather extreme assumptions are necessary, which would not be compatible with observations of other stars.

%
%
\subsection{Chemical mixing}

Rotational mixing in massive stars is presently one of the most uncertain processes influencing stellar evolution. A large number of observations show discrepancies with current model predictions \citep[e.g.,][]{hunter2009,martins2017,cazorla2017b,markova2018}. 

Potential sources of these discrepancies are likely many-fold. Importantly, stellar evolution model computations remain restricted to a one-dimensional treatment. Therefore, necessarily, scaling factors are introduced to model chemical mixing \citep{pin1989} and some mixing prescriptions are derived based on order-of-magnitude estimates \citep{brott2011}. Even more, the relevant physical processes and their interactions are not yet fully understood \citep[e.g.,][]{Maeder2009a} and some processes are not yet ubiquitously modelled and included in the computations, such as mixing by internal gravity waves \citep{dec2009,rogers2013,mathis2013,rogers2017,aerts2015,bowman2019b,bowman2019a,bowman2020a}. 

Despite the notable uncertainties, a drastic increase in the efficiency of the present-day prescriptions of rotational mixing by a factor of 3 (in the initially fast-rotator \textsc{mesa} model) or a factor of 10 (in the initially moderate-rotator \textsc{mesa} model) seem excessive in a single-star model. We estimate that (along with an increase in magnetic braking efficiency) a similar increase in chemical mixing would also be required in the "D11" \textsc{genec} model to obtain a self-consistent solution. 

Let us recall here that the chemical mixing efficiency in rotating stellar evolution models is calibrated such that with an initial mean rotation rate, they reproduce the mean surface nitrogen abundances of B-type stars at the end of the main sequence phase \citep[e.g.,][]{brott2011}. If we change this mixing efficiency for the purpose of fitting the data of $\tau$~Sco, then an appropriate physical reason should be given. At the moment, we are unaware of a physical cause which could be invoked in the single-star channel. Nevertheless, adopting a factor of 10 increase is not unprecedented in evolutionary modelling, for example, \cite{ad2020} use such an increased mixing efficiency in their approach. 

%
%
%

%
Interestingly, $\tau$ Sco stands out from the sample of magnetic massive stars as being a putative blue straggler. Blue stragglers are often associated with stellar mergers or quasi-chemically homogeneously evolving stars. Although quasi-chemically homogeneous evolution via long-term rapid rotation may help explain the surface enrichment of $\tau$~Sco, it is unclear if at all the star could suddenly become a very slow rotator as observed, therefore this evolutionary channel does not seem very favourable. Magnetic OB stars are typically found to be consistent with a usual "redward" evolution after their initial spin down (\PaperI{}and \PaperII). 
%
Furthermore, being a magnetic blue straggler poses the interesting question whether all blue stragglers would have a detectable magnetic field. \cite{schneider2016}
suggested that since the merger rate may be higher in blue stragglers, consequently the incidence rate of magnetism may also be higher. However, observational efforts dedicated to hot stars have not found such hints yet \citep{mathys1988,grunhut2017}.

%
\cite{morel2008,aerts2014} and \cite{martins2012,martins2015} analysed the nitrogen enrichment of magnetic OB stars and identified a number of cases with nitrogen excess (including $\tau$~Sco). Presently, it remains elusive why some stars show excess and others do not. In \PaperI, we proposed that the observable nitrogen enrichment of magnetic stars largely depends on presently unconstrained mixing processes (and their efficiencies), therefore the outcome is a multivariate function of a number of parameters \citep{aerts2014,maeder2014a}, and not only a function of the magnetic field strength.

%
%
\subsection{Magnetic field evolution and spin down}\label{sec:42}

Regarding its magnetic field characteristics, $\tau$ Sco clearly stands out of the known non-chemically peculiar magnetic B-type stars (this sample is discussed by \citealt{shultz32019} and confronted with evolutionary models in \PaperII). 
%
The complexity of $\tau$ Sco's magnetic field is unique. In general, a dipole-dominated model allows for reproducing magnetic field measurements (even if other higher order harmonics are present), although a number of cases indeed point to a quadruple-dominated geometry \citep{shultz2018,shultz22019}. Recently, \cite{daviduraz2021} found that NGC~1624~-~2's magnetic field is more complex than the typically assumed pure dipole. This evidence suggests that deviations from the pure dipole geometry may not be uncommon but require extensive monitoring to identify it.

The evolution of surface fossil magnetic fields remains largely uncertain. Several observational studies are consistent with magnetic flux conservation, whereas other empirical evidence points to a more rapid decline in field strength over time, suggesting a flux decay scenario (see \PaperI{} and \PaperII, and references therein). 

From a theoretical standpoint, fossil fields are expected to slowly dissipate on an Ohmic timescale, given by the induction equation of non-ideal MHD: 
\begin{equation}
    \eta \nabla^{2} \mathbf{B} = \frac{\partial \mathbf{B}}{\partial t} \, ,
\end{equation}
\noindent where $\eta$ is the magnetic diffusivity (assumed to be constant in space), $\mathbf{B}$ is the magnetic field vector, $t$ is the time, and we assume that the fluid velocity is zero, accounting for an equilibrium fossil magnetic field \citep{braithwaite2017}. 

For a fully ionised plasma, the diffusion timescale can be approximated as the ratio of the square of a characteristic length scale and the magnetic diffusivity:
\begin{equation}
    t_{\rm diff} \sim \frac{R^2}{\eta}  \, .
\end{equation}
\noindent For the Sun and low-mass stars, this formula typically leads to estimates of diffusion timescales longer than the main sequence lifetime \citep[e.g.,][]{cowling1945,feiden2014}. Here, we refrain from providing an estimate for more massive stars (and $\tau$~Sco in particular) since the uncertainty in an appropriate magnetic diffusivity is several orders of magnitudes \citep[e.g.,][]{charbonneau2001}. Since the outward diffusion of the magnetic flux is a possible mechanism to explain why flux decay may need to be invoked as a field evolution scenario, in this case we would require the diffusion timescale to be shorter than the $\sim$~12 Myr main sequence lifetime of our models. In conclusion, the scenario of magnetic flux decay still requires rigorous theoretical considerations \citep{braithwaite2017}.

The magnetic braking formula by \cite{ud2009} is robust but relies on important assumptions such as the field geometry and alignment with the rotation axis. In principle, the complexity of $\tau$ Sco's magnetic field geometry is expected to decrease, and not increase, the spin-down efficiency since the lowest order harmonic (i.e., the dipole component) has the most relevant contribution to spin down. Nevertheless, the impact of a strong azimuthal field remains uncharacterised. Recent simulations (ud Doula et al., in prep.) show that oblique rotation does not significantly affect the spin-down scaling either\footnote{Deviations in the rate of angular momentum loss by approximately up to 30\% can be reached for oblique rotators (ud Doula, priv.comm.). However, these deviations are expected to decrease the rate of angular momentum loss.}.

%

%
%
\subsection{Alternative evolutionary scenarios}

Although the single-star and the merger scenario have been suggested to explain the observables of $\tau$ Sco, nature may potentially offer different ways to explain them. 

Two main channels might be considered. \textit{i)} Higher mixing efficiency may be induced by a close companion star. Recent works have indeed shown that tidally-induced mixing can supersede the efficiency compared to a single star \citep[e.g.,][]{song2013}. This would require the presence of a main sequence companion star (see below). \textit{ii)} The surface enrichment of $\tau$ Sco might have been caused by an earlier mass-transfer event from a more massive companion (but without a merger event), leaving behind a present-day helium star or a compact object. Since the proper motion and radial velocity of $\tau$~Sco are low, it is unlikely that it could have been ejected from such a system as a run-away star during the supernova explosion creating the compact remnant. However, in this hypothetical scenario, the system's orbit may have widened sufficiently that it is not a close binary system anymore.

\begin{figure}
\begin{center}
\includegraphics[width=7cm]{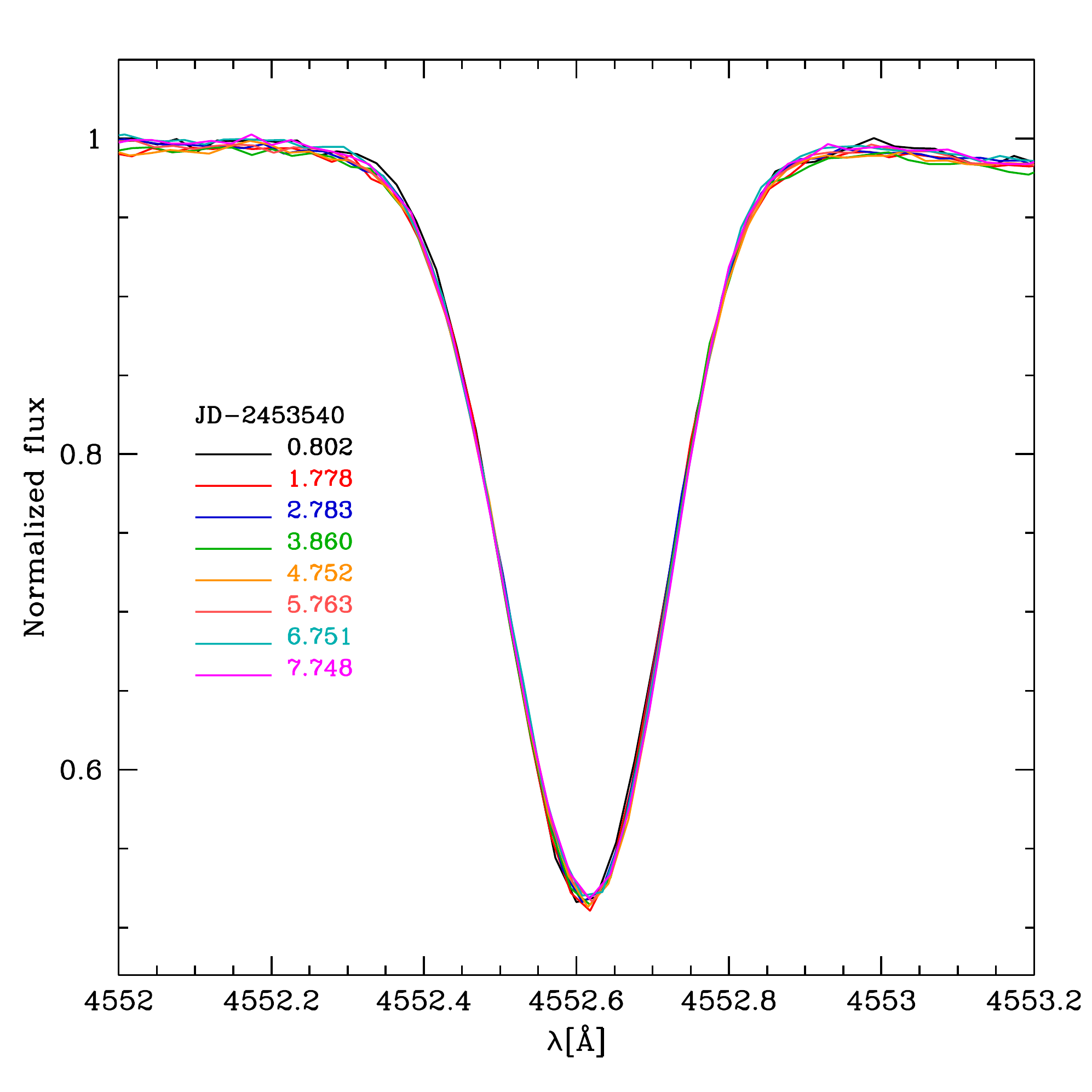}
\caption{The \ion{Si}{iii}~4553 line in the spectra of $\tau$ Sco taken over 8 days with the CHFT/ESPaDOnS instrument.}\label{fig:rv}
\end{center}
\end{figure}

In order to check the possibility that $\tau$ Sco is in a close binary system (with an orbital period of less than 20 days), we retrieved CFHT/ESPaDOnS spectra\footnote{The spectra were collected from the Polarbase database of the CFHT/ESPaDOnS and TBL/NARVAL spectropolarimeters, accesible at: \url{http://polarbase.irap.omp.eu/}} taken in 2005 between June 19$^{\rm th}$ and June 26$^{\rm th}$. Figure~\ref{fig:rv} displays these spectra around the \ion{Si}{iii}~4553 line. The radial velocity is remarkably stable over a week, with variations estimated to be less than 1 km\,s$^{-1}$. Other lines show the same behaviour. We also checked that no variability is present in spectra taken within the same night, a few hours apart. This hints at the absence of a massive companion on a short period orbit. Hence, effects of close binarity on the present-day properties and evolution of $\tau$ Sco are not found. The probability that we would view such a binary system at an inclination angle low enough to accommodate a $<$~1 km\,s$^{-1}$ projected orbital velocity is extremely low.

There is no clear consensus whether a long-period ($>$ 20 d) companion is present or not. The study of \cite{rizzuto2013} identifies a possible companion with a mass of roughly 1/3 that of the primary at an orbital separation of 2.84 AU; however, the literature is not conclusive about such a possible companion \citep{eggleton2008,pecaut2012,laf2014,grellmann2015}.

While it is difficult to observationally identify a main sequence companion on a long-period orbit, it remains to be worked out if it could produce a sufficient boost in chemical mixing, providing a viable alternative explanation for $\tau$~Sco's observables.

%
%
%
%

%
%
\subsection{Testing the single star vs. the merger scenario}

The major difficulty to evaluate the evolutionary history of $\tau$ Sco is that the presently considered scenarios (single star or merger) both lead to similar evolutionary tracks as the post-merger object is expected to evolve as a look-alike single star \citep{schneider2020}. However, a few important points can be addressed. \textit{i)} It is expected that the merger event leads to an abrupt mass removal from the system, and \citealt{glebbeek2013} calculate that this mass-shedding could amount to about 10 per cent of the total mass of the system in the merger of equal mass stars. Such mass-shedding may be linked to the transient phenomenon known as luminous red nova \citep{blago2017}, ultimately leading to a young nebula around the star \citep{langer2012,schneider2019,schneider2020}. A young nebula is not seen around $\tau$ Sco, however, an asymmetric bow-shock appears present in the 22~$\mu$m \textit{WISE} observations \citep{gv2019}. The origin of this bow-shock remains to be characterised. \textit{ii)} Asteroseismology could provide strong evidence regarding the near core chemical composition and rotation profile of the star \citep{aerts2019,bowman2020b} and these predicted characteristics are distinguishable between the post-merger model of \citet{schneider2020} and our single-star models\footnote{$\tau$ Sco was observed with the Transiting Exoplanet Survey Satellite (TESS). Unfortunately, from the TESS data no clear non-radial oscillations are detectable which prevents seismic modelling (Buzasi, Aerts, Bowman, priv.comm.).}. \textit{iii)} The surface helium content is a strong indicator of evolutionary status and we now discuss this point further.

%
%
%
%
%
\begin{figure*}
\includegraphics[width=6cm]{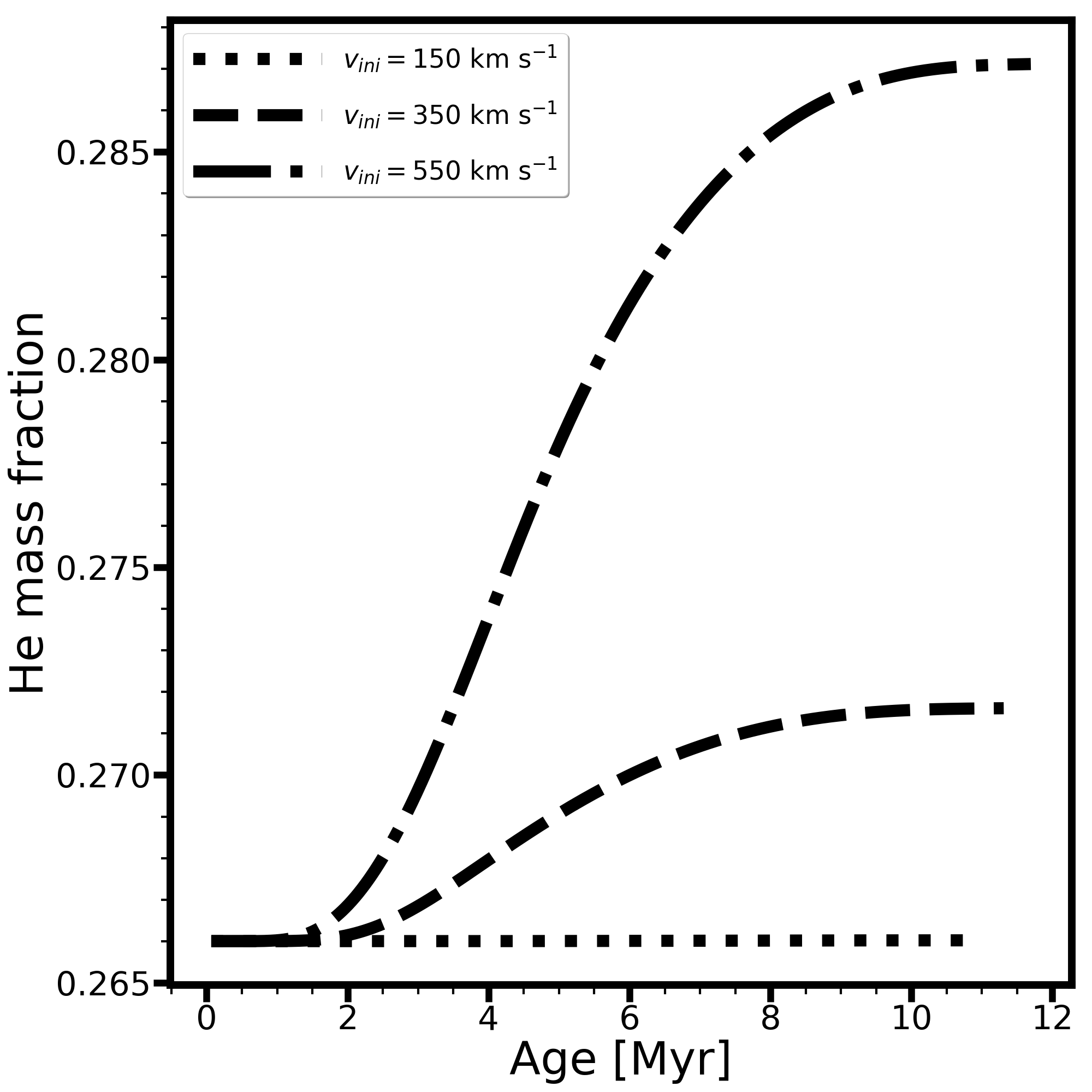}\includegraphics[width=6cm]{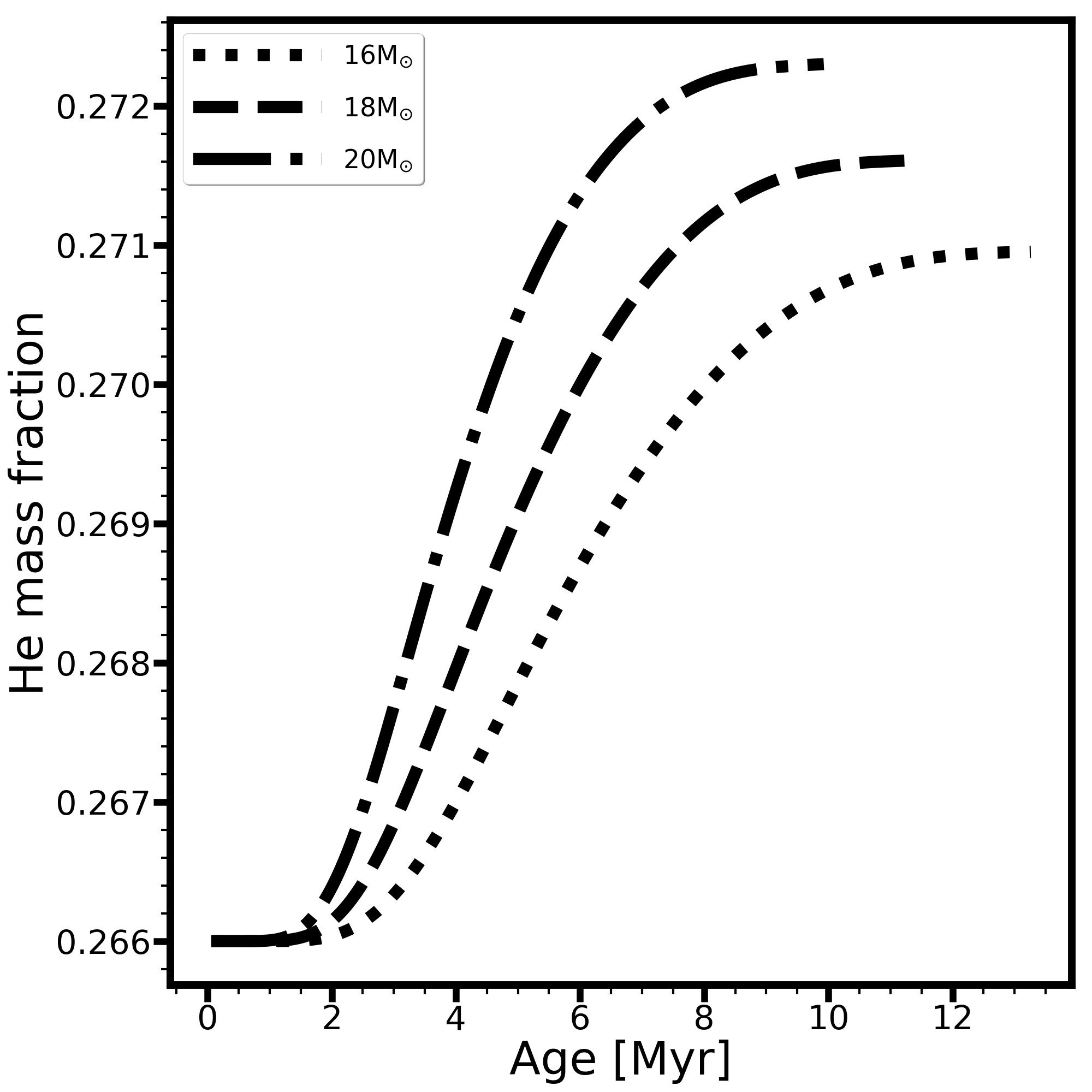}\includegraphics[width=6cm]{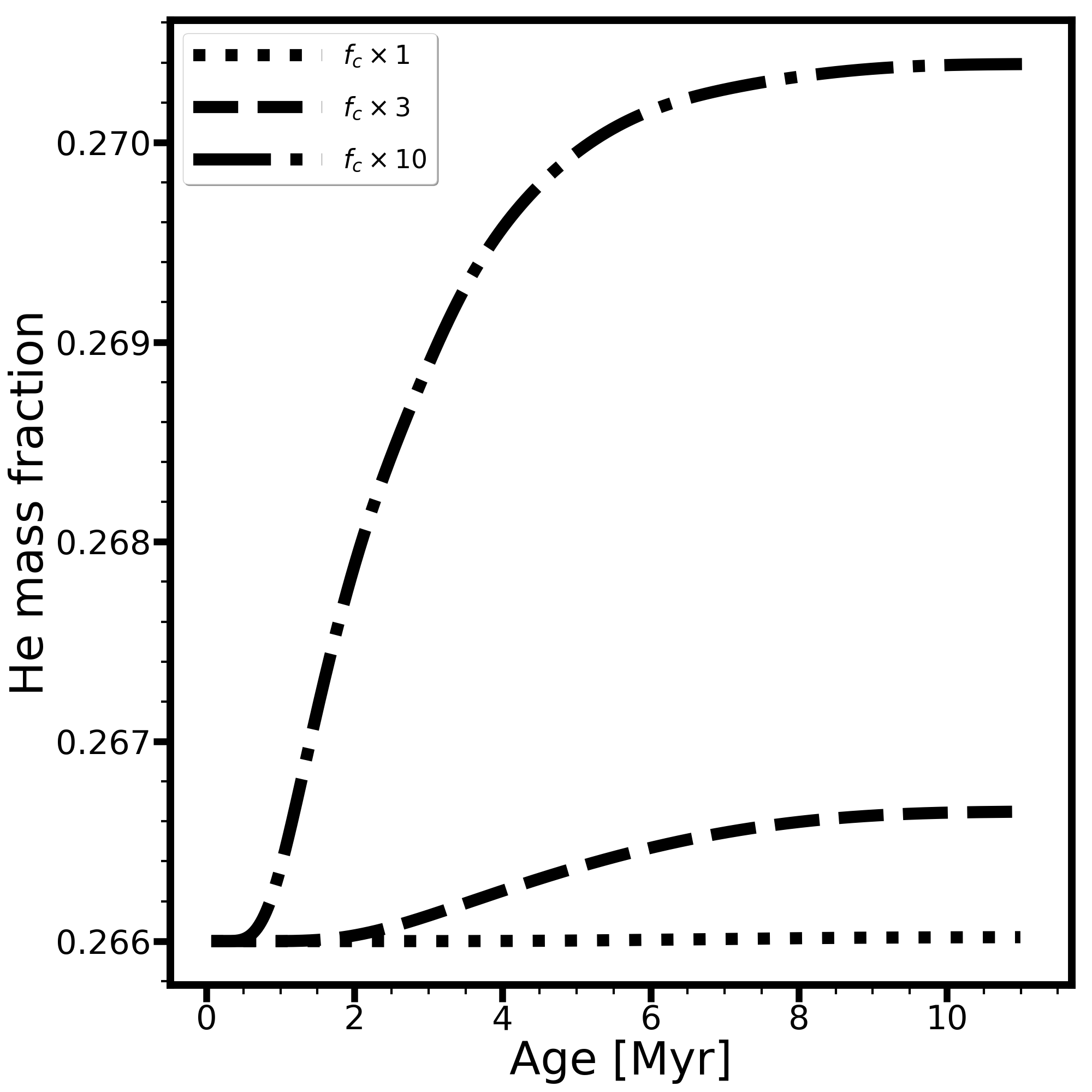}
\includegraphics[width=6cm]{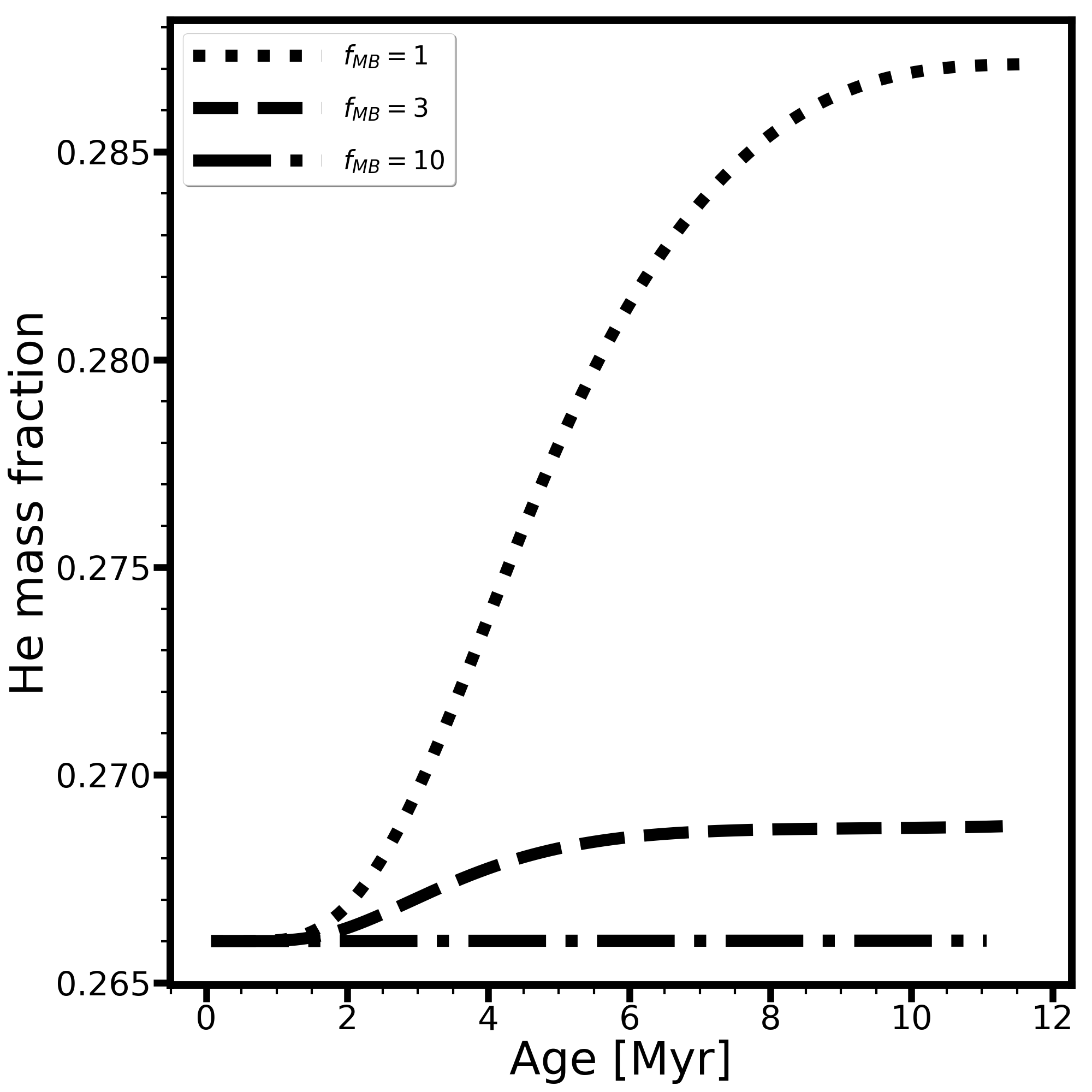}\includegraphics[width=6cm]{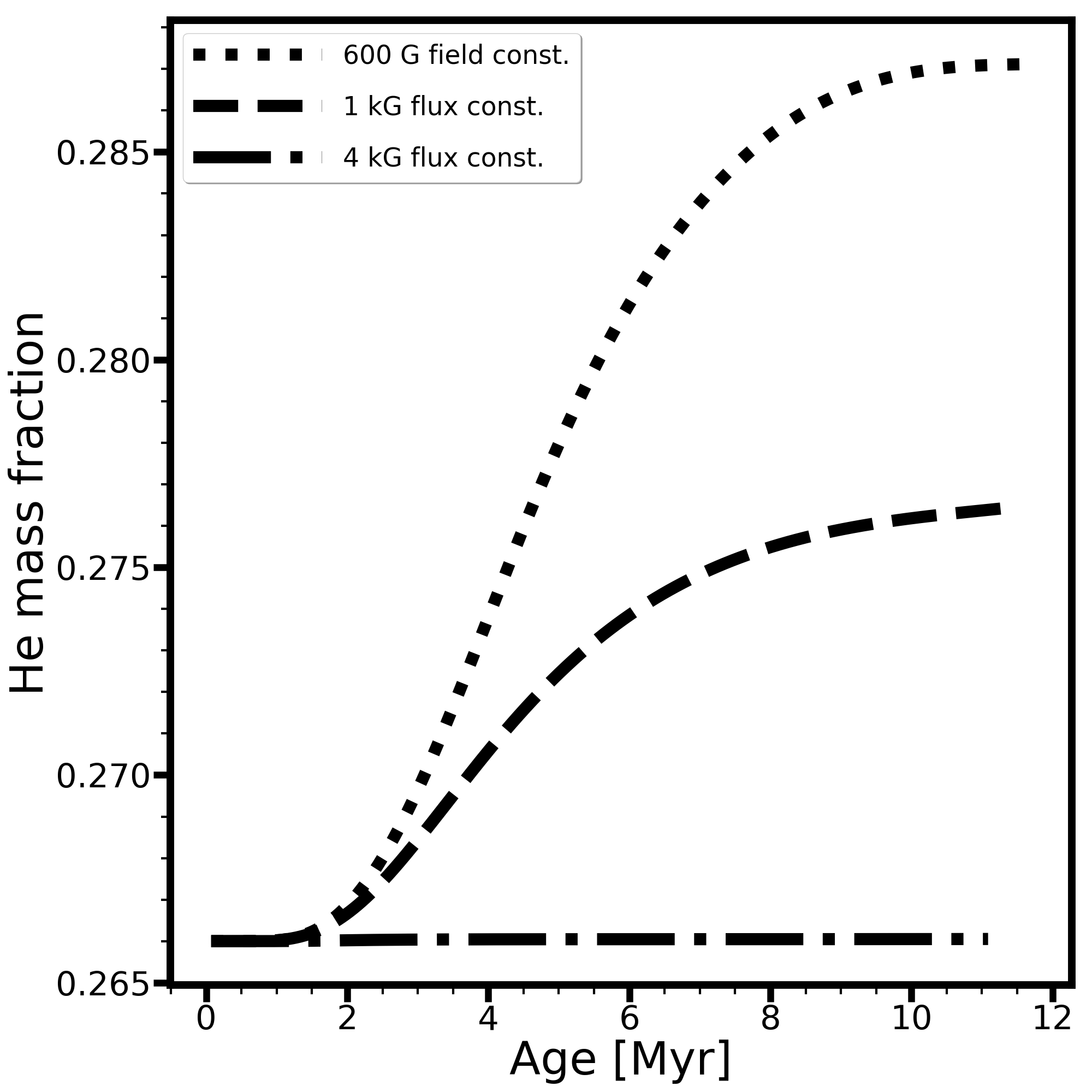}\includegraphics[width=6cm]{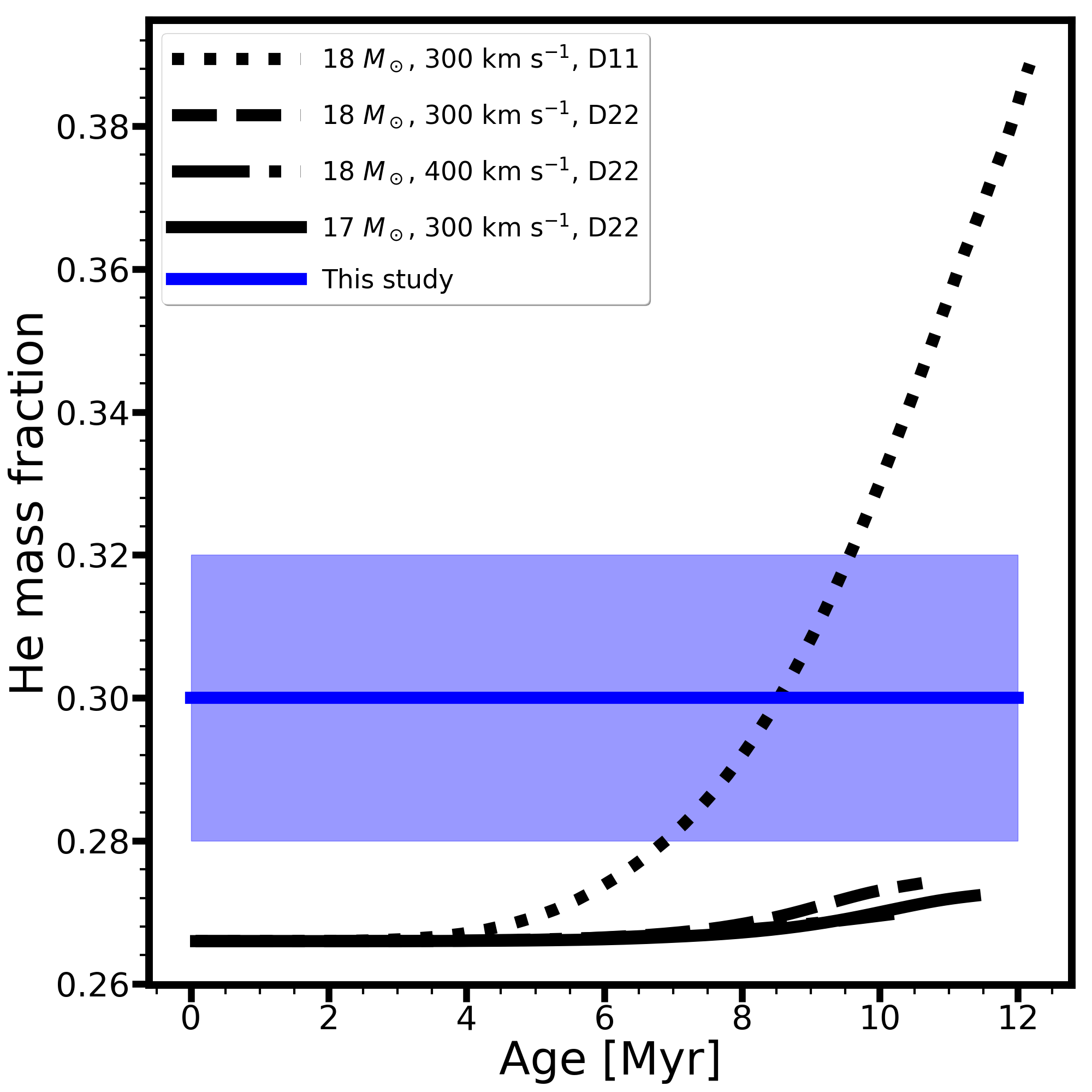}
\caption{Surface helium abundance vs. time. The same \textsc{mesa} models are shown as in Figures \ref{fig:f2}, and the top panels of Figure \ref{fig:f3} and \ref{fig:f5}, from top left to lower middle. The lower right panel shows the same \textsc{genec} models as in Figure \ref{fig:f6}, including our measured helium abundance with blue (however, see Sections~\ref{sec:intro} and Appendix~\ref{sec:app2} for a range of measurements by other authors). Note the different scales on the vertical axes.}\label{fig:he}
\end{figure*}

\subsubsection{Helium abundance}

Figure \ref{fig:he} shows those of our evolutionary models that have some helium enrichment over the main sequence. The models with increased magnetic braking efficiency and the models with field decay are not shown as they rapidly brake the rotation and do not lead to any notable helium enrichment. We find that within 6 Myr, none of the models produce a surface helium abundance compatible with our measured helium abundance of $Y = 0.30 \pm 0.02$ (Section \ref{sec:sec4}). Some models, particularly the "D11" \textsc{genec} model, may reach sufficiently high values but only close to the TAMS. 

In fact, in single stars a large amount of helium is only expected to be mixed to the surface at the end of the main sequence evolution, while, for example, the build-up of nitrogen is much more rapid.

\noindent The reason is that for nitrogen a strong gradient is built very early between the core (rich in nitrogen) and the envelope (with only the initial nitrogen), and the velocity of diffusion increases with the gradient of the considered element (Eq. 3 of \citealt{meynet2004}).
\noindent During the early evolution, the contrast between the
helium abundance in the core and envelope is much weaker thus the diffusion is much less rapid. Eventually, the gradient in helium composition will increase and, consequently, will allow for some surface enrichment.

Although in \textsc{mesa} we adopted a commonly used formalism with a scaling factor $f_\mu$ to mitigate the impact of composition gradients \citep{yoon2006, brott2011}, the rapid decrease of the rotational velocity means that these models cannot mix a sufficient amount of helium to the stellar surface. Of course, an important provision is that the effect of the magnetic field on chemical mixing still needs to be explored. Especially for $\tau $ Sco's complex magnetic field, this remains an open question which warrants further investigation.\\ 

In summary, when adopting a baseline of $Y_{\rm ini} = 0.266$ in our models, the measured helium abundance is at odds with single-star model predictions at a 1-$\sigma$ level. This clue suggests that invoking binarity may be necessary.

\subsection{Age and abundance of Upper Sco and $\tau$ Sco's membership}

The proper motion and radial velocity of $\tau$~Sco are broadly consistent with other stars identified as members of the Upper Sco association. The \textit{Gaia} DR2 parallax places $\tau$~Sco near the edge of the association with a distance of 195 pc, compared to the association's mean distance of 145 pc \citep{gaia2018}. %


The age of the Upper Sco association is constrained by main-sequence turn-off fitting at about 11 Myr by several authors \citep{sartori2003,pecaut2012,feiden2016,trevor2019}. Some authors find a much lower value of about 5 Myr \citep{degeus1989,blaauw1991,preibisch2002,dahm2009}. The differences in reported ages are attributed to modelling uncertainties and observational bias but not to an actual age spread in the association \citep{fang2017, donaldson2017}.

However, it is important to realise that Upper Sco's age determination plays a vital role in determining the star's evolutionary history. If Upper Sco's age is indeed about 11 Myr, then a rejuvenating mechanism (disfavouring single-star evolution) must be invoked since the apparent age of the star is less than 6 Myr. A detailed kinematic study of $\tau$ Sco is much needed and tracing back its birthplace could help to better constrain its evolution. 

Previous studies have found that the elemental abundance of Upper Sco is close to solar \citep{mamajek2013}. This lends support to choosing the solar metallicity (and nitrogen abundance) as the baseline to model the chemical enrichment. Therefore, we do not expect that the observed nitrogen excess of $\tau$~Sco could somehow be due to an initially higher nitrogen abundance. We further test the impact of choosing various initial abundances in Appendix \ref{sec:app1}. 

In summary, there is no evidence suggesting that $\tau$ Sco does not belong to Upper Sco. Likewise, there is no evidence that an unusual star formation could have led to a late formation of $\tau$~Sco. However, the age of Upper Sco plays a decisive role in establishing $\tau$~Sco's evolutionary history. Therefore further work on constraining Upper Sco's age and chemical composition with similar modelling assumptions as taken here would be a logical next step.
%

%
%
\section{Conclusions}\label{sec:sec7}

$\tau$ Scorpii is a very special and unusual magnetic star. In this work, we address the question if it is possible at all to reconcile the observable properties of $\tau$ Sco with single-star evolution.

%
%
We perform atmospheric modelling and determine the CNO and helium abundances. The measured nitrogen enrichment (along with carbon depletion) implies that the stellar surface is mixed with core-processed material. The derived helium abundance is close to the baseline value but might hint a slight excess. This rules out that $\tau$ Sco could be an evolved helium star. Equal-mass mergers are expected to lead to a substantial increase in the surface helium abundance \citep{glebbeek2013,schneider2016}, whereas in single stars the surface helium abundance should only increase towards the end of the main sequence.

We present experiments with stellar evolution models from which we draw the following conclusions: 
\begin{itemize}
    \item Considering observational and modelling uncertainties, we estimate a maximum age of 6 Myr for $\tau$~Sco in the framework of the single-star scenario.
    \item In order to produce surface enrichment by rotational mixing, an initial spin velocity higher than 150~km~s$^{-1}$ is required. This disfavours that $\tau$ Sco could have been a very slow rotator at the ZAMS if it was a single star. 
    \item We can only reconcile the spin-down of the stellar surface with the observed 41 day rotational period \citep{donati2006} if we increase the magnetic braking efficiency by a factor of 10 compared to a standard prescription.
    \item Alternatively, an increase in magnetic braking efficiency may not need to be invoked if one assumes an initially much stronger field (30~kG) which has rapidly decayed over time. There exists tentative evidence for magnetic flux decay from observations \citep[e.g.,][]{fossati2016,shultz32019}, however, the decay rate remains to be quantified. Interestingly, the flux decay could potentially be caused by a companion star \citep{vidal2019}.
    \item In a typical rapidly-rotating massive star model the observed nitrogen excess may be reached within 6 Myr. However, due to the necessary rapid spin-down by magnetic braking, in both of the above scenarios one needs to invoke a subsequent increase in the efficiency of rotational mixing by at least a factor of 3.
\end{itemize}

%

Presently, neither the single-star nor the post-merger models can satisfactorily explain the simultaneous slow rotation and nitrogen excess of $\tau$ Sco. Neither can the rotation be explained with a present-day sub-kG magnetic field strength, unless a seemingly extreme magnetic field evolution is invoked. Although there are ways to reconcile these discrepancies in both channels, binary evolution may provide a more versatile scenario to achieve this. We find no evidence for a present-day close companion star and the literature remains inconclusive regarding a possible companion on a long-period orbit.


A number of follow-up observations and modelling efforts can help shed more light on the nature of $\tau$~Sco.
\begin{itemize}
\item Extended series of Stokes~\textit{Q},~\textit{U} profiles over the rotation period of the star would allow to measure the magnetic field modulus, aiding to constrain the surface field geometry. 
\item A detailed observational investigation of a possible companion object (a long-period main sequence star, or a compact remnant) would be of great value to assess binary-evolution scenarios which may not lead to a stellar merger.
\item A theoretical explanation of tidally-induced mixing from a highly eccentric, long-period companion or whether previous binary mass transfer could produce self-consistent solutions should be scrutinised in the context of $\tau$~Sco.
\end{itemize}

Finally, we emphasise that $\tau$ Sco does not seem to be a representative magnetic B-type star and its magnetic field is much more complex than those inferred for other B-type stars. Consequently, whichever evolutionary channel produced this special star, may not be able to provide a general explanation for the origin of fossil magnetic fields in massive stars. However, it is an intriguing prospect if unusually complex magnetic field geometries of OBA stars may be linked to stellar multiplicity.

\section*{Acknowledgements}

We thank the anonymous referee for a constructive report which helped improve the manuscript. We thank John Hillier for making \textsc{cmfgen} available to the community and for constant help with it. We thank the \textsc{mesa} developers for making their code publicly available. We appreciate discussions with Fabian Schneider, Henk Spruit, Conny Aerts, Derek Buzasi, Dominic Bowman, Cole Johnston, Jo Puls, Paco Najarro, Asif ud-Doula, Difeng Guo, and Sam Geen.
GM acknowledges support from the Swiss National Science Foundation (project number 200020-172505). G.M. has received funding from the European Research Council (ERC) under the European Union's Horizon 2020 research and innovation programme (grant agreement No 833925, project STAREX). 
ADU gratefully acknowledges the support of the
Natural Science and Engineering Research Council of Canada (NSERC). This work is supported by NASA under award number 80GSFC17M0002.
This work made use of the Polarbase database (developed and maintained by CNRS/INSU, Observatoire Midi-Pyr\'en\'ees and Universit\'e Toulouse III).
This work was carried out on the Dutch national e-infrastructure with the support of SURF Cooperative.

\section*{Data availability}

A full reproduction package is shared on zenodo in accordance with the Research Data Management plan of the Anton Pannekoek Institute for Astronomy at the University of Amsterdam: \nobreak{\url{10.5281/zenodo.4633408}}

%
%

\bibliographystyle{mnras}
\bibliography{ref}

\begin{thebibliography}{}
\makeatletter
\relax
\def\mn@urlcharsother{\let\do\@makeother \do\$\do\&\do\#\do\^\do\_\do\%\do\~}
\def\mn@doi{\begingroup\mn@urlcharsother \@ifnextchar [ {\mn@doi@}
  {\mn@doi@[]}}
\def\mn@doi@[#1]#2{\def\@tempa{#1}\ifx\@tempa\@empty \href
  {http://dx.doi.org/#2} {doi:#2}\else \href {http://dx.doi.org/#2} {#1}\fi
  \endgroup}
\def\mn@eprint#1#2{\mn@eprint@#1:#2::\@nil}
\def\mn@eprint@arXiv#1{\href {http://arxiv.org/abs/#1} {{\tt arXiv:#1}}}
\def\mn@eprint@dblp#1{\href {http://dblp.uni-trier.de/rec/bibtex/#1.xml}
  {dblp:#1}}
\def\mn@eprint@#1:#2:#3:#4\@nil{\def\@tempa {#1}\def\@tempb {#2}\def\@tempc
  {#3}\ifx \@tempc \@empty \let \@tempc \@tempb \let \@tempb \@tempa \fi \ifx
  \@tempb \@empty \def\@tempb {arXiv}\fi \@ifundefined
  {mn@eprint@\@tempb}{\@tempb:\@tempc}{\expandafter \expandafter \csname
  mn@eprint@\@tempb\endcsname \expandafter{\@tempc}}}

\bibitem[\protect\citeauthoryear{{Aerts} \& {Rogers}}{{Aerts} \&
  {Rogers}}{2015}]{aerts2015}
{Aerts} C.,  {Rogers} T.~M.,  2015, \mn@doi [\apjl]
  {10.1088/2041-8205/806/2/L33}, \href
  {https://ui.adsabs.harvard.edu/abs/2015ApJ...806L..33A} {806, L33}

\bibitem[\protect\citeauthoryear{{Aerts}, {Molenberghs}, {Kenward}  \&
  {Neiner}}{{Aerts} et~al.}{2014}]{aerts2014}
{Aerts} C.,  {Molenberghs} G.,  {Kenward} M.~G.,   {Neiner} C.,  2014, \mn@doi
  [\apj] {10.1088/0004-637X/781/2/88}, \href
  {http://adsabs.harvard.edu/abs/2014ApJ...781...88A} {781, 88}

\bibitem[\protect\citeauthoryear{{Aerts}, {Mathis}  \& {Rogers}}{{Aerts}
  et~al.}{2019}]{aerts2019}
{Aerts} C.,  {Mathis} S.,   {Rogers} T.~M.,  2019, \mn@doi [\araa]
  {10.1146/annurev-astro-091918-104359}, \href
  {https://ui.adsabs.harvard.edu/abs/2019ARA&A..57...35A} {57, 35}

\bibitem[\protect\citeauthoryear{{Aguilera-Dena}, {Langer}, {Antoniadis}  \&
  {M{\"u}ller}}{{Aguilera-Dena} et~al.}{2020}]{ad2020}
{Aguilera-Dena} D.~R.,  {Langer} N.,  {Antoniadis} J.,   {M{\"u}ller} B.,
  2020, \mn@doi [\apj] {10.3847/1538-4357/abb138}, \href
  {https://ui.adsabs.harvard.edu/abs/2020ApJ...901..114A} {901, 114}

\bibitem[\protect\citeauthoryear{{Alfv{\'e}n}}{{Alfv{\'e}n}}{1942}]{alfven1942}
{Alfv{\'e}n} H.,  1942, \mn@doi [\nat] {10.1038/150405d0}, \href
  {http://adsabs.harvard.edu/abs/1942Natur.150..405A} {150, 405}

\bibitem[\protect\citeauthoryear{{Aller}, {Faulkner}  \& {Norton}}{{Aller}
  et~al.}{1966}]{aller1966}
{Aller} L.~H.,  {Faulkner} D.~J.,   {Norton} R.~H.,  1966, \mn@doi [\apj]
  {10.1086/148704}, \href
  {https://ui.adsabs.harvard.edu/abs/1966ApJ...144.1073A} {144, 1073}

\bibitem[\protect\citeauthoryear{{Anders} \& {Grevesse}}{{Anders} \&
  {Grevesse}}{1989}]{anders1989}
{Anders} E.,  {Grevesse} N.,  1989, \mn@doi [\gca]
  {10.1016/0016-7037(89)90286-X}, \href
  {https://ui.adsabs.harvard.edu/abs/1989GeCoA..53..197A} {53, 197}

\bibitem[\protect\citeauthoryear{{Asplund}, {Grevesse}  \& {Sauval}}{{Asplund}
  et~al.}{2005}]{asplund2005}
{Asplund} M.,  {Grevesse} N.,   {Sauval} A.~J.,  2005, in {Barnes} III T.~G.,
  {Bash} F.~N.,  eds,  Astronomical Society of the Pacific Conference Series
  Vol. 336, Cosmic Abundances as Records of Stellar Evolution and
  Nucleosynthesis. p.~25

\bibitem[\protect\citeauthoryear{{Asplund}, {Grevesse}, {Sauval}  \&
  {Scott}}{{Asplund} et~al.}{2009}]{asplund2009}
{Asplund} M.,  {Grevesse} N.,  {Sauval} A.~J.,   {Scott} P.,  2009, \mn@doi
  [\araa] {10.1146/annurev.astro.46.060407.145222}, \href
  {http://adsabs.harvard.edu/abs/2009ARA%26A..47..481A} {47, 481}

\bibitem[\protect\citeauthoryear{{Blaauw}}{{Blaauw}}{1991}]{blaauw1991}
{Blaauw} A.,  1991, in {Lada} C.~J.,  {Kylafis} N.~D.,  eds,  NATO Advanced
  Study Institute (ASI) Series C Vol. 342, The Physics of Star Formation and
  Early Stellar Evolution. p.~125

\bibitem[\protect\citeauthoryear{{Blagorodnova} et~al.,}{{Blagorodnova}
  et~al.}{2017}]{blago2017}
{Blagorodnova} N.,  et~al., 2017, \mn@doi [\apj] {10.3847/1538-4357/834/2/107},
  \href {https://ui.adsabs.harvard.edu/abs/2017ApJ...834..107B} {834, 107}

\bibitem[\protect\citeauthoryear{{Bowman}}{{Bowman}}{2020}]{bowman2020b}
{Bowman} D.~M.,  2020, arXiv e-prints, \href
  {https://ui.adsabs.harvard.edu/abs/2020arXiv200811162B} {p. arXiv:2008.11162}

\bibitem[\protect\citeauthoryear{{Bowman} et~al.,}{{Bowman}
  et~al.}{2019a}]{bowman2019b}
{Bowman} D.~M.,  et~al., 2019a, \mn@doi [Nature Astronomy]
  {10.1038/s41550-019-0768-1}, \href
  {https://ui.adsabs.harvard.edu/abs/2019NatAs...3..760B} {3, 760}

\bibitem[\protect\citeauthoryear{{Bowman} et~al.,}{{Bowman}
  et~al.}{2019b}]{bowman2019a}
{Bowman} D.~M.,  et~al., 2019b, \mn@doi [\aap] {10.1051/0004-6361/201833662},
  \href {https://ui.adsabs.harvard.edu/abs/2019A&A...621A.135B} {621, A135}

\bibitem[\protect\citeauthoryear{{Bowman}, {Burssens}, {Sim{\'o}n-D{\'\i}az},
  {Edelmann}, {Rogers}, {Horst}, {R{\"o}pke}  \& {Aerts}}{{Bowman}
  et~al.}{2020}]{bowman2020a}
{Bowman} D.~M.,  {Burssens} S.,  {Sim{\'o}n-D{\'\i}az} S.,  {Edelmann}
  P.~V.~F.,  {Rogers} T.~M.,  {Horst} L.,  {R{\"o}pke} F.~K.,   {Aerts} C.,
  2020, \mn@doi [\aap] {10.1051/0004-6361/202038224}, \href
  {https://ui.adsabs.harvard.edu/abs/2020A&A...640A..36B} {640, A36}

\bibitem[\protect\citeauthoryear{{Bragan{\c{c}}a}, {Daflon}, {Cunha}, {Bensby},
  {Oey}  \& {Walth}}{{Bragan{\c{c}}a} et~al.}{2012}]{2012AJ....144..130B}
{Bragan{\c{c}}a} G.~A.,  {Daflon} S.,  {Cunha} K.,  {Bensby} T.,  {Oey} M.~S.,
   {Walth} G.,  2012, \mn@doi [\aj] {10.1088/0004-6256/144/5/130}, \href
  {https://ui.adsabs.harvard.edu/abs/2012AJ....144..130B} {144, 130}

\bibitem[\protect\citeauthoryear{{Braithwaite}}{{Braithwaite}}{2008}]{braithwaite2008}
{Braithwaite} J.,  2008, \mn@doi [\mnras] {10.1111/j.1365-2966.2008.13218.x},
  \href {https://ui.adsabs.harvard.edu/abs/2008MNRAS.386.1947B} {386, 1947}

\bibitem[\protect\citeauthoryear{{Braithwaite} \& {Spruit}}{{Braithwaite} \&
  {Spruit}}{2017}]{braithwaite2017}
{Braithwaite} J.,  {Spruit} H.~C.,  2017, \mn@doi [Royal Society Open Science]
  {10.1098/rsos.160271}, \href
  {http://adsabs.harvard.edu/abs/2017RSOS....460271B} {4, 160271}

\bibitem[\protect\citeauthoryear{{Brott} et~al.,}{{Brott}
  et~al.}{2011}]{brott2011}
{Brott} I.,  et~al., 2011, \mn@doi [\aap] {10.1051/0004-6361/201016113}, \href
  {http://adsabs.harvard.edu/abs/2011A%26A...530A.115B} {530, A115}

\bibitem[\protect\citeauthoryear{{Cazorla}, {Morel}, {Naz{\'e}}, {Rauw},
  {Semaan}, {Daflon}  \& {Oey}}{{Cazorla} et~al.}{2017a}]{cazorla2017a}
{Cazorla} C.,  {Morel} T.,  {Naz{\'e}} Y.,  {Rauw} G.,  {Semaan} T.,  {Daflon}
  S.,   {Oey} M.~S.,  2017a, \mn@doi [\aap] {10.1051/0004-6361/201629841},
  \href {http://adsabs.harvard.edu/abs/2017A%26A...603A..56C} {603, A56}

\bibitem[\protect\citeauthoryear{{Cazorla}, {Naz{\'e}}, {Morel}, {Georgy},
  {Godart}  \& {Langer}}{{Cazorla} et~al.}{2017b}]{cazorla2017b}
{Cazorla} C.,  {Naz{\'e}} Y.,  {Morel} T.,  {Georgy} C.,  {Godart} M.,
  {Langer} N.,  2017b, \mn@doi [\aap] {10.1051/0004-6361/201730680}, \href
  {http://adsabs.harvard.edu/abs/2017A%26A...604A.123C} {604, A123}

\bibitem[\protect\citeauthoryear{{Charbonneau} \& {MacGregor}}{{Charbonneau} \&
  {MacGregor}}{2001}]{charbonneau2001}
{Charbonneau} P.,  {MacGregor} K.~B.,  2001, \mn@doi [\apj] {10.1086/322417},
  \href {https://ui.adsabs.harvard.edu/abs/2001ApJ...559.1094C} {559, 1094}

\bibitem[\protect\citeauthoryear{{Cohen}, {Cassinelli}  \& {Waldron}}{{Cohen}
  et~al.}{1997}]{cohen1997}
{Cohen} D.~H.,  {Cassinelli} J.~P.,   {Waldron} W.~L.,  1997, \mn@doi [\apj]
  {10.1086/304678}, \href
  {https://ui.adsabs.harvard.edu/abs/1997ApJ...488..397C} {488, 397}

\bibitem[\protect\citeauthoryear{{Cohen}, {de Messi{\`e}res}, {MacFarlane},
  {Miller}, {Cassinelli}, {Owocki}  \& {Liedahl}}{{Cohen}
  et~al.}{2003}]{cohen2003}
{Cohen} D.~H.,  {de Messi{\`e}res} G.~E.,  {MacFarlane} J.~J.,  {Miller} N.~A.,
   {Cassinelli} J.~P.,  {Owocki} S.~P.,   {Liedahl} D.~A.,  2003, \mn@doi
  [\apj] {10.1086/367553}, \href
  {https://ui.adsabs.harvard.edu/abs/2003ApJ...586..495C} {586, 495}

\bibitem[\protect\citeauthoryear{{Cowley} \& {Merritt}}{{Cowley} \&
  {Merritt}}{1987}]{cowley1987}
{Cowley} C.~R.,  {Merritt} D.~R.,  1987, \mn@doi [\apj] {10.1086/165651}, \href
  {https://ui.adsabs.harvard.edu/abs/1987ApJ...321..553C} {321, 553}

\bibitem[\protect\citeauthoryear{{Cowling}}{{Cowling}}{1945}]{cowling1945}
{Cowling} T.~G.,  1945, \mn@doi [\mnras] {10.1093/mnras/105.3.166}, \href
  {http://adsabs.harvard.edu/abs/1945MNRAS.105..166C} {105, 166}

\bibitem[\protect\citeauthoryear{{Dahm} \& {Carpenter}}{{Dahm} \&
  {Carpenter}}{2009}]{dahm2009}
{Dahm} S.~E.,  {Carpenter} J.~M.,  2009, \mn@doi [\aj]
  {10.1088/0004-6256/137/4/4024}, \href
  {https://ui.adsabs.harvard.edu/abs/2009AJ....137.4024D} {137, 4024}

\bibitem[\protect\citeauthoryear{{David-Uraz} et~al.,}{{David-Uraz}
  et~al.}{2019}]{2019MNRAS.483.2814D}
{David-Uraz} A.,  et~al., 2019, \mn@doi [\mnras] {10.1093/mnras/sty3227}, \href
  {https://ui.adsabs.harvard.edu/abs/2019MNRAS.483.2814D} {483, 2814}

\bibitem[\protect\citeauthoryear{{David-Uraz}, {Petit}, {Shultz}, {Fullerton},
  {Erba}, {Keszthelyi}, {Seadrow}  \& {Wade}}{{David-Uraz}
  et~al.}{2021}]{daviduraz2021}
{David-Uraz} A.,  {Petit} V.,  {Shultz} M.~E.,  {Fullerton} A.~W.,  {Erba} C.,
  {Keszthelyi} Z.,  {Seadrow} S.,   {Wade} G.~A.,  2021, \mn@doi [\mnras]
  {10.1093/mnras/staa3768}, \href
  {https://ui.adsabs.harvard.edu/abs/2021MNRAS.501.2677D} {501, 2677}

\bibitem[\protect\citeauthoryear{{David}, {Hillenbrand}, {Gillen}, {Cody},
  {Howell}, {Isaacson}  \& {Livingston}}{{David} et~al.}{2019}]{trevor2019}
{David} T.~J.,  {Hillenbrand} L.~A.,  {Gillen} E.,  {Cody} A.~M.,  {Howell}
  S.~B.,  {Isaacson} H.~T.,   {Livingston} J.~H.,  2019, \mn@doi [\apj]
  {10.3847/1538-4357/aafe09}, \href
  {https://ui.adsabs.harvard.edu/abs/2019ApJ...872..161D} {872, 161}

\bibitem[\protect\citeauthoryear{{Decressin}, {Mathis}, {Palacios}, {Siess},
  {Talon}, {Charbonnel}  \& {Zahn}}{{Decressin} et~al.}{2009}]{dec2009}
{Decressin} T.,  {Mathis} S.,  {Palacios} A.,  {Siess} L.,  {Talon} S.,
  {Charbonnel} C.,   {Zahn} J.~P.,  2009, \mn@doi [\aap]
  {10.1051/0004-6361:200810665}, \href
  {https://ui.adsabs.harvard.edu/abs/2009A&A...495..271D} {495, 271}

\bibitem[\protect\citeauthoryear{{Donaldson}, {Weinberger}, {Gagn{\'e}}, {Boss}
   \& {Keiser}}{{Donaldson} et~al.}{2017}]{donaldson2017}
{Donaldson} J.~K.,  {Weinberger} A.~J.,  {Gagn{\'e}} J.,  {Boss} A.~P.,
  {Keiser} S.~A.,  2017, \mn@doi [\apj] {10.3847/1538-4357/aa9117}, \href
  {https://ui.adsabs.harvard.edu/abs/2017ApJ...850...11D} {850, 11}

\bibitem[\protect\citeauthoryear{{Donati} \& {Landstreet}}{{Donati} \&
  {Landstreet}}{2009}]{donati2009}
{Donati} J.-F.,  {Landstreet} J.~D.,  2009, \mn@doi [\araa]
  {10.1146/annurev-astro-082708-101833}, \href
  {http://adsabs.harvard.edu/abs/2009ARA%26A..47..333D} {47, 333}

\bibitem[\protect\citeauthoryear{{Donati} et~al.,}{{Donati}
  et~al.}{2006}]{donati2006}
{Donati} J.~F.,  et~al., 2006, \mn@doi [\mnras]
  {10.1111/j.1365-2966.2006.10558.x}, \href
  {https://ui.adsabs.harvard.edu/abs/2006MNRAS.370..629D} {370, 629}

\bibitem[\protect\citeauthoryear{{Eddington}}{{Eddington}}{1925}]{eddington1925}
{Eddington} A.~S.,  1925, The Observatory, \href
  {http://adsabs.harvard.edu/abs/1925Obs....48...73E} {48, 73}

\bibitem[\protect\citeauthoryear{{Eggenberger}, {Meynet}, {Maeder}, {Hirschi},
  {Charbonnel}, {Talon}  \& {Ekstr{\"o}m}}{{Eggenberger}
  et~al.}{2008}]{eggenberger2008}
{Eggenberger} P.,  {Meynet} G.,  {Maeder} A.,  {Hirschi} R.,  {Charbonnel} C.,
  {Talon} S.,   {Ekstr{\"o}m} S.,  2008, \mn@doi [\apss]
  {10.1007/s10509-007-9511-y}, \href
  {http://adsabs.harvard.edu/abs/2008Ap%26SS.316...43E} {316, 43}

\bibitem[\protect\citeauthoryear{{Eggleton} \& {Tokovinin}}{{Eggleton} \&
  {Tokovinin}}{2008}]{eggleton2008}
{Eggleton} P.~P.,  {Tokovinin} A.~A.,  2008, \mn@doi [\mnras]
  {10.1111/j.1365-2966.2008.13596.x}, \href
  {https://ui.adsabs.harvard.edu/abs/2008MNRAS.389..869E} {389, 869}

\bibitem[\protect\citeauthoryear{{Ekstr{\"o}m} et~al.,}{{Ekstr{\"o}m}
  et~al.}{2012}]{ekstroem2012}
{Ekstr{\"o}m} S.,  et~al., 2012, \mn@doi [\aap] {10.1051/0004-6361/201117751},
  \href {http://adsabs.harvard.edu/abs/2012A%26A...537A.146E} {537, A146}

\bibitem[\protect\citeauthoryear{{Endal} \& {Sofia}}{{Endal} \&
  {Sofia}}{1978}]{endal1978}
{Endal} A.~S.,  {Sofia} S.,  1978, \mn@doi [\apj] {10.1086/155904}, \href
  {http://adsabs.harvard.edu/abs/1978ApJ...220..279E} {220, 279}

\bibitem[\protect\citeauthoryear{{Fang}, {Herczeg}  \& {Rizzuto}}{{Fang}
  et~al.}{2017}]{fang2017}
{Fang} Q.,  {Herczeg} G.~J.,   {Rizzuto} A.,  2017, \mn@doi [\apj]
  {10.3847/1538-4357/aa74ca}, \href
  {https://ui.adsabs.harvard.edu/abs/2017ApJ...842..123F} {842, 123}

\bibitem[\protect\citeauthoryear{{Feiden}}{{Feiden}}{2016}]{feiden2016}
{Feiden} G.~A.,  2016, \mn@doi [\aap] {10.1051/0004-6361/201527613}, \href
  {https://ui.adsabs.harvard.edu/abs/2016A&A...593A..99F} {593, A99}

\bibitem[\protect\citeauthoryear{{Feiden} \& {Chaboyer}}{{Feiden} \&
  {Chaboyer}}{2014}]{feiden2014}
{Feiden} G.~A.,  {Chaboyer} B.,  2014, \mn@doi [\apj]
  {10.1088/0004-637X/789/1/53}, \href
  {https://ui.adsabs.harvard.edu/abs/2014ApJ...789...53F} {789, 53}

\bibitem[\protect\citeauthoryear{{Ferrario}, {Pringle}, {Tout}  \&
  {Wickramasinghe}}{{Ferrario} et~al.}{2009}]{ferrario2009}
{Ferrario} L.,  {Pringle} J.~E.,  {Tout} C.~A.,   {Wickramasinghe} D.~T.,
  2009, \mn@doi [\mnras] {10.1111/j.1745-3933.2009.00765.x}, \href
  {http://adsabs.harvard.edu/abs/2009MNRAS.400L..71F} {400, L71}

\bibitem[\protect\citeauthoryear{{Fletcher}}{{Fletcher}}{2018}]{2018PhDT........18F}
{Fletcher} C.~L.,  2018, PhD thesis, Department of Physics and Space Sciences,
  Florida Institute of Technology, Melbourne, Florida

\bibitem[\protect\citeauthoryear{{Fletcher}, {Petit}, {Cohen}, {Townsend}  \&
  {Wade}}{{Fletcher} et~al.}{2018}]{fletcher2018}
{Fletcher} C.~L.,  {Petit} V.,  {Cohen} D.~H.,  {Townsend} R.~H.,   {Wade}
  G.~A.,  2018, Contributions of the Astronomical Observatory Skalnate Pleso,
  \href {https://ui.adsabs.harvard.edu/abs/2018CoSka..48..144F} {48, 144}

\bibitem[\protect\citeauthoryear{{Fossati} et~al.,}{{Fossati}
  et~al.}{2016}]{fossati2016}
{Fossati} L.,  et~al., 2016, \mn@doi [\aap] {10.1051/0004-6361/201628259},
  \href {http://adsabs.harvard.edu/abs/2016A%26A...592A..84F} {592, A84}

\bibitem[\protect\citeauthoryear{{Fricke}}{{Fricke}}{1968}]{fricke1968}
{Fricke} K.,  1968, \zap, \href
  {http://adsabs.harvard.edu/abs/1968ZA.....68..317F} {68, 317}

\bibitem[\protect\citeauthoryear{{Gaia Collaboration} et~al.,}{{Gaia
  Collaboration} et~al.}{2018}]{gaia2018}
{Gaia Collaboration} et~al., 2018, \mn@doi [\aap]
  {10.1051/0004-6361/201833051}, \href
  {https://ui.adsabs.harvard.edu/abs/2018A&A...616A...1G} {616, A1}

\bibitem[\protect\citeauthoryear{{Georgy} et~al.,}{{Georgy}
  et~al.}{2013}]{georgy2013}
{Georgy} C.,  et~al., 2013, \mn@doi [\aap] {10.1051/0004-6361/201322178}, \href
  {http://adsabs.harvard.edu/abs/2013A%26A...558A.103G} {558, A103}

\bibitem[\protect\citeauthoryear{{Georgy}, {Meynet}, {Ekstr{\"o}m}, {Wade},
  {Petit}, {Keszthelyi}  \& {Hirschi}}{{Georgy} et~al.}{2017}]{georgy2017}
{Georgy} C.,  {Meynet} G.,  {Ekstr{\"o}m} S.,  {Wade} G.~A.,  {Petit} V.,
  {Keszthelyi} Z.,   {Hirschi} R.,  2017, \mn@doi [\aap]
  {10.1051/0004-6361/201730401}, \href
  {http://adsabs.harvard.edu/abs/2017A%26A...599L...5G} {599, L5}

\bibitem[\protect\citeauthoryear{{Glebbeek}, {Gaburov}, {Portegies Zwart}  \&
  {Pols}}{{Glebbeek} et~al.}{2013}]{glebbeek2013}
{Glebbeek} E.,  {Gaburov} E.,  {Portegies Zwart} S.,   {Pols} O.~R.,  2013,
  \mn@doi [\mnras] {10.1093/mnras/stt1268}, \href
  {https://ui.adsabs.harvard.edu/abs/2013MNRAS.434.3497G} {434, 3497}

\bibitem[\protect\citeauthoryear{{Goldreich} \& {Schubert}}{{Goldreich} \&
  {Schubert}}{1967}]{goldreich1967}
{Goldreich} P.,  {Schubert} G.,  1967, \mn@doi [\apj] {10.1086/149360}, \href
  {http://adsabs.harvard.edu/abs/1967ApJ...150..571G} {150, 571}

\bibitem[\protect\citeauthoryear{{Grellmann}, {Ratzka}, {K{\"o}hler},
  {Preibisch}  \& {Mucciarelli}}{{Grellmann} et~al.}{2015}]{grellmann2015}
{Grellmann} R.,  {Ratzka} T.,  {K{\"o}hler} R.,  {Preibisch} T.,
  {Mucciarelli} P.,  2015, \mn@doi [\aap] {10.1051/0004-6361/201219577}, \href
  {https://ui.adsabs.harvard.edu/abs/2015A&A...578A..84G} {578, A84}

\bibitem[\protect\citeauthoryear{{Grevesse}, {Noels}  \& {Sauval}}{{Grevesse}
  et~al.}{1996}]{grevesse1996}
{Grevesse} N.,  {Noels} A.,   {Sauval} A.~J.,  1996, in {Holt} S.~S.,
  {Sonneborn} G.,  eds,  Astronomical Society of the Pacific Conference Series
  Vol. 99, Cosmic Abundances. p.~117

\bibitem[\protect\citeauthoryear{{Groh} et~al.,}{{Groh}
  et~al.}{2019}]{groh2019}
{Groh} J.~H.,  et~al., 2019, \mn@doi [\aap] {10.1051/0004-6361/201833720},
  \href {https://ui.adsabs.harvard.edu/abs/2019A&A...627A..24G} {627, A24}

\bibitem[\protect\citeauthoryear{{Grunhut} et~al.,}{{Grunhut}
  et~al.}{2017}]{grunhut2017}
{Grunhut} J.~H.,  et~al., 2017, \mn@doi [\mnras] {10.1093/mnras/stw2743}, \href
  {http://adsabs.harvard.edu/abs/2017MNRAS.465.2432G} {465, 2432}

\bibitem[\protect\citeauthoryear{{Gvaramadze}, {Maryeva}, {Kniazev},
  {Alexashov}, {Castro}, {Langer}  \& {Katkov}}{{Gvaramadze}
  et~al.}{2019}]{gv2019}
{Gvaramadze} V.~V.,  {Maryeva} O.~V.,  {Kniazev} A.~Y.,  {Alexashov} D.~B.,
  {Castro} N.,  {Langer} N.,   {Katkov} I.~Y.,  2019, \mn@doi [\mnras]
  {10.1093/mnras/sty2987}, \href
  {https://ui.adsabs.harvard.edu/abs/2019MNRAS.482.4408G} {482, 4408}

\bibitem[\protect\citeauthoryear{{Hardorp} \& {Scholz}}{{Hardorp} \&
  {Scholz}}{1970}]{hardorp1970}
{Hardorp} J.,  {Scholz} M.,  1970, \mn@doi [\apjs] {10.1086/190205}, \href
  {https://ui.adsabs.harvard.edu/abs/1970ApJS...19..193H} {19, 193}

\bibitem[\protect\citeauthoryear{{Heger}, {Langer}  \& {Woosley}}{{Heger}
  et~al.}{2000}]{heger2000}
{Heger} A.,  {Langer} N.,   {Woosley} S.~E.,  2000, \mn@doi [\apj]
  {10.1086/308158}, \href {http://adsabs.harvard.edu/abs/2000ApJ...528..368H}
  {528, 368}

\bibitem[\protect\citeauthoryear{{Herwig}}{{Herwig}}{2000}]{herwig2000}
{Herwig} F.,  2000, \aap, \href
  {http://adsabs.harvard.edu/abs/2000A%26A...360..952H} {360, 952}

\bibitem[\protect\citeauthoryear{{Hillier} \& {Miller}}{{Hillier} \&
  {Miller}}{1998}]{hm98}
{Hillier} D.~J.,  {Miller} D.~L.,  1998, \mn@doi [\apj] {10.1086/305350}, \href
  {http://cdsads.u-strasbg.fr/abs/1998ApJ...496..407H} {496, 407}

\bibitem[\protect\citeauthoryear{{H{\o}iland}}{{H{\o}iland}}{1941}]{hoiland1941}
{H{\o}iland} K.,  1941, Avhandliger Norske Videnskaps-Akademi i Oslo,
  I,math.-naturv. Klasse, 11, 1

\bibitem[\protect\citeauthoryear{{Howk}, {Cassinelli}, {Bjorkman}  \&
  {Lamers}}{{Howk} et~al.}{2000}]{howk2000}
{Howk} J.~C.,  {Cassinelli} J.~P.,  {Bjorkman} J.~E.,   {Lamers} H.
  J.~G.~L.~M.,  2000, \mn@doi [\apj] {10.1086/308730}, \href
  {https://ui.adsabs.harvard.edu/abs/2000ApJ...534..348H} {534, 348}

\bibitem[\protect\citeauthoryear{{Hunter} et~al.,}{{Hunter}
  et~al.}{2009}]{hunter2009}
{Hunter} I.,  et~al., 2009, \mn@doi [\aap] {10.1051/0004-6361/200809925}, \href
  {http://adsabs.harvard.edu/abs/2009A%26A...496..841H} {496, 841}

\bibitem[\protect\citeauthoryear{{Ignace}, {Oskinova}, {Jardine}, {Cassinelli},
  {Cohen}, {Donati}, {Townsend}  \& {ud-Doula}}{{Ignace}
  et~al.}{2010}]{ignace2010}
{Ignace} R.,  {Oskinova} L.~M.,  {Jardine} M.,  {Cassinelli} J.~P.,  {Cohen}
  D.~H.,  {Donati} J.~F.,  {Townsend} R.~H.~D.,   {ud-Doula} A.,  2010, \mn@doi
  [\apj] {10.1088/0004-637X/721/2/1412}, \href
  {https://ui.adsabs.harvard.edu/abs/2010ApJ...721.1412I} {721, 1412}

\bibitem[\protect\citeauthoryear{{Kaiser}, {Hirschi}, {Arnett}, {Georgy},
  {Scott}  \& {Cristini}}{{Kaiser} et~al.}{2020}]{kaiser2020}
{Kaiser} E.~A.,  {Hirschi} R.,  {Arnett} W.~D.,  {Georgy} C.,  {Scott} L.
  J.~A.,   {Cristini} A.,  2020, \mn@doi [\mnras] {10.1093/mnras/staa1595},
  \href {https://ui.adsabs.harvard.edu/abs/2020MNRAS.496.1967K} {496, 1967}

\bibitem[\protect\citeauthoryear{{Keszthelyi}, {Puls}  \& {Wade}}{{Keszthelyi}
  et~al.}{2017}]{keszthelyi2017b}
{Keszthelyi} Z.,  {Puls} J.,   {Wade} G.~A.,  2017, \mn@doi [\aap]
  {10.1051/0004-6361/201629468}, \href
  {http://adsabs.harvard.edu/abs/2017A%26A...598A...4K} {598, A4}

\bibitem[\protect\citeauthoryear{{Keszthelyi}, {Meynet}, {Georgy}, {Wade},
  {Petit}  \& {David-Uraz}}{{Keszthelyi} et~al.}{2019}]{keszthelyi2019}
{Keszthelyi} Z.,  {Meynet} G.,  {Georgy} C.,  {Wade} G.~A.,  {Petit} V.,
  {David-Uraz} A.,  2019, \mn@doi [\mnras] {10.1093/mnras/stz772}, \href
  {http://adsabs.harvard.edu/abs/2019MNRAS.485.5843K} {485, 5843}

\bibitem[\protect\citeauthoryear{{Keszthelyi} et~al.,}{{Keszthelyi}
  et~al.}{2020}]{keszthelyi2020}
{Keszthelyi} Z.,  et~al., 2020, \mn@doi [\mnras] {10.1093/mnras/staa237}, \href
  {https://ui.adsabs.harvard.edu/abs/2020MNRAS.493..518K} {493, 518}

\bibitem[\protect\citeauthoryear{{Khan} \& {Shulyak}}{{Khan} \&
  {Shulyak}}{2006}]{khan2006}
{Khan} S.~A.,  {Shulyak} D.~V.,  2006, \mn@doi [\aap]
  {10.1051/0004-6361:20054269}, \href
  {https://ui.adsabs.harvard.edu/abs/2006A&A...448.1153K} {448, 1153}

\bibitem[\protect\citeauthoryear{{Kilian}}{{Kilian}}{1992}]{kilian1992}
{Kilian} J.,  1992, \aap, \href
  {https://ui.adsabs.harvard.edu/abs/1992A&A...262..171K} {262, 171}

\bibitem[\protect\citeauthoryear{{Kilian}, {Becker}, {Gehren}  \&
  {Nissen}}{{Kilian} et~al.}{1991}]{1991A&A...244..419K}
{Kilian} J.,  {Becker} S.~R.,  {Gehren} T.,   {Nissen} P.~E.,  1991, \aap,
  \href {https://ui.adsabs.harvard.edu/abs/1991A&A...244..419K} {244, 419}

\bibitem[\protect\citeauthoryear{{Kochukhov} \& {Wade}}{{Kochukhov} \&
  {Wade}}{2016}]{kochukhov2016}
{Kochukhov} O.,  {Wade} G.~A.,  2016, \mn@doi [\aap]
  {10.1051/0004-6361/201527454}, \href
  {https://ui.adsabs.harvard.edu/abs/2016A&A...586A..30K} {586, A30}

\bibitem[\protect\citeauthoryear{{Krti{\v{c}}ka}}{{Krti{\v{c}}ka}}{2018}]{krticka2018}
{Krti{\v{c}}ka} J.,  2018, \mn@doi [\aap] {10.1051/0004-6361/201834097}, \href
  {https://ui.adsabs.harvard.edu/abs/2018A&A...620A.176K} {620, A176}

\bibitem[\protect\citeauthoryear{{Lafreni{\`e}re}, {Jayawardhana}, {van
  Kerkwijk}, {Brandeker}  \& {Janson}}{{Lafreni{\`e}re} et~al.}{2014}]{laf2014}
{Lafreni{\`e}re} D.,  {Jayawardhana} R.,  {van Kerkwijk} M.~H.,  {Brandeker}
  A.,   {Janson} M.,  2014, \mn@doi [\apj] {10.1088/0004-637X/785/1/47}, \href
  {https://ui.adsabs.harvard.edu/abs/2014ApJ...785...47L} {785, 47}

\bibitem[\protect\citeauthoryear{{Lamers} \& {Rogerson}}{{Lamers} \&
  {Rogerson}}{1978}]{lamers1978}
{Lamers} H.~J.~G.~L.~M.,  {Rogerson} J.~B. J.,  1978, \aap, \href
  {https://ui.adsabs.harvard.edu/abs/1978A&A....66..417L} {66, 417}

\bibitem[\protect\citeauthoryear{{Langer}}{{Langer}}{2012}]{langer2012}
{Langer} N.,  2012, \mn@doi [\araa] {10.1146/annurev-astro-081811-125534},
  \href {http://adsabs.harvard.edu/abs/2012ARA%26A..50..107L} {50, 107}

\bibitem[\protect\citeauthoryear{{Lodders}}{{Lodders}}{2003}]{lodders2003}
{Lodders} K.,  2003, \mn@doi [\apj] {10.1086/375492}, \href
  {http://adsabs.harvard.edu/abs/2003ApJ...591.1220L} {591, 1220}

\bibitem[\protect\citeauthoryear{{Macfarlane} \& {Cassinelli}}{{Macfarlane} \&
  {Cassinelli}}{1989}]{macfarlane1989}
{Macfarlane} J.~J.,  {Cassinelli} J.~P.,  1989, \mn@doi [\apj]
  {10.1086/168197}, \href
  {https://ui.adsabs.harvard.edu/abs/1989ApJ...347.1090M} {347, 1090}

\bibitem[\protect\citeauthoryear{{Maeder}}{{Maeder}}{1997}]{maeder1997}
{Maeder} A.,  1997, \aap, \href
  {https://ui.adsabs.harvard.edu/abs/1997A&A...321..134M} {321, 134}

\bibitem[\protect\citeauthoryear{{Maeder}}{{Maeder}}{2003}]{maeder2003b}
{Maeder} A.,  2003, \mn@doi [\aap] {10.1051/0004-6361:20021731}, \href
  {https://ui.adsabs.harvard.edu/abs/2003A&A...399..263M} {399, 263}

\bibitem[\protect\citeauthoryear{{Maeder}}{{Maeder}}{2009}]{Maeder2009a}
{Maeder} A.,  2009, {Physics, Formation and Evolution of Rotating Stars},
  \mn@doi{10.1007/978-3-540-76949-1.
}

\bibitem[\protect\citeauthoryear{{Maeder}, {Przybilla}, {Nieva}, {Georgy},
  {Meynet}, {Ekstr{\"o}m}  \& {Eggenberger}}{{Maeder}
  et~al.}{2014}]{maeder2014a}
{Maeder} A.,  {Przybilla} N.,  {Nieva} M.-F.,  {Georgy} C.,  {Meynet} G.,
  {Ekstr{\"o}m} S.,   {Eggenberger} P.,  2014, \mn@doi [\aap]
  {10.1051/0004-6361/201220602}, \href
  {https://ui.adsabs.harvard.edu/abs/2014A&A...565A..39M} {565, A39}

\bibitem[\protect\citeauthoryear{{Mamajek}, {Pecaut}, {Nguyen}  \&
  {Bubar}}{{Mamajek} et~al.}{2013}]{mamajek2013}
{Mamajek} E.~E.,  {Pecaut} M.~J.,  {Nguyen} D.~C.,   {Bubar} E.~J.,  2013, in
  Protostars and Planets VI Posters.

\bibitem[\protect\citeauthoryear{{Markova}, {Puls}  \& {Langer}}{{Markova}
  et~al.}{2018}]{markova2018}
{Markova} N.,  {Puls} J.,   {Langer} N.,  2018, \mn@doi [\aap]
  {10.1051/0004-6361/201731361}, \href
  {http://adsabs.harvard.edu/abs/2018A%26A...613A..12M} {613, A12}

\bibitem[\protect\citeauthoryear{{Martins} \& {Plez}}{{Martins} \&
  {Plez}}{2006}]{mp06}
{Martins} F.,  {Plez} B.,  2006, \mn@doi [\aap] {10.1051/0004-6361:20065753},
  \href {https://ui.adsabs.harvard.edu/abs/2006A&A...457..637M} {457, 637}

\bibitem[\protect\citeauthoryear{{Martins}, {Escolano}, {Wade}, {Donati},
  {Bouret}  \& {Mimes Collaboration}}{{Martins} et~al.}{2012}]{martins2012}
{Martins} F.,  {Escolano} C.,  {Wade} G.~A.,  {Donati} J.~F.,  {Bouret} J.~C.,
   {Mimes Collaboration} 2012, \mn@doi [\aap] {10.1051/0004-6361/201118039},
  \href {https://ui.adsabs.harvard.edu/abs/2012A%26A...538A..29M} {538, A29}

\bibitem[\protect\citeauthoryear{{Martins} et~al.,}{{Martins}
  et~al.}{2015}]{martins2015}
{Martins} F.,  et~al., 2015, \mn@doi [\aap] {10.1051/0004-6361/201425173},
  \href {http://adsabs.harvard.edu/abs/2015A%26A...575A..34M} {575, A34}

\bibitem[\protect\citeauthoryear{{Martins}, {Sim{\'o}n-D{\'\i}az}, {Barb{\'a}},
  {Gamen}  \& {Ekstr{\"o}m}}{{Martins} et~al.}{2017}]{martins2017}
{Martins} F.,  {Sim{\'o}n-D{\'\i}az} S.,  {Barb{\'a}} R.~H.,  {Gamen} R.~C.,
  {Ekstr{\"o}m} S.,  2017, \mn@doi [\aap] {10.1051/0004-6361/201629548}, \href
  {https://ui.adsabs.harvard.edu/abs/2017A&A...599A..30M} {599, A30}

\bibitem[\protect\citeauthoryear{{Mathis}, {Decressin}, {Eggenberger}  \&
  {Charbonnel}}{{Mathis} et~al.}{2013}]{mathis2013}
{Mathis} S.,  {Decressin} T.,  {Eggenberger} P.,   {Charbonnel} C.,  2013,
  \mn@doi [\aap] {10.1051/0004-6361/201321934}, \href
  {https://ui.adsabs.harvard.edu/abs/2013A&A...558A..11M} {558, A11}

\bibitem[\protect\citeauthoryear{{Mathys}}{{Mathys}}{1988}]{mathys1988}
{Mathys} G.,  1988, The Messenger, \href
  {https://ui.adsabs.harvard.edu/abs/1988Msngr..53...39M} {53, 39}

\bibitem[\protect\citeauthoryear{Meakin \& Arnett}{Meakin \&
  Arnett}{2007}]{meakin2007}
Meakin C.~A.,  Arnett D.,  2007, \mn@doi [The Astrophysical Journal]
  {10.1086/520318}, 667, 448

\bibitem[\protect\citeauthoryear{{Mewe}, {Raassen}, {Cassinelli}, {van der
  Hucht}, {Miller}  \& {G{\"u}del}}{{Mewe} et~al.}{2003}]{mewe2003}
{Mewe} R.,  {Raassen} A.~J.~J.,  {Cassinelli} J.~P.,  {van der Hucht} K.~A.,
  {Miller} N.~A.,   {G{\"u}del} M.,  2003, \mn@doi [\aap]
  {10.1051/0004-6361:20021577}, \href
  {https://ui.adsabs.harvard.edu/abs/2003A&A...398..203M} {398, 203}

\bibitem[\protect\citeauthoryear{{Meynet}, {Maeder}  \& {Mowlavi}}{{Meynet}
  et~al.}{2004}]{meynet2004}
{Meynet} G.,  {Maeder} A.,   {Mowlavi} N.,  2004, \mn@doi [\aap]
  {10.1051/0004-6361:20031735}, \href
  {https://ui.adsabs.harvard.edu/abs/2004A&A...416.1023M} {416, 1023}

\bibitem[\protect\citeauthoryear{{Meynet}, {Eggenberger}  \& {Maeder}}{{Meynet}
  et~al.}{2011}]{meynet2011}
{Meynet} G.,  {Eggenberger} P.,   {Maeder} A.,  2011, \mn@doi [\aap]
  {10.1051/0004-6361/201016017}, \href
  {http://adsabs.harvard.edu/abs/2011A%26A...525L..11M} {525, L11}

\bibitem[\protect\citeauthoryear{{Meynet}, {Ekstrom}, {Maeder}, {Eggenberger},
  {Saio}, {Chomienne}  \& {Haemmerl{\'e}}}{{Meynet} et~al.}{2013}]{meynet2013}
{Meynet} G.,  {Ekstrom} S.,  {Maeder} A.,  {Eggenberger} P.,  {Saio} H.,
  {Chomienne} V.,   {Haemmerl{\'e}} L.,  2013, in {Goupil} M.,  {Belkacem} K.,
  {Neiner} C.,  {Ligni{\`e}res} F.,   {Green} J.~J.,  eds,  Lecture Notes in
  Physics, Berlin Springer Verlag Vol. 865, Lecture Notes in Physics, Berlin
  Springer Verlag. p.~3 (\mn@eprint {arXiv} {1301.2487}),
  \mn@doi{10.1007/978-3-642-33380-4_1}

\bibitem[\protect\citeauthoryear{{Mokiem}, {de Koter}, {Puls}, {Herrero},
  {Najarro}  \& {Villamariz}}{{Mokiem} et~al.}{2005}]{mokiem2005}
{Mokiem} M.~R.,  {de Koter} A.,  {Puls} J.,  {Herrero} A.,  {Najarro} F.,
  {Villamariz} M.~R.,  2005, \mn@doi [\aap] {10.1051/0004-6361:20053522}, \href
  {https://ui.adsabs.harvard.edu/abs/2005A&A...441..711M} {441, 711}

\bibitem[\protect\citeauthoryear{{Moravveji}, {Aerts}, {P{\'a}pics}, {Triana}
  \& {Vandoren}}{{Moravveji} et~al.}{2015}]{moravveji2015}
{Moravveji} E.,  {Aerts} C.,  {P{\'a}pics} P.~I.,  {Triana} S.~A.,   {Vandoren}
  B.,  2015, \mn@doi [\aap] {10.1051/0004-6361/201425290}, \href
  {http://adsabs.harvard.edu/abs/2015A%26A...580A..27M} {580, A27}

\bibitem[\protect\citeauthoryear{{Morel}, {Hubrig}  \& {Briquet}}{{Morel}
  et~al.}{2008}]{morel2008}
{Morel} T.,  {Hubrig} S.,   {Briquet} M.,  2008, \mn@doi [\aap]
  {10.1051/0004-6361:20078999}, \href
  {http://adsabs.harvard.edu/abs/2008A%26A...481..453M} {481, 453}

\bibitem[\protect\citeauthoryear{{Murphy} et~al.,}{{Murphy}
  et~al.}{2021}]{murphy2021}
{Murphy} L.~J.,  et~al., 2021, \mn@doi [\mnras] {10.1093/mnras/staa3803}, \href
  {https://ui.adsabs.harvard.edu/abs/2021MNRAS.501.2745M} {501, 2745}

\bibitem[\protect\citeauthoryear{{Naz{\'e}}, {Petit}, {Rinbrand}, {Cohen},
  {Owocki}, {ud-Doula}  \& {Wade}}{{Naz{\'e}} et~al.}{2014}]{naze2014}
{Naz{\'e}} Y.,  {Petit} V.,  {Rinbrand} M.,  {Cohen} D.,  {Owocki} S.,
  {ud-Doula} A.,   {Wade} G.~A.,  2014, \mn@doi [\apjs]
  {10.1088/0067-0049/215/1/10}, \href
  {https://ui.adsabs.harvard.edu/abs/2014ApJS..215...10N} {215, 10}

\bibitem[\protect\citeauthoryear{{Nieva} \& {Przybilla}}{{Nieva} \&
  {Przybilla}}{2012a}]{nieva2012b}
{Nieva} M.~F.,  {Przybilla} N.,  2012a, \mn@doi [\aap]
  {10.1051/0004-6361/201118158}, \href
  {https://ui.adsabs.harvard.edu/abs/2012A&A...539A.143N} {539, A143}

\bibitem[\protect\citeauthoryear{{Nieva} \& {Przybilla}}{{Nieva} \&
  {Przybilla}}{2012b}]{nieva2012}
{Nieva} M.-F.,  {Przybilla} N.,  2012b, \mn@doi [\aap]
  {10.1051/0004-6361/201118158}, \href
  {http://adsabs.harvard.edu/abs/2012A%26A...539A.143N} {539, A143}

\bibitem[\protect\citeauthoryear{{Nieva} \& {Przybilla}}{{Nieva} \&
  {Przybilla}}{2014}]{nieva2014}
{Nieva} M.-F.,  {Przybilla} N.,  2014, \mn@doi [\aap]
  {10.1051/0004-6361/201423373}, \href
  {http://adsabs.harvard.edu/abs/2014A%26A...566A...7N} {566, A7}

\bibitem[\protect\citeauthoryear{{Owocki}, {ud-Doula}, {Sundqvist}, {Petit},
  {Cohen}  \& {Townsend}}{{Owocki} et~al.}{2016}]{owocki2016}
{Owocki} S.~P.,  {ud-Doula} A.,  {Sundqvist} J.~O.,  {Petit} V.,  {Cohen}
  D.~H.,   {Townsend} R.~H.~D.,  2016, \mn@doi [\mnras]
  {10.1093/mnras/stw1894}, \href
  {http://adsabs.harvard.edu/abs/2016MNRAS.462.3830O} {462, 3830}

\bibitem[\protect\citeauthoryear{{P{\'a}pics} et~al.,}{{P{\'a}pics}
  et~al.}{2017}]{papics2017}
{P{\'a}pics} P.~I.,  et~al., 2017, \mn@doi [\aap]
  {10.1051/0004-6361/201629814}, \href
  {http://adsabs.harvard.edu/abs/2017A26A...598A..74P} {598, A74}

\bibitem[\protect\citeauthoryear{{Paxton}, {Bildsten}, {Dotter}, {Herwig},
  {Lesaffre}  \& {Timmes}}{{Paxton} et~al.}{2011}]{paxton2011}
{Paxton} B.,  {Bildsten} L.,  {Dotter} A.,  {Herwig} F.,  {Lesaffre} P.,
  {Timmes} F.,  2011, \mn@doi [\apjs] {10.1088/0067-0049/192/1/3}, \href
  {http://adsabs.harvard.edu/abs/2011ApJS..192....3P} {192, 3}

\bibitem[\protect\citeauthoryear{{Paxton} et~al.,}{{Paxton}
  et~al.}{2013}]{paxton2013}
{Paxton} B.,  et~al., 2013, \mn@doi [\apjs] {10.1088/0067-0049/208/1/4}, \href
  {http://adsabs.harvard.edu/abs/2013ApJS..208....4P} {208, 4}

\bibitem[\protect\citeauthoryear{{Paxton} et~al.,}{{Paxton}
  et~al.}{2015}]{paxton2015}
{Paxton} B.,  et~al., 2015, \mn@doi [\apjs] {10.1088/0067-0049/220/1/15}, \href
  {http://adsabs.harvard.edu/abs/2015ApJS..220...15P} {220, 15}

\bibitem[\protect\citeauthoryear{{Paxton} et~al.,}{{Paxton}
  et~al.}{2018}]{paxton2018}
{Paxton} B.,  et~al., 2018, \mn@doi [\apjs] {10.3847/1538-4365/aaa5a8}, \href
  {http://adsabs.harvard.edu/abs/2018ApJS..234...34P} {234, 34}

\bibitem[\protect\citeauthoryear{{Paxton} et~al.,}{{Paxton}
  et~al.}{2019}]{paxton2019}
{Paxton} B.,  et~al., 2019, \mn@doi [\apjs] {10.3847/1538-4365/ab2241}, \href
  {https://ui.adsabs.harvard.edu/abs/2019ApJS..243...10P} {243, 10}

\bibitem[\protect\citeauthoryear{{Pecaut}, {Mamajek}  \& {Bubar}}{{Pecaut}
  et~al.}{2012}]{pecaut2012}
{Pecaut} M.~J.,  {Mamajek} E.~E.,   {Bubar} E.~J.,  2012, \mn@doi [\apj]
  {10.1088/0004-637X/746/2/154}, \href
  {https://ui.adsabs.harvard.edu/abs/2012ApJ...746..154P} {746, 154}

\bibitem[\protect\citeauthoryear{{Peters} \& {Polidan}}{{Peters} \&
  {Polidan}}{1985}]{peters1985}
{Peters} G.~J.,  {Polidan} R.~S.,  1985, in {Hayes} D.~S.,  {Pasinetti} L.~E.,
   {Philip} A.~G.~D.,  eds,  IAU Symposium Vol. 111, Calibration of Fundamental
  Stellar Quantities. pp 417--421

\bibitem[\protect\citeauthoryear{{Petit} et~al.,}{{Petit}
  et~al.}{2013}]{petit2013}
{Petit} V.,  et~al., 2013, \mn@doi [\mnras] {10.1093/mnras/sts344}, \href
  {http://adsabs.harvard.edu/abs/2013MNRAS.429..398P} {429, 398}

\bibitem[\protect\citeauthoryear{{Petit} et~al.,}{{Petit}
  et~al.}{2017}]{petit2017}
{Petit} V.,  et~al., 2017, \mn@doi [\mnras] {10.1093/mnras/stw3126}, \href
  {http://adsabs.harvard.edu/abs/2017MNRAS.466.1052P} {466, 1052}

\bibitem[\protect\citeauthoryear{{Pinsonneault}, {Kawaler}, {Sofia}  \&
  {Demarque}}{{Pinsonneault} et~al.}{1989}]{pin1989}
{Pinsonneault} M.~H.,  {Kawaler} S.~D.,  {Sofia} S.,   {Demarque} P.,  1989,
  \mn@doi [\apj] {10.1086/167210}, \href
  {http://adsabs.harvard.edu/abs/1989ApJ...338..424P} {338, 424}

\bibitem[\protect\citeauthoryear{{Preibisch}, {Brown}, {Bridges}, {Guenther}
  \& {Zinnecker}}{{Preibisch} et~al.}{2002}]{preibisch2002}
{Preibisch} T.,  {Brown} A. G.~A.,  {Bridges} T.,  {Guenther} E.,   {Zinnecker}
  H.,  2002, \mn@doi [\aj] {10.1086/341174}, \href
  {https://ui.adsabs.harvard.edu/abs/2002AJ....124..404P} {124, 404}

\bibitem[\protect\citeauthoryear{{Przybilla} \& {Butler}}{{Przybilla} \&
  {Butler}}{2004}]{2004ApJ...609.1181P}
{Przybilla} N.,  {Butler} K.,  2004, \mn@doi [\apj] {10.1086/421316}, \href
  {https://ui.adsabs.harvard.edu/abs/2004ApJ...609.1181P} {609, 1181}

\bibitem[\protect\citeauthoryear{{Przybilla}, {Nieva}  \& {Butler}}{{Przybilla}
  et~al.}{2008}]{2008ApJ...688L.103P}
{Przybilla} N.,  {Nieva} M.-F.,   {Butler} K.,  2008, \mn@doi [\apjl]
  {10.1086/595618}, \href
  {https://ui.adsabs.harvard.edu/abs/2008ApJ...688L.103P} {688, L103}

\bibitem[\protect\citeauthoryear{{Przybilla}, {Firnstein}, {Nieva}, {Meynet}
  \& {Maeder}}{{Przybilla} et~al.}{2010}]{przybilla2010}
{Przybilla} N.,  {Firnstein} M.,  {Nieva} M.~F.,  {Meynet} G.,   {Maeder} A.,
  2010, \mn@doi [\aap] {10.1051/0004-6361/201014164}, \href
  {https://ui.adsabs.harvard.edu/abs/2010A&A...517A..38P} {517, A38}

\bibitem[\protect\citeauthoryear{{Puls}, {Urbaneja}, {Venero}, {Repolust},
  {Springmann}, {Jokuthy}  \& {Mokiem}}{{Puls}
  et~al.}{2005}]{2005A&A...435..669P}
{Puls} J.,  {Urbaneja} M.~A.,  {Venero} R.,  {Repolust} T.,  {Springmann} U.,
  {Jokuthy} A.,   {Mokiem} M.~R.,  2005, \mn@doi [\aap]
  {10.1051/0004-6361:20042365}, \href
  {https://ui.adsabs.harvard.edu/abs/2005A&A...435..669P} {435, 669}

\bibitem[\protect\citeauthoryear{{Repolust}, {Puls}, {Hanson}, {Kudritzki}  \&
  {Mokiem}}{{Repolust} et~al.}{2005}]{repolust2005}
{Repolust} T.,  {Puls} J.,  {Hanson} M.~M.,  {Kudritzki} R.~P.,   {Mokiem}
  M.~R.,  2005, \mn@doi [\aap] {10.1051/0004-6361:20052739}, \href
  {https://ui.adsabs.harvard.edu/abs/2005A&A...440..261R} {440, 261}

\bibitem[\protect\citeauthoryear{{Rizzuto} et~al.,}{{Rizzuto}
  et~al.}{2013}]{rizzuto2013}
{Rizzuto} A.~C.,  et~al., 2013, \mn@doi [\mnras] {10.1093/mnras/stt1690}, \href
  {https://ui.adsabs.harvard.edu/abs/2013MNRAS.436.1694R} {436, 1694}

\bibitem[\protect\citeauthoryear{{Rogers} \& {Iglesias}}{{Rogers} \&
  {Iglesias}}{1992}]{rogers1992}
{Rogers} F.~J.,  {Iglesias} C.~A.,  1992, \mn@doi [\apjs] {10.1086/191659},
  \href {http://adsabs.harvard.edu/abs/1992ApJS...79..507R} {79, 507}

\bibitem[\protect\citeauthoryear{{Rogers} \& {McElwaine}}{{Rogers} \&
  {McElwaine}}{2017}]{rogers2017}
{Rogers} T.~M.,  {McElwaine} J.~N.,  2017, \mn@doi [\apjl]
  {10.3847/2041-8213/aa8d13}, \href
  {https://ui.adsabs.harvard.edu/abs/2017ApJ...848L...1R} {848, L1}

\bibitem[\protect\citeauthoryear{{Rogers}, {Lin}, {McElwaine}  \&
  {Lau}}{{Rogers} et~al.}{2013}]{rogers2013}
{Rogers} T.~M.,  {Lin} D.~N.~C.,  {McElwaine} J.~N.,   {Lau} H.~H.~B.,  2013,
  \mn@doi [\apj] {10.1088/0004-637X/772/1/21}, \href
  {https://ui.adsabs.harvard.edu/abs/2013ApJ...772...21R} {772, 21}

\bibitem[\protect\citeauthoryear{{Rogerson} \& {Ewell}}{{Rogerson} \&
  {Ewell}}{1985}]{rogerson1985}
{Rogerson} J.~B. J.,  {Ewell} M.~W. J.,  1985, \mn@doi [\apjs]
  {10.1086/191041}, \href
  {https://ui.adsabs.harvard.edu/abs/1985ApJS...58..265R} {58, 265}

\bibitem[\protect\citeauthoryear{{Sartori}, {L{\'e}pine}  \& {Dias}}{{Sartori}
  et~al.}{2003}]{sartori2003}
{Sartori} M.~J.,  {L{\'e}pine} J.~R.~D.,   {Dias} W.~S.,  2003, \mn@doi [\aap]
  {10.1051/0004-6361:20030581}, \href
  {https://ui.adsabs.harvard.edu/abs/2003A&A...404..913S} {404, 913}

\bibitem[\protect\citeauthoryear{{Schneider}, {Podsiadlowski}, {Langer},
  {Castro}  \& {Fossati}}{{Schneider} et~al.}{2016}]{schneider2016}
{Schneider} F.~R.~N.,  {Podsiadlowski} P.,  {Langer} N.,  {Castro} N.,
  {Fossati} L.,  2016, \mn@doi [\mnras] {10.1093/mnras/stw148}, \href
  {http://adsabs.harvard.edu/abs/2016MNRAS.457.2355S} {457, 2355}

\bibitem[\protect\citeauthoryear{{Schneider}, {Ohlmann}, {Podsiadlowski},
  {R{\"o}pke}, {Balbus}, {Pakmor}  \& {Springel}}{{Schneider}
  et~al.}{2019}]{schneider2019}
{Schneider} F. R.~N.,  {Ohlmann} S.~T.,  {Podsiadlowski} P.,  {R{\"o}pke}
  F.~K.,  {Balbus} S.~A.,  {Pakmor} R.,   {Springel} V.,  2019, \mn@doi [\nat]
  {10.1038/s41586-019-1621-5}, \href
  {https://ui.adsabs.harvard.edu/abs/2019Natur.574..211S} {574, 211}

\bibitem[\protect\citeauthoryear{{Schneider}, {Ohlmann}, {Podsiadlowski},
  {R{\"o}pke}, {Balbus}  \& {Pakmor}}{{Schneider} et~al.}{2020}]{schneider2020}
{Schneider} F.~R.~N.,  {Ohlmann} S.~T.,  {Podsiadlowski} P.,  {R{\"o}pke}
  F.~K.,  {Balbus} S.~A.,   {Pakmor} R.,  2020, \mn@doi [\mnras]
  {10.1093/mnras/staa1326}, \href
  {https://ui.adsabs.harvard.edu/abs/2020MNRAS.495.2796S} {495, 2796}

\bibitem[\protect\citeauthoryear{{Shultz} et~al.,}{{Shultz}
  et~al.}{2018}]{shultz2018}
{Shultz} M.~E.,  et~al., 2018, \mn@doi [\mnras] {10.1093/mnras/sty103}, \href
  {http://adsabs.harvard.edu/abs/2018MNRAS.475.5144S} {475, 5144}

\bibitem[\protect\citeauthoryear{{Shultz}, {Wade}, {Rivinius}, {Alecian},
  {Neiner}, {Petit}  \& {Wisniewski}}{{Shultz} et~al.}{2019a}]{shultz22019}
{Shultz} M.~E.,  {Wade} G.~A.,  {Rivinius} T.,  {Alecian} E.,  {Neiner} C.,
  {Petit} V.,   {Wisniewski} J.~P.,  2019a, \mn@doi [\mnras]
  {10.1093/mnras/stz416}, \href
  {http://adsabs.harvard.edu/abs/2019MNRAS.485.1508S} {485, 1508}

\bibitem[\protect\citeauthoryear{{Shultz} et~al.,}{{Shultz}
  et~al.}{2019b}]{shultz32019}
{Shultz} M.~E.,  et~al., 2019b, \mn@doi [\mnras] {10.1093/mnras/stz2551}, \href
  {https://ui.adsabs.harvard.edu/abs/2019MNRAS.490..274S} {490, 274}

\bibitem[\protect\citeauthoryear{{Sim{\'o}n-D{\'\i}az} \&
  {Herrero}}{{Sim{\'o}n-D{\'\i}az} \& {Herrero}}{2007}]{ss07}
{Sim{\'o}n-D{\'\i}az} S.,  {Herrero} A.,  2007, \mn@doi [\aap]
  {10.1051/0004-6361:20066060}, \href
  {https://ui.adsabs.harvard.edu/abs/2007A&A...468.1063S} {468, 1063}

\bibitem[\protect\citeauthoryear{{Sim{\'o}n-D{\'\i}az}, {Herrero}, {Esteban}
  \& {Najarro}}{{Sim{\'o}n-D{\'\i}az} et~al.}{2006}]{simondiaz2006}
{Sim{\'o}n-D{\'\i}az} S.,  {Herrero} A.,  {Esteban} C.,   {Najarro} F.,  2006,
  \mn@doi [\aap] {10.1051/0004-6361:20053066}, \href
  {https://ui.adsabs.harvard.edu/abs/2006A&A...448..351S} {448, 351}

\bibitem[\protect\citeauthoryear{{Sim{\'o}n-D{\'\i}az}, {Godart}, {Castro},
  {Herrero}, {Aerts}, {Puls}, {Telting}  \&
  {Grassitelli}}{{Sim{\'o}n-D{\'\i}az} et~al.}{2017}]{2017AA...597A..22S}
{Sim{\'o}n-D{\'\i}az} S.,  {Godart} M.,  {Castro} N.,  {Herrero} A.,  {Aerts}
  C.,  {Puls} J.,  {Telting} J.,   {Grassitelli} L.,  2017, \mn@doi [\aap]
  {10.1051/0004-6361/201628541}, \href
  {https://ui.adsabs.harvard.edu/abs/2017A&A...597A..22S} {597, A22}

\bibitem[\protect\citeauthoryear{{Snow}, {Lamers}, {Lindholm}  \&
  {Odell}}{{Snow} et~al.}{1994}]{snow1994}
{Snow} T.~P.,  {Lamers} H. J.~G.~L.~M.,  {Lindholm} D.~M.,   {Odell} A.~P.,
  1994, \mn@doi [\apjs] {10.1086/192099}, \href
  {https://ui.adsabs.harvard.edu/abs/1994ApJS...95..163S} {95, 163}

\bibitem[\protect\citeauthoryear{{Solberg}}{{Solberg}}{1936}]{solberg1936}
{Solberg} H.,  1936, Astrophysica Norvegica, \href
  {https://ui.adsabs.harvard.edu/abs/1936ApNr....1..237S} {1, 237}

\bibitem[\protect\citeauthoryear{{Song}, {Maeder}, {Meynet}, {Huang},
  {Ekstr{\"o}m}  \& {Granada}}{{Song} et~al.}{2013}]{song2013}
{Song} H.~F.,  {Maeder} A.,  {Meynet} G.,  {Huang} R.~Q.,  {Ekstr{\"o}m} S.,
  {Granada} A.,  2013, \mn@doi [\aap] {10.1051/0004-6361/201321870}, \href
  {https://ui.adsabs.harvard.edu/abs/2013A&A...556A.100S} {556, A100}

\bibitem[\protect\citeauthoryear{{Spruit}}{{Spruit}}{2002}]{spruit2002}
{Spruit} H.~C.,  2002, \mn@doi [\aap] {10.1051/0004-6361:20011465}, \href
  {http://adsabs.harvard.edu/abs/2002A%26A...381..923S} {381, 923}

\bibitem[\protect\citeauthoryear{{Stibbs}}{{Stibbs}}{1950}]{stibbs1950}
{Stibbs} D.~W.~N.,  1950, \mn@doi [\mnras] {10.1093/mnras/110.4.395}, \href
  {https://ui.adsabs.harvard.edu/abs/1950MNRAS.110..395S} {110, 395}

\bibitem[\protect\citeauthoryear{{Struve} \& {Dunham}}{{Struve} \&
  {Dunham}}{1933}]{struve1933}
{Struve} O.,  {Dunham} T. J.,  1933, \mn@doi [\apj] {10.1086/143475}, \href
  {https://ui.adsabs.harvard.edu/abs/1933ApJ....77..321S} {77, 321}

\bibitem[\protect\citeauthoryear{{Sweet}}{{Sweet}}{1950}]{sweet1950}
{Sweet} P.~A.,  1950, \mn@doi [\mnras] {10.1093/mnras/110.6.548}, \href
  {http://adsabs.harvard.edu/abs/1950MNRAS.110..548S} {110, 548}

\bibitem[\protect\citeauthoryear{{Talon} \& {Zahn}}{{Talon} \&
  {Zahn}}{1997}]{talon1997}
{Talon} S.,  {Zahn} J.-P.,  1997, \aap, \href
  {http://adsabs.harvard.edu/abs/1997A%26A...317..749T} {317, 749}

\bibitem[\protect\citeauthoryear{{Tayler}}{{Tayler}}{1973}]{tayler73}
{Tayler} R.~J.,  1973, \mn@doi [\mnras] {10.1093/mnras/161.4.365}, \href
  {http://adsabs.harvard.edu/abs/1973MNRAS.161..365T} {161, 365}

\bibitem[\protect\citeauthoryear{{Traving}}{{Traving}}{1955}]{traving1955}
{Traving} G.,  1955, \zap, \href
  {https://ui.adsabs.harvard.edu/abs/1955ZA.....36....1T} {36, 1}

\bibitem[\protect\citeauthoryear{{Uns{\"o}ld}}{{Uns{\"o}ld}}{1942}]{unsold1942}
{Uns{\"o}ld} A.,  1942, \zap, \href
  {https://ui.adsabs.harvard.edu/abs/1942ZA.....21....1U} {21, 1}

\bibitem[\protect\citeauthoryear{{Vidal}, {C{\'e}bron}, {ud-Doula}  \&
  {Alecian}}{{Vidal} et~al.}{2019}]{vidal2019}
{Vidal} J.,  {C{\'e}bron} D.,  {ud-Doula} A.,   {Alecian} E.,  2019, \mn@doi
  [\aap] {10.1051/0004-6361/201935658}, \href
  {https://ui.adsabs.harvard.edu/abs/2019A&A...629A.142V} {629, A142}

\bibitem[\protect\citeauthoryear{{Vink}, {de Koter}  \& {Lamers}}{{Vink}
  et~al.}{2000}]{vink2000}
{Vink} J.~S.,  {de Koter} A.,   {Lamers} H.~J.~G.~L.~M.,  2000, \aap, \href
  {http://adsabs.harvard.edu/abs/2000A%26A...362..295V} {362, 295}

\bibitem[\protect\citeauthoryear{{Vink}, {de Koter}  \& {Lamers}}{{Vink}
  et~al.}{2001}]{vink2001}
{Vink} J.~S.,  {de Koter} A.,   {Lamers} H.~J.~G.~L.~M.,  2001, \mn@doi [\aap]
  {10.1051/0004-6361:20010127}, \href
  {http://adsabs.harvard.edu/abs/2001A%26A...369..574V} {369, 574}

\bibitem[\protect\citeauthoryear{{Walborn} \& {Panek}}{{Walborn} \&
  {Panek}}{1984}]{walborn1984}
{Walborn} N.~R.,  {Panek} R.~J.,  1984, \mn@doi [\apj] {10.1086/162647}, \href
  {https://ui.adsabs.harvard.edu/abs/1984ApJ...286..718W} {286, 718}

\bibitem[\protect\citeauthoryear{{Waters}, {Marlborough}, {Geballe},
  {Oosterbroek}  \& {Zaal}}{{Waters} et~al.}{1993}]{waters1993}
{Waters} L.~B.~F.~M.,  {Marlborough} J.~M.,  {Geballe} T.~R.,  {Oosterbroek}
  T.,   {Zaal} P.,  1993, \aap, \href
  {https://ui.adsabs.harvard.edu/abs/1993A&A...272L...9W} {272, L9}

\bibitem[\protect\citeauthoryear{{Weber} \& {Davis}}{{Weber} \&
  {Davis}}{1967}]{weber1967}
{Weber} E.~J.,  {Davis} Jr. L.,  1967, \mn@doi [\apj] {10.1086/149138}, \href
  {http://adsabs.harvard.edu/abs/1967ApJ...148..217W} {148, 217}

\bibitem[\protect\citeauthoryear{{Wolff} \& {Heasley}}{{Wolff} \&
  {Heasley}}{1985}]{wolff1985}
{Wolff} S.~C.,  {Heasley} J.~N.,  1985, \mn@doi [\apj] {10.1086/163191}, \href
  {https://ui.adsabs.harvard.edu/abs/1985ApJ...292..589W} {292, 589}

\bibitem[\protect\citeauthoryear{{Yoon}, {Langer}  \& {Norman}}{{Yoon}
  et~al.}{2006}]{yoon2006}
{Yoon} S.~C.,  {Langer} N.,   {Norman} C.,  2006, \mn@doi [\aap]
  {10.1051/0004-6361:20065912}, \href
  {https://ui.adsabs.harvard.edu/abs/2006A&A...460..199Y} {460, 199}

\bibitem[\protect\citeauthoryear{{Zaal}, {de Koter}, {Waters}, {Marlborough},
  {Geballe}, {Oliveira}  \& {Foing}}{{Zaal} et~al.}{1999}]{zaal1999}
{Zaal} P.~A.,  {de Koter} A.,  {Waters} L.~B.~F.~M.,  {Marlborough} J.~M.,
  {Geballe} T.~R.,  {Oliveira} J.~M.,   {Foing} B.~H.,  1999, \aap, \href
  {https://ui.adsabs.harvard.edu/abs/1999A&A...349..573Z} {349, 573}

\bibitem[\protect\citeauthoryear{{Zahn}}{{Zahn}}{1992}]{zahn92}
{Zahn} J.-P.,  1992, \aap, \href
  {http://adsabs.harvard.edu/abs/1992A%26A...265..115Z} {265, 115}

\bibitem[\protect\citeauthoryear{{Zorec}, {Cidale}, {Arias}, {Fr{\'e}mat},
  {Muratore}, {Torres}  \& {Martayan}}{{Zorec}
  et~al.}{2009}]{2009AA...501..297Z}
{Zorec} J.,  {Cidale} L.,  {Arias} M.~L.,  {Fr{\'e}mat} Y.,  {Muratore} M.~F.,
  {Torres} A.~F.,   {Martayan} C.,  2009, \mn@doi [\aap]
  {10.1051/0004-6361/200811147}, \href
  {https://ui.adsabs.harvard.edu/abs/2009A&A...501..297Z} {501, 297}

\bibitem[\protect\citeauthoryear{{de Geus}, {de Zeeuw}  \& {Lub}}{{de Geus}
  et~al.}{1989}]{degeus1989}
{de Geus} E.~J.,  {de Zeeuw} P.~T.,   {Lub} J.,  1989, \aap, \href
  {https://ui.adsabs.harvard.edu/abs/1989A&A...216...44D} {216, 44}

\bibitem[\protect\citeauthoryear{{de Jager}, {Nieuwenhuijzen}  \& {van der
  Hucht}}{{de Jager} et~al.}{1988}]{dej1988}
{de Jager} C.,  {Nieuwenhuijzen} H.,   {van der Hucht} K.~A.,  1988, \aaps,
  \href {http://adsabs.harvard.edu/abs/1988A%26AS...72..259D} {72, 259}

\bibitem[\protect\citeauthoryear{{de Messieres}, {Cardamone}, {Cohen},
  {MacFarlane}, {Owocki}  \& {ud-Doula}}{{de Messieres} et~al.}{2001}]{dem2001}
{de Messieres} G.~E.,  {Cardamone} C.,  {Cohen} D.~H.,  {MacFarlane} J.~J.,
  {Owocki} S.~P.,   {ud-Doula} A.,  2001, in American Astronomical Society
  Meeting Abstracts. p. 135.12

\bibitem[\protect\citeauthoryear{{ud-Doula}}{{ud-Doula}}{2017}]{ud2017}
{ud-Doula} A.,  2017, \mn@doi [Astronomische Nachrichten]
  {10.1002/asna.201713394}, \href
  {https://ui.adsabs.harvard.edu/abs/2017AN....338..944U} {338, 944}

\bibitem[\protect\citeauthoryear{{ud-Doula}, {Owocki}  \&
  {Townsend}}{{ud-Doula} et~al.}{2009}]{ud2009}
{ud-Doula} A.,  {Owocki} S.~P.,   {Townsend} R.~H.~D.,  2009, \mn@doi [\mnras]
  {10.1111/j.1365-2966.2008.14134.x}, \href
  {http://adsabs.harvard.edu/abs/2009MNRAS.392.1022U} {392, 1022}

\makeatother
\end{thebibliography}


\appendix

\section{Further tests}\label{sec:app1}

%
%
%
%
%
\begin{figure*}
\includegraphics[width=7cm]{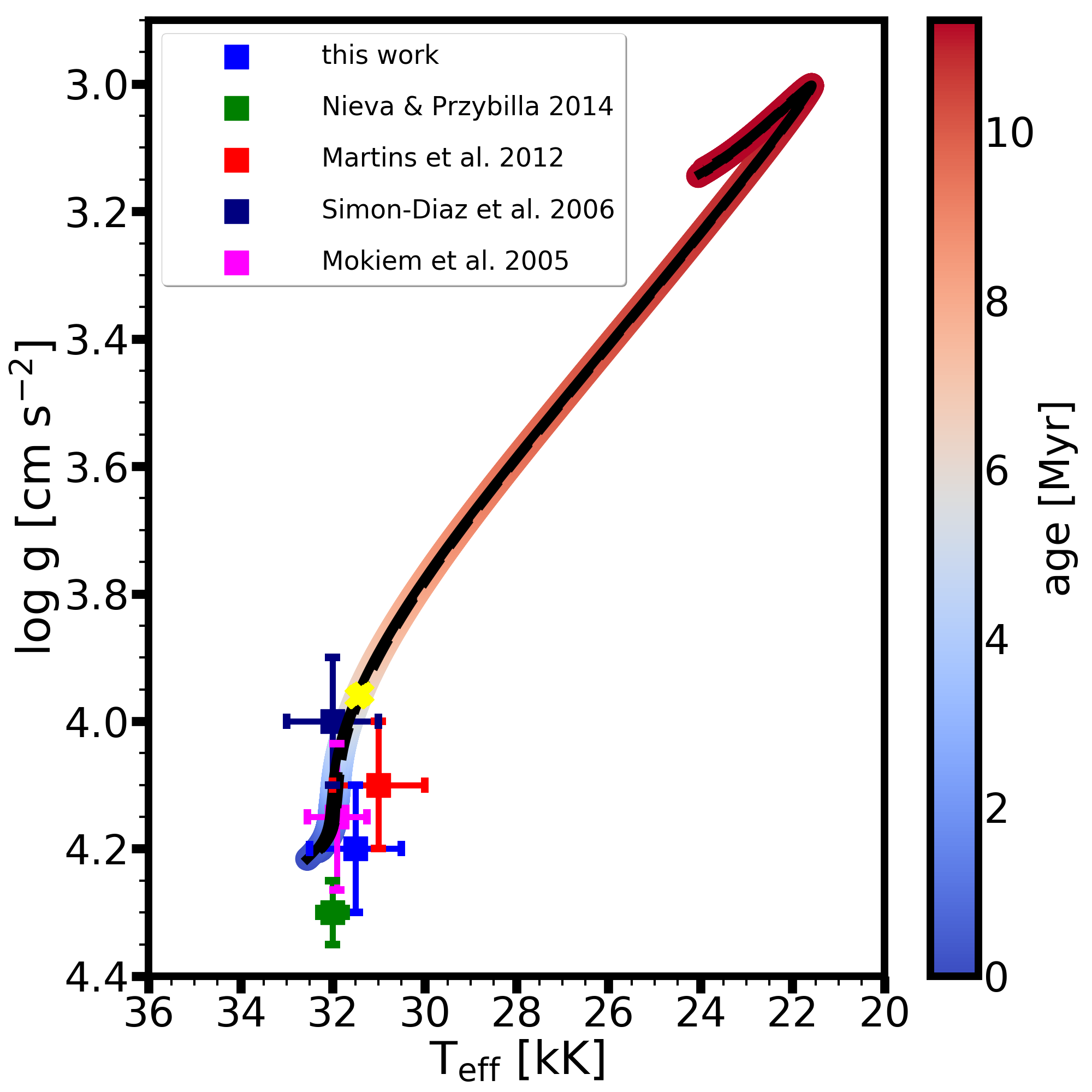}\hspace{1em}\includegraphics[width=7cm]{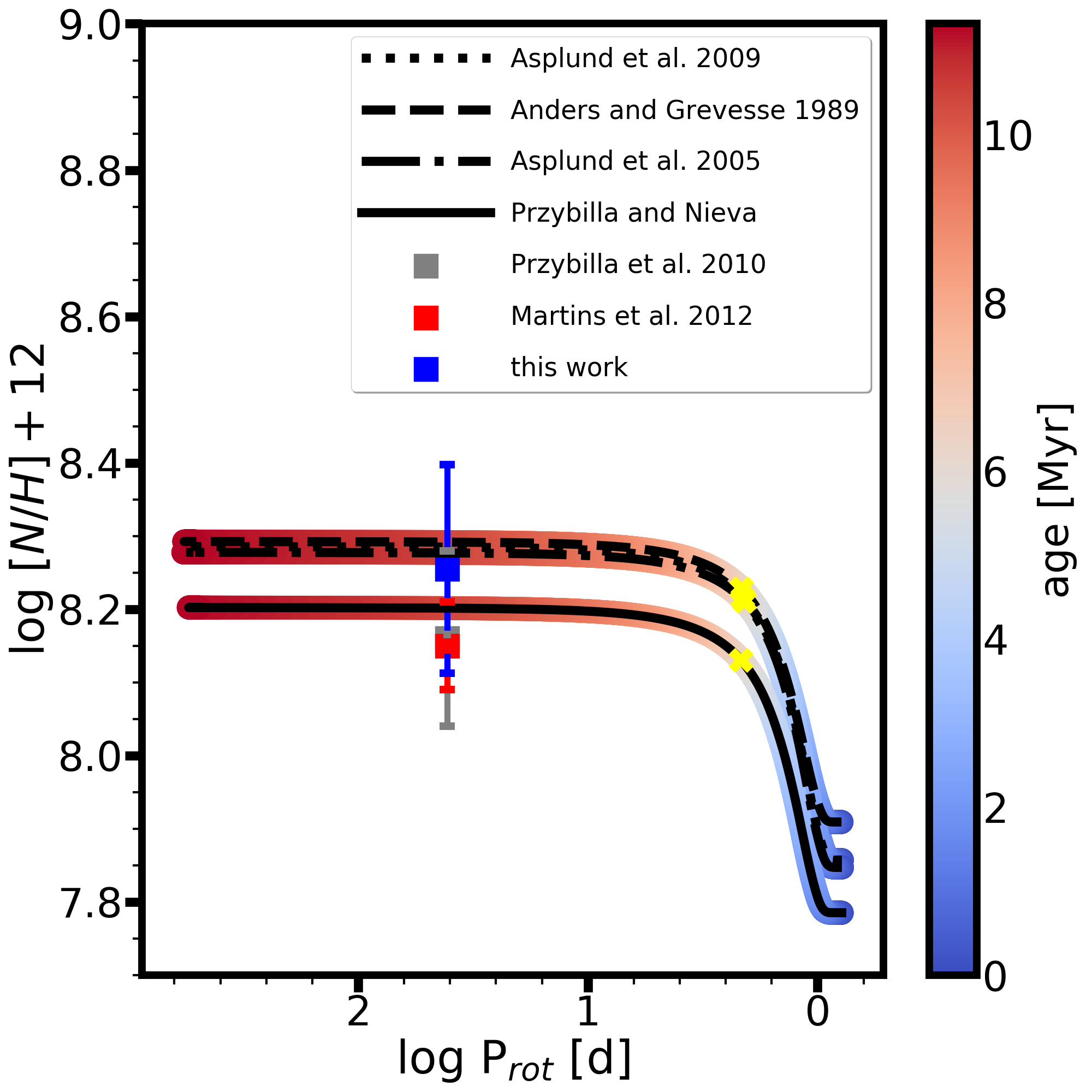}
\includegraphics[width=7cm]{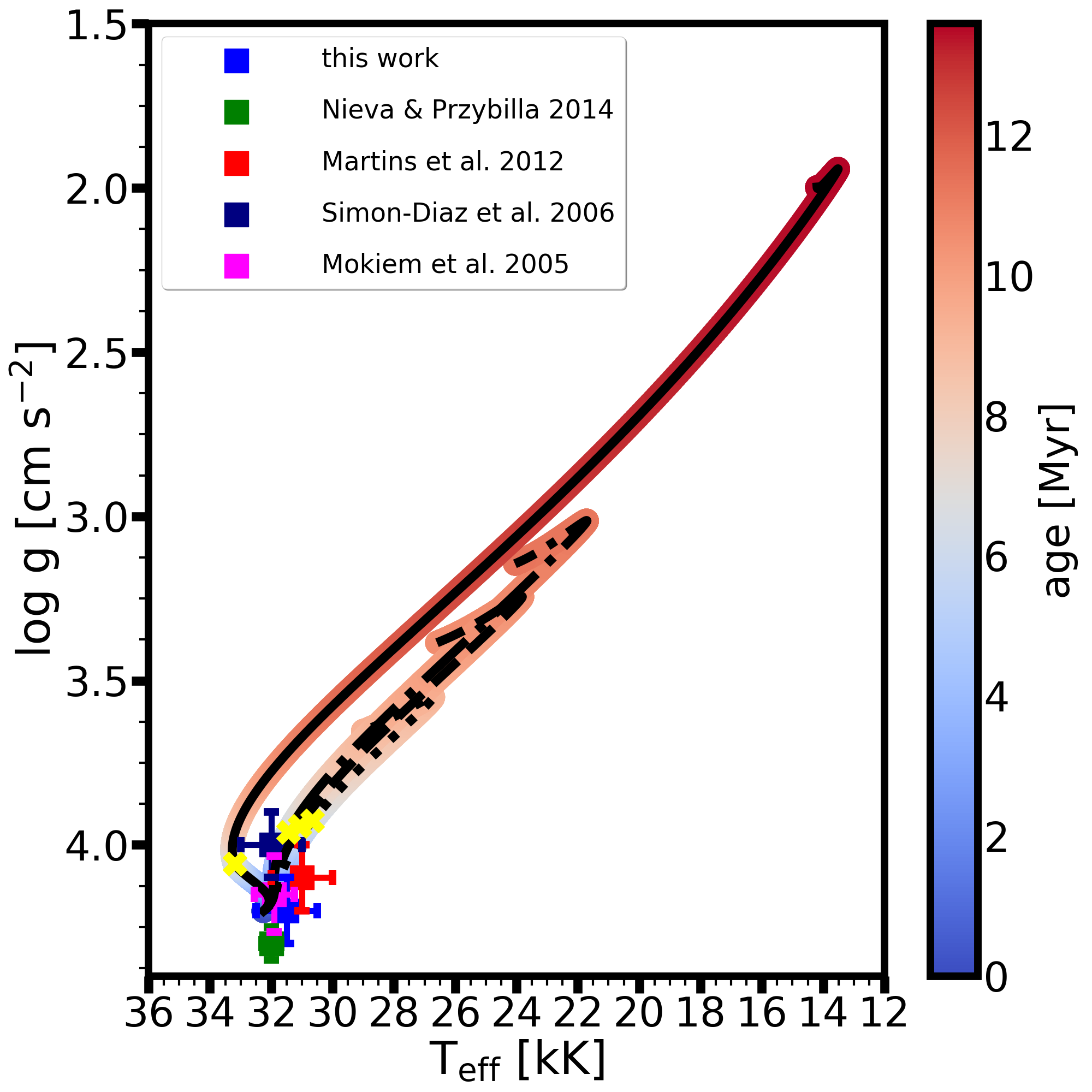}\hspace{1em}\includegraphics[width=7cm]{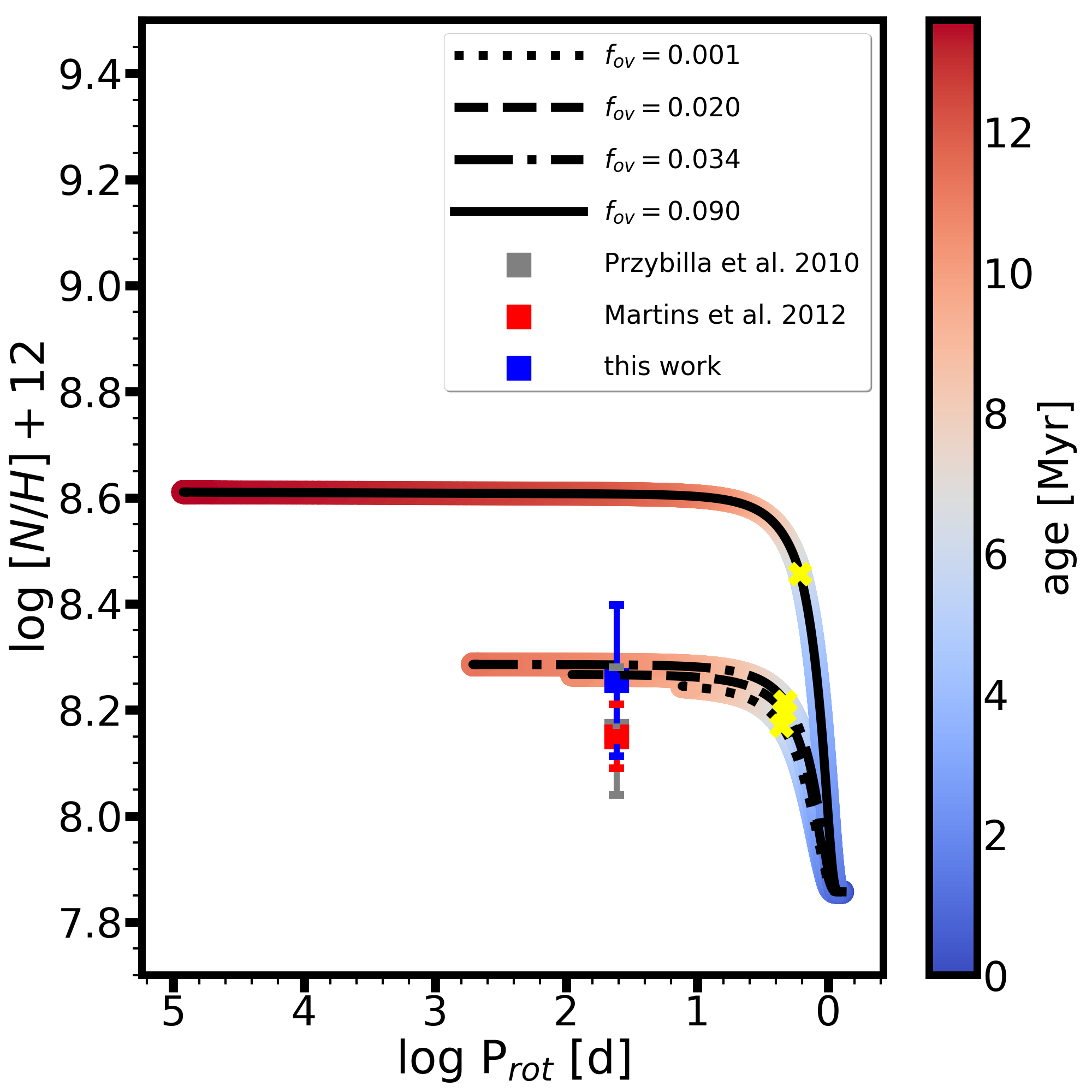}
\caption{Upper panels: Test with different initial metal fractions adopted from various authors. Lower panels: Test with different exponential overshooting parameters. The reference \textsc{mesa} model in both cases has $M_{\rm ini} = 18$~M$_\odot$, $\varv_{\rm ini} = 350$~km\,s$^{-1}$, $B_{\rm p} = 600 $~G, $f_{\rm c} = 0.033$, $f_{\rm MB}= 1$, $f_{\rm ov} = 0.034$, and uses the \citet{asplund2009} mixture of metals.}\label{fig:appfig1}
\end{figure*}

There are a number of modelling assumptions which may affect the quantitative results. 
On the top panels of Figure \ref{fig:appfig1} we first test how adopting different mixtures of metals, that is, altering the initial baseline value of nitrogen, modifies our findings. To this extent, we use a reference \textsc{mesa} model ($M_{\rm ini} = 18$~M$_\odot$, $\varv_{\rm ini} = 350$~km\,s$^{-1}$, $B_{\rm p} = 600 $~G, $f_{\rm c} = 0.033$, $f_{\rm MB}= 1$) and change the initial abundances which are now adopted from \citet{anders1989}, \citet{asplund2005,asplund2009} and the latter one updated by \citet{przybilla2010} and \citet{nieva2012b}. (In \textsc{mesa} this correspond to choosing \texttt{initial\_zfracs}$=$1,5,6, and 8, respectively.) We find that the initial baseline value does not play a significant role.

The lower panels of Figure \ref{fig:appfig1} show the results of adopting different overshooting values. The exponential overshooting values tested here are $f_{\rm ov} = 0.001, 0.020, 0.034, 0.090$ together with $f_{0} = 0.006$.

The lowest value formally leads to decreasing the convective core size since $f_{\rm ov}$ is applied from the nominal convective core boundary minus a distance $f_{0}$ of the local pressure scale height. The latter values would correspond to extending the convective core by roughly $\alpha_{\rm ov} =$~0.1, 0.25, ~and~0.8 of the local pressure scale height in non-rotating models. 

An extreme value of overshooting as shown with $f_{\rm ov} = 0.090$ could, in principle, mitigate the need for a much more efficient envelope mixing (which is controlled by $f_{\rm c}$ in our approach). However, overshooting only mixes core material to the base of the envelope. Therefore, if envelope mixing is inefficient, even an extreme overshooting value will fail to produce a model which accounts for surface enrichment.

\section{Previous determinations of atmospheric, rotational, wind and magnetic properties}\label{sec:app2}

In this section, we report a collection of previous determinations of stellar properties that are found in the recent (in the last $\sim$20 yr) literature. This modern list (see Table~\ref{tab:app_t1}) is not meant to be exhaustive, but rather serves to illustrate typical ranges and uncertainties existing in the determination of these parameters, and that can affect the interpretation of our evolutionary modelling results. The listed works use a range of photometric and spectroscopic methods to determine parameters, and therefore provide a good sense of the variance across methods.

We note that some authors contrast different methods, which yield, for example, $T_{\rm eff} = 29.6 \pm 0.15$\,kK in one case and $T_{\rm eff} = 31.4 \pm 2$\,kK in the other by \cite{2009AA...501..297Z}.

\begin{table*}
\caption{Stellar parameters determined in previous studies: spectroscopic distance, effective temperature, surface gravity, luminosity, projected rotational velocity, microturbulent velocity, macroturbulent velocity, and He and CNO abundances in number fractions.}
\label{tab:app_t1}
\begin{tabular}{l|c|c|c|c|c|c|c|c}  
\hline\hline
Sources & M05 & P08 & M08 & P12 & B12 & NP12 & C17 & SD17 \\[2pt]
\hline\hline
%
%
Distance [pc] & -- & 152$\pm$20 &-- & -- & -- & 143$\pm$9 &-- &--  \\[2pt] 
%
%
$T_\textrm{eff}$ [kK] & 31.9$^{+0.5}_{-0.8}$ & 32$\pm$0.3 & 31.5 & 29.9$\pm$5 & $\sim$29.9 & 32$\pm$0.3 &  31.2$\pm$1.5 & $\sim$32.4$\pm$0.16  \\[2pt]
%
%
$\log g$ [cm s$^{-2}$] & 4.15$^{+0.09}_{-0.14}$ & 4.30$\pm$0.05 & 4.05$\pm$0.15 & - & - & 4.30$\pm$0.05 & 4.3$\pm$0.15 & --  \\[2pt]
%
%
$\log (L/L_\odot)$ & 4.39$\pm$ 0.09 & -- &-- & 4.31$\pm$0.16 & -- & -- & --  & --  \\[2pt]
%
%
$v \sin i$ [km\ s$^{-1}$] & $\sim$ 5 & 4$\pm$4 & 8$\pm$2 & 10$\pm$2 & 3$\pm$2 & 4$\pm$1 & 8 & 7 and 9 \\[2pt]
%
%
$\varv_\textrm{mic}$ [km\ s$^{-1}$] & -- & 5$\pm$1 & 2$\pm$2 & -- & -- & 5$\pm$1 & --&-- \\[2pt]
%
%
$\varv_\textrm{mac}$ [km\ s$^{-1}$] & -- & 4$\pm$4 &-- & -- & -- & 4$\pm$1 & --&10 \\[2pt]
%
%
He/H  & 0.12$^{+0.04}_{-0.02}$ & 0.098$\pm$0.01 & -- &-- & -- & -- & 0.083$\pm$0.025 &-- \\[2pt]
%
%
log(C/H) + 12 & - & 8.30$\pm$0.12  & 8.19$\pm$0.14& -- & -- & 8.30$\pm$0.12 & 8.18$\pm$0.12 & -- \\[2pt]
%
%
log(N/H) + 12 & - & 8.16$\pm$0.12  & 8.15$\pm$0.20& - & -- & 8.16$\pm$0.12 &7.90$\pm$0.13& -- \\[2pt]
%
%
log(O/H) + 12 & - & 8.77$\pm$0.08  & 8.62$\pm$0.20 & -- & -- & 8.77$\pm$0.08 &8.39$\pm$0.21 & -- \\[2pt]
\hline
\multicolumn{9}{l}{M05 = \citet{mokiem2005}; P08 = \citet{2008ApJ...688L.103P}; M08 = \citet{morel2008}; P12 = \citet{pecaut2012};}\\
\multicolumn{9}{l}{B12 = \citet{2012AJ....144..130B}; NP12 = \citet{nieva2012}; C17 = \citet{cazorla2017a}; SD17 = \citet{2017AA...597A..22S}}\\
\end{tabular} 
\end{table*}

The wind and magnetic properties have a significant impact on the modelling presented here, as their interplay determines how quickly the stellar surface spins down. However, certain complications arise specifically as a result of this interaction. Any empirical determination of the wind mass-loss rate $\dot{M}$ based on spherically symmetric models will not properly account for the structure and anisotropy of the outflow induced by the field, so theoretical determinations of the \textit{wind-feeding rate} (see e.g. discussion by \citealt{2019MNRAS.483.2814D} on this topic), such as that adopted in this study, might be more straightforward. Previous determinations in the literature provide a range of values. On the lower end, \citet{repolust2005} derive either $\dot{M} = 2 \times 10^{-8}$ M$_\odot$~yr$^{-1}$ using \textsc{fastwind} \citep{2005A&A...435..669P} to fit $H$- and $K$-band spectra, or values varying between 9 $\times 10^{-9}$ and 2-6 $\times 10^{-8}$ M$_\odot$~yr$^{-1}$ using optical spectra, for parameters derived respectively by \citet{1991A&A...244..419K} and \citet{2004ApJ...609.1181P}. Similarly, on the higher end, \citet{mokiem2005} derive $\dot{M}$~= ~6.14 $\times 10^{-8}$ M$_\odot$\ yr$^{-1}$ by coupling optical spectra with their fitting algorithm, also using \textsc{fastwind} models. These values are roughly consistent with the adopted mass-loss rates in the evolutionary models.

Furthermore, $\tau$ Sco possesses a complex magnetic field, which cannot be simply parametrised using the usual conventions for dipolar fields. The most comprehensive description of $\tau$ Sco's field is provided by \citet{donati2006}. Their longitudinal field measurements vary between -51.9 and +87.8 G, and they also derive an average field modulus over the entire surface of the star of roughly 300 G. Nevertheless, the surface mapping of the field (performed using Zeeman Doppler Imaging) offers the best description of its geometry. Other measurements of this star's field are expressed in a variety of ways, including reporting the coefficients of a multi-harmonic fit to the longitudinal field curve (\citealt{shultz2018} obtain the following coefficients: $B_0$ = 0.0094 $\pm$ 0.0009 kG, $B_1$ = 0.049 $\pm$ 0.001 kG, $B_2$ = 0.032 $\pm$ 0.001 kG, $B_3$ = 0.019 $\pm$ 0.001 kG), and reporting the strength of the dipolar component of the field (\citealt{shultz32019} obtain $B_p$ = 0.31 $\pm$ 0.02 kG).

Finally, to extrapolate the magnetic field structure into the wind, \citet{donati2006} derive a \textit{source radius}\footnote{Similar to the Alfv\'{e}n radius for a dipolar field but applied to a more complex geometry, this is related to the radius of closure of the largest closed magnetic field loop.} of 2 $R_*$, based on their adopted value of $\dot{M} = 2 \times 10^{-8}$ M$_\odot$\ yr$^{-1}$. However, X-ray modelling using an arbitrary (non-dipolar) implementation of the Analytic Dynamical Magnetosphere model \citep{owocki2016} finds that a factor of 2 larger source radius (and hence a lower wind-feeding rate), leading to higher shock temperatures, is required to properly reproduce the features of $\tau$ Sco's X-ray spectrum \citep{fletcher2018, 2018PhDT........18F}.


\bsp	
\label{lastpage}
\end{document}